\documentclass[numbers=noendperiod]{scrartcl}

\usepackage[utf8]{inputenc}
\usepackage[T1]{fontenc}
\usepackage[margin=1.2cm]{geometry}
\usepackage[svgnames]{xcolor}
\usepackage{multicol,microtype,enumitem,lmodern}

\usepackage{mathtools,amssymb}

\usepackage[colorlinks=true,allcolors=DarkBlue]{hyperref}
\usepackage{cleveref}
\Crefname{equation}{Eq.}{Eqs.}
\crefname{section}{sec.}{secs.}
\crefname{appendix}{app.}{apps.}

\usepackage[backend=biber,sorting=none,citestyle=numeric-comp,giveninits=true,doi=false,date=year,url=false]{biblatex}
\renewbibmacro{in:}{}
\AtEveryBibitem{\clearfield{number}\clearfield{issue}\clearfield{note}}
\DeclareFieldFormat[article]{volume}{\mkbibbold{#1}}

\DeclareFieldFormat{pages}{\mkfirstpage{#1}}
\DeclareFieldFormat{eprint:arxiv}{\href{http://arxiv.org/abs/#1}{#1}}

\usepackage[titletoc,title]{appendix}
\makeatletter
\def\@hangfrom#1{\setbox\@tempboxa\hbox{{#1}}%
	\hangindent 0pt
	\noindent\box\@tempboxa}
\makeatother

\newcommand{\dif}{\mathrm{d}}
\DeclareMathOperator{\tr}{tr}
\DeclareMathOperator{\Tr}{Tr}
\DeclareMathOperator{\Det}{Det}
\DeclareMathOperator{\diag}{diag}
\DeclareMathOperator{\sign}{sign}
\newcommand{\tran}[1]{#1^\mathrm{T}}
\newcommand{\csqrt}[1]{\sqrt{\smash[b]{#1}}}
\newenvironment{alignedeqn}{\begin{equation}\begin{aligned}}{\end{aligned}\end{equation}\ignorespacesafterend}

\begin{document}

\title{Information transport in classical statistical systems}

\author{C. Wetterich\\
	\small\href{eqn:mailto:c.wetterich@thphys.uni-heidelberg.de}{c.wetterich@thphys.uni-heidelberg.de}}
\date{\small Institut für Theoretische Physik, Universität Heidelberg, Philosophenweg 16, 69120 Heidelberg, Germany}
\maketitle

\begin{abstract}
	\noindent For ``static memory materials'' the bulk properties depend on boundary conditions. Such materials can be realized by classical statistical systems which admit no unique equilibrium state. We describe the propagation of information from the boundary to the bulk by classical wave functions. The dependence of wave functions on the location of hypersurfaces in the bulk is governed by a linear evolution equation that can be viewed as a generalized Schrödinger equation. Classical wave functions obey the superposition principle, with local probabilities realized as bilinears of wave functions. For static memory materials the evolution within a subsector is unitary, as characteristic for the time evolution in quantum mechanics. The space-dependence in static memory materials can be used as an analogue representation of the time evolution in quantum mechanics - such materials are ``quantum simulators''. For example, an asymmetric Ising model on a Euclidean two-dimensional lattice represents the time evolution of free relativistic fermions in two-dimensional Minkowski space.
\end{abstract}

\BeforeStartingTOC[toc]{\begin{multicols}{2}}
\AfterStartingTOC[toc]{\end{multicols}}
{\hypersetup{hidelinks}\tableofcontents}

\begin{multicols}{2}

\section{Introduction}
\label{sec:introduction}

Memory and information transport are key issues in information technology. The study of statistical systems of Ising spins or bits has shaped the conceptual advances for the role of information \cite{SHA}. The understanding of computations with materials able to conserve memory \cite{LTP,CWEL,NP,ALL} may well influence future information processing. In this paper we propose a formalism for the problem of (static) information transport based on the concept of classical wave functions. It resembles the derivation of the wave function from the path integral in quantum mechanics by Feynman \cite{FEY}. However, our approach remains entirely rooted in classical statistics, describing classical Ising spins in thermal equilibrium or more elaborate static states in classical statistical systems.

We investigate probability distributions in classical statistics for systems with boundaries and a ``bulk'' limited by the boundaries. How is a change in the boundary conditions reflected by the observables within the bulk? This amounts to the question how a signal propagates from the boundary into regions within the bulk or how information is transported within the bulk. In turn, this issue is directly related to the question how information can be transported from one boundary to another, say between the two ends of a wire. We address these questions in a static context, for example thermal equilibrium, without any dependence on time. The absence of a genuine time evolution associates the bulk of such static states to a generalized notion of ``equilibrium state'', even for situations where the latter is not unique.

One finds a rather rich variety of different possible behaviors for the information transport. For the most common situation the boundary properties are not relevant for the bulk. Boundary information is lost within a finite correlation length, either monotonically or as damped oscillations. This situation is realized if the bulk equilibrium state is unique. A neighboring case is the power-like decay in case of an infinite correlation length, as for critical phenomena. Most interesting in our context are static memory materials for which expectation values of observables in the bulk depend on the boundary conditions. This becomes possible if no unique bulk equilibrium state exists, as in case of spontaneous symmetry breaking. Information can now be transported from the boundary to the bulk, imprinted on the properties of the degenerate ``equilibrium states.'' For example, we find models with oscillating local probabilities in the bulk, with details of the oscillations depending on the boundary conditions.

In general, local probabilities and therefore the expectation values of local observables $A(t)$ will depend on the location $t$ of some hypersurface in the bulk. For an Ising model with a finite correlation length $\xi$ the boundary information will be exponentially erased for $\Delta t = \min(t_\text{f} - t,t - t_\text{in})$ larger than $\xi$
\begin{equation}\label{eqn:I8-1}
	\langle A(t)\rangle
	= \bar A + c_A \exp(-\Delta t/\xi),
\end{equation}
with $\bar A$ the ``bulk expectation value'' or equilibrium value. At a phase transition $\xi$ diverges and $\langle A\rangle$ is approached with a power law in $\Delta t$. We will see that this loss of memory of boundary conditions is characteristic for all systems with a unique equilibrium state.

The loss of memory of boundary information is, however, not the only possibility. One may ask under which circumstances memory of boundary conditions is kept, for example by an oscillating behavior as
\begin{equation}\label{eqn:I6}
	\langle A(t)\rangle
	= a_0 \cos(\omega \, t + \alpha),
\end{equation}
with $\alpha$ depending on the boundary conditions. We find such a behavior for highly interesting ``static memory materials'' where bulk observables keep high sensitivity to boundary conditions. Static memory materials can be realized if no unique equilibrium state for the bulk exists. Degenerate generalized equilibrium states occur, in particular, in case of spontaneous symmetry breaking or in presence of conserved quantities.

A simple example for a static memory material is the asymmetric diagonal two-dimensional Ising model, with action
\begin{equation}\label{eqn:6XA}
	S_\text{cl}
	= -\frac\beta2\sum_{t,x}
	\bigl[s(t + \epsilon,x + \epsilon) \, s(t,x) - 1\bigr],
\end{equation}
taken in the limit $\beta \to \infty$. The points $(t,x)$ are on a quadratic lattice, and local observables at given $t$ correspond to Ising spins $s(x) = \pm 1$ or occupation numbers $n(x) = \bigl( s(x) + 1\bigr)/2 = (1,0)$. The interactions are only on one diagonal, and $\beta$ may be considered as a combination of interaction strength and inverse temperature. For suitable boundary conditions the expectation values $\langle s(t,x)\rangle$ at fixed $x$ oscillate with $t$, similar to \cref{eqn:I6}.

For $\beta \to \infty$ the asymmetric diagonal Ising model constitutes a simple cellular automaton \cite{CA,ICJ,TH,TH2,EL}. For the evolution from a hypersurface at $t$ to the next hypersurface at $t + \epsilon$ each spin up or particle at $x$ hops to the next side on the right at $t + \epsilon$, and each spin down or hole does the same. The system can be visualized as propagating fermions, with occupation numbers $n(t,x)$ either one or zero. Cellular automata may be realized by experimental setups or as computing architectures. Finite $\beta$ or the addition of other interactions may be considered as probabilistic perturbations to the deterministic cellular automata. Also a probabilistic distribution of boundary values turns this model into a genuine classical statistical system. Our general formalism embeds the deterministic cellular automata into a classical statistical probabilistic setting. We will further be interested in the conditions for the realization of general static memory materials that do not need to be deterministic cellular automata.

Our investigation addresses a particular form of ``equilibrium signal transport'' for which the usual dynamics of a system plays no role. In particular, the signal is available ``simultaneously'' at all locations $t$. This could be an interesting aspect for numerical computations, in particular since reading out the information at $t$ by measuring some local observable $A(t)$ would influence the probability distribution, similar to quantum mechanics. An experimental realization of static memory materials may use recent advances in spin based information technology \cite{KWC,LBL,MMW}. In our context it is important that the system obeys classical statistics, at least to a good approximation. This requires the control of the quantum fluctuations in the real world \cite{BRH}.

It is the aim of our paper to develop a general formalism for the problem of information transport. It is based on the central concept of classical wave functions. While Feynman's path integral for quantum mechanics involves a complex weight factor, our investigation concerns a real positive weight factor $\sim \exp(-S_\text{cl})$. We therefore expect analogies, but also important differences as compared to quantum mechanics. Similar to quantum mechanics the classical wave functions obey a linear evolution law realizing the superposition principle. Probabilities are bilinears in the wave functions, such that interference is, in principle, possible. As in quantum mechanics the expectation values of observables involve bilinears of the wave functions. The main differences to quantum mechanics are twofold. Instead of the complex wave function in quantum mechanics the classical statistical setting involves two different real wave functions. While their evolution is linear, it is, in general, not unitary. The norm of the wave functions can change. Only for particular static memory materials the evolution becomes unitary and the two wave functions can be identified. In this situation the system obeys all axioms of quantum mechanics.

The transfer matrix formalism \cite{TM,MS,FU} introduces the notion of non-commuting operators into classical statistics. In a sense it can be viewed as the analog of the Heisenberg picture in quantum mechanics. Our approach supplements a Schrödinger picture for classical statistics by implementing classical wave functions. Their derivation from the partition function or functional integral shows many analogies to Feynman's derivation \cite{FEY} of the quantum wave function from the path integral.

We will demonstrate our general considerations with several examples for which the boundary value problem can be solved explicitly. The first is the well-known one-dimensional Ising model \cite{IM,IS} in a homogeneous magnetic field. This model serves as a demonstration how the concept of a classical wave function can be used in practice. We recover the known results of the exact solution \cite{On} directly from the evolution of the classical wave function. The expectation values of local spins for arbitrary boundary conditions are described explicitly.

As a second explicit solution we consider the ``four-state oscillator chain''. This is again a one-dimensional Ising type model, now with two species of spins $s_1$ and $s_2$. For a particular choice of next-neighbor interactions this system shows oscillatory behavior of expectation values of local observables. These oscillations are damped, however, since they occur in a sector for which the eigenvalues of the step evolution operator obey $|\lambda| < 1$. Only for a particular limit this system becomes a memory material with all four eigenvalues of $S$ obeying $|\lambda| = 1$. It is then a ``unique jump chain'', corresponding to a simple cellular automaton \cite{CA,ICJ,TH,TH2,EL}.

Our third example is the asymmetric diagonal two-dimensional Ising model \labelcref{eqn:6XA}, as well as neighboring models. For $\beta \to \infty$ we find the explicit solution of the boundary value problem in terms of the free propagation of relativistic Weyl, Majorana, Majorana-Weyl or Dirac fermions in two dimensions. The limit $\beta \to \infty$ of this model can again be associated with cellular automata. This model describes a genuine memory material and we briefly discuss a possible realization in experiments or by numerical simulation. This model constitutes an existence proof for static memory materials within the setting of classical statistical systems. It is a quantum simulator with non-trivial Hamiltonian and associated oscillations of local probabilities and expectation values of local observables.

The aim of the present paper is mainly on the conceptual side. We only briefly discuss some aspects of an experimental realization of static memory materials, or of computer realizations of such materials. It is our hope that the concepts developed here are helpful for a future practical realization of static memory materials, and that such objects will reveal interesting new aspects for information processing.

This paper is organized as follows. In \cref{sec:functional integral} the wave function $\tilde q(t)$ and conjugate wave function $\bar q(t)$ are defined as suitable ``functional integrals'' over variables at $t^\prime < t$ or $t^\prime > t$, respectively. Thus $\tilde q(t)$ depends on the ``initial boundary factor'' $f_\text{in}$, while $\bar q(t)$ involves the ``final boundary factor'' $\bar f_\text{f}$. The local probability distribution $p(t)$ can be expressed as a bilinear in $\tilde q(t)$ and $\bar q(t)$. In \cref{sec:evolution} we discuss the evolution $(t$-dependence) of $\tilde q(t)$ and $\bar q(t)$, and therefore of $p(t)$. The explicit solution for the classical wave function $\tilde q(t)$ of the one-dimensional Ising model in a homogeneous magnetic field contains the full information about the equilibrium properties in the bulk, as well as the quantitative approach to equilibrium as one moves from the boundary to the bulk. We show that the monotonic information loss in the Ising model is not the only possible behavior. This is done by the explicit construction of classical probability distributions for which the local probability distribution $p(t)$ oscillates with $t$. The evolution of quantum mechanics is found if the step evolution operator $S$ is a rotation. This generalizes to subsectors of $S$ acting on a suitable subspace of wave functions. In \cref{sec:generalized schrödinger equation} we discuss the generalized Schrödinger equation. This linear differential evolution equation can be obtained if a suitable continuum limit $\epsilon \to 0$ can be realized. We also introduce the density matrix for mixed state boundary conditions. We summarize the basic concepts and the relation to the quantum formalism in \cref{sec:basic concepts}. To gain a first impression of the content of the formalism the reader could start with this section.

\Cref{sec:simple static memory materials} is devoted to the discussion of simple static memory materials. We first discuss ``unique jump chains'' with properties similar to cellular automata. This is extended in \cref{sec:ising models} to Ising type models describing the propagation of free fermions in one space and one time dimension. They realize oscillating local probabilities and expectation values of the type of \cref{eqn:I6}. We briefly discuss the possibility of an experimental realization. In \cref{sec:general static memory materials} we comment on the general conditions for memory materials. For memory materials the wave function, or a subsector of the density matrix, follows a quantum evolution. We discuss some conceptual implications of our results in the concluding \cref{sec:conclusions}, in particular a possible impact on the foundations of quantum mechanics and ideas of an emergent time \cite{Wi,Wh,CWPT}.

\section{Functional integral for occupation numbers and classical wave functions}
\label{sec:functional integral}

In \crefrange{sec:functional integral}{sec:generalized schrödinger equation} we discuss the basic formalism for our discussion of information transport in static classical statistical systems. Since the formalism of classical wave functions is unfamiliar to most readers we proceed in detail. An overview of the formalism is provided in \cref{sec:basic concepts}, to which a reader mainly interested in examples for static memory materials may jump. We employ a discrete setting such that all quantities are mathematically well defined as long as the number of degrees of freedom remains finite. The continuum limit of an infinite number of degrees of freedom does not pose a particular problem for the investigations of this paper.

\subsection{Spin chains}

In order to be specific we consider a one-dimensional wire with locations in the wire labeled by a discrete position variable $t$, and endpoints at both sides denoted by $t_\text{in}$ and $t_\text{f},t_\text{in}\leq t\leq t_\text{f}$. For the particular set of observables used to define the probability distribution we take here as ``variables'' the occupation numbers $n_\gamma(t)$. They can take the values zero or one, implying the relation 
\begin{equation}\label{eqn:73}
	n_\gamma^2
	= n_\gamma.
\end{equation}
The occupation numbers could also express bits of information. They can be directly related to Ising spins $s_\gamma = 2 n_\gamma - 1$ that take the values $\pm 1$. For our purposes it will be more convenient to work directly with the occupation numbers, but is is evident that a configuration sum or functional integral for occupation numbers can be translated directly to a functional integral for Ising spins. In this sense we discuss here generalized Ising models. In analogy to the Ising model we mean by ``functional integral'' a sum over distributions of discrete numbers. For a finite number of degrees of freedom this is a finite sum. All expressions are therefore regularized.

The index $\gamma$ labels different species of occupation numbers at a given $t$. It could comprise location or momentum labels. Discrete variables with more than two values can be constructed by combining several $n_\gamma$. For example, integer values from zero to three obtain as binary code from two occupation numbers. The generalization to infinitely many discrete values or continuous values is straightforward. In this sense our setting will correspond to the most general quasilocal classical statistical system. A ``local configuration'' or set of occupation numbers $\bigl[n_\gamma(t)\bigr]$ at a given $t$ can also be expressed in a fermionic language. For $n_\gamma = 1$ a fermion of species $\gamma$ is present, while for $n_\gamma = 0$ it is absent.

The overall classical statistical probability distribution $p\bigl(\{n\}\bigr)$ is specified by associating to every sequence of values of the occupation numbers $n_\gamma(t)$, e.g. $\{n\} = \bigl\{n_{\gamma_1}(t_1),n_{\gamma_2}(t_1),\dots n_{\gamma_1}(t_2),\dots\bigr\}$, a probability,
\begin{equation}\label{eqn:I1}
	p\bigl(\{n\}\bigr)\geq 0,
	\qquad
	\sum_{\{n\}}p\bigl(\{n\}\bigr)
	= 1.
\end{equation}
We investigate ``quasi-local'' probability distributions that can be written in the form
\begin{equation}\label{eqn:I2a}
	p\bigl(\{n\}\bigr)
	= \bar f_\text{f}\prod_t \mathcal{K} (t) f_\text{in}.
\end{equation}
Here $\mathcal{K}(t)$ involves only variables $n_\gamma(t^\prime)$ with $t^\prime$ in the neighborhood of $t$, while the boundary terms $f_\text{in}$ and $\bar f_\text{f}$ involve only $n(t^\prime)$ in the vicinity of $t_\text{in}$ or $t_\text{f}$, respectively. This is a generic class of statistical systems. It comprises Ising models and all statistical systems with local interactions. Mostly we consider only interactions between neighboring $t$-layers, with $\mathcal{K}(t)$ depending on $n_\gamma(t)$ and $n_\gamma(t + \epsilon)$. Also $f_\text{in}$ will depend only on $n_\gamma(t_\text{in})$, and $\bar f_\text{f}$ on $n_\gamma(t_\text{f})$.

This setting is easily generalized to multi-dimensional systems. In this case $t$ labels a sequence of hypersurfaces, and the index $\gamma$ comprises a label for the location within the particular hypersurface at a given $t$. For the two-dimensional Ising models the observables $n_\gamma(t)$ may be associated with Ising spins $s(t,x) = \pm 1$, and $\mathcal{K}(t)$ contains the information on next-neighbor or diagonal interactions.

We employ the label $t$ for the different hypersurfaces in analogy with time, but we stress again that we have not introduced any a priori concept of time or dynamics. The wire is seen as a timeless or static object. Its description is given by a static probability distribution depending on boundary conditions. Thus our systems do not represent dynamical systems with time dependent boundary conditions as studied in the wide field of signal transmission. Nevertheless, the dependence of expectation values of observables on $t$ shares many features with a time evolution. We therefore associate sometimes $t$ with an emergent time and use the corresponding language.

The boundary conditions can be collected in a ``boundary matrix''
\begin{equation}\label{eqn:I3}
	b
	= f_\text{in} \, \bar f_\text{f}.
\end{equation}
This may be extended to a statistical distribution of boundary conditions
\begin{equation}\label{eqn:I2}
	b
	= \sum_\alpha w_\alpha f^{(\alpha)}_\text{in}\bar f_\text{f}^{(\alpha)},
\end{equation}
with $w_\alpha$ appropriate weights.

General local observables $A(t)$ can be constructed from $n_\gamma(t)$. In our context the central question of information transport asks how the expectation value of a local observable in the bulk, $\langle A(t)\rangle$, responds to a change in the boundary matrix $b$. The question of ``equilibrium signal transport'' from one boundary to another can be explored as a special case. The boundary matrix $b$ is varied only by varying the initial condition $f_\text{in}$ at one end (``initial time''). One then investigates observables $A(t_\text{f})$ at the other end (``final time'') in their response to the variation of $b$. For the asymmetric diagonal Ising model \labelcref{eqn:6XA} the initial signal is completely transmitted to the final boundary.

We start by introducing the concepts of a local probability distribution and classical wave function, and discuss subsequently how the local probability distribution and wave functions are obtained from the ``overall probability distribution'' $p\bigl(\{n\}\bigr)$. We employ a normalized $p\bigl(\{n\}\bigr)$, corresponding to a partition function $Z = 1$. The partition function is expressed as a product of step evolution operators, multiplied by boundary terms. Splitting the functional integral for $Z$ into parts with $t^\prime < t$ and $t^\prime > t$ introduces the concept of the classical wave function $\tilde q(t)$ and conjugate wave function $\bar q(t)$.

\subsection{Local probabilities and wave functions}

We begin with the simple case where $\gamma$ takes only two values, $n_1$ and $n_2$. The species can be associated with spin up and spin down of a fermion. There are four different local states at a given $t$ that we label by $\tau = 1\dots 4$. For $\tau = 1$ two fermions are present, $n_1 = 1$, $n_2 = 1$. For $\tau = 2$ only one fermion with spin down is present, $n_1 = 0$, $n_2 = 1$, while for $\tau = 3$ one has only a spin up fermion, $n_1 = 1$, $n_2 = 0$. For $\tau = 4$ no fermion is present, $n_1 = n_2 = 0$. Local probabilities $p_\tau(t)$ for the four states obey the usual requirements, $p_\tau(t) \geq 0$, $\sum_\tau p_\tau(t) = 1$. At a given $t$ the expectation values of local occupation numbers or local observables $A(t)$ follow the standard classical statistical rule
\begin{equation}\label{eqn:1A}
	\langle A(t)\rangle
	= \sum_\tau p_\tau(t)A_\tau(t),
\end{equation}
with $A_\tau(t)$ the value of the observable in the state $\tau$.

A general local observable $A(t)$ can be expressed as a linear combination of four basis observables constructed from products of occupation numbers, as $1$, $n_1$, $n_2$, $n_1 n_2$. For a given $t$ the expectation values are given by 
\begin{equation}\label{eqn:97A1}
	\langle n_1\rangle
	= p_1 + p_3,
	\quad
	\langle n_2\rangle
	= p_1 + p_2,
	\quad
	\langle n_1n_2\rangle
	= p_1.
\end{equation}
We will discuss later how the local probabilities $p_\tau(t)$ can be computed from the overall probability distribution $p\bigl(\{n\}\bigr)$ which depends on the sequence of occupation numbers $\{n\}$ for all $t$.

We will express the set of local probabilities $[p_\tau(t)]$ in terms of a ``classical wave function'' $[\tilde q_\tau(t)]$ and ``conjugate classical wave function'' $[\bar q_\tau(t)]$ as (no sum over $\tau$ here)
\begin{equation}\label{eqn:97A2}
	p_\tau(t)
	= \bar q_\tau(t) \tilde q_\tau(t).
\end{equation}
In the next section we will formulate the $t$-dependence of the local probabilities $p_\tau(t)$ in terms of the wave function and its conjugate. The four real numbers $\tilde q_\tau$ can be seen as the coefficients of an abstract wave function in the occupation number basis, and similar for the conjugate wave function.

As local basis observables we employ four basis functions $h_\tau(n)$ given by
\begin{alignedeqn}\label{eqn:74}
	&h_1
	= n_1 \, n_2,
	\quad
	&&h_2
	= (1 - n_1)n_2,\\
	&h_3
	= n_1(1 - n_2),
	\quad
	&&h_4
	= (1 - n_1) (1 - n_2).
\end{alignedeqn}
In this short notation we have suppressed the $t$-argument. It will always be understood that local basis observables $h_\tau(t)$ depend on occupation numbers $n_\gamma(t)$. The local basis observables are eigenfunctions of occupation numbers with eigenvalues one or zero, in the sense
\begin{alignedeqn}\label{eqn:75}
	&n_1 \, h_1
	= h_1,
	\quad
	n_1 \, h_2
	= 0,
	\quad
	n_1 \, h_3
	= h_3,
	\quad
	n_1 \, h_4
	= 0,\\
	&n_2 \, h_1
	= h_1,
	\quad
	n_2 \, h_2
	= h_2,
	\quad
	n_2 \, h_3
	= 0,
	\quad
	n_2 \, h_4
	= 0.
\end{alignedeqn}

With $n_\gamma (1 - n_\gamma) = 0$, $(1 - n_\gamma) (1 - n_\gamma) = (1 - n_\gamma)$, one easily verifies three central relations: the first two are the product rule
\begin{equation}\label{eqn:76}
	h_\tau \, h_\rho
	= h_\tau \, \delta_{\tau\rho},
\end{equation}
and the integration rule
\begin{equation}\label{eqn:76A}
	\int \dif n \, h_\tau
	= \prod_\gamma \smashoperator[r]{\sum_{n_\gamma=0,1}} h_\tau
	= 1,
\end{equation}
while the third is the sum rule stating that the sum of the basis functions obeys
\begin{equation}\label{eqn:77A}
	\sum_\tau h_\tau
	= 1.
\end{equation}

The wave function $f(t)$ and the conjugate wave function $\bar f(t)$ are defined as
\begin{equation}\label{eqn:102A1}
	f(t)
	= \tilde q_\tau(t) h_\tau(t),
	\quad
	\bar f(t)
	= \bar q_\tau(t) h_\tau(t).
\end{equation}
(Sums over repeated indices are implied unless stated otherwise.) Similarly, the local observables $A(t)$ can be written as
\begin{equation}\label{eqn:18YA}
	A(t)
	= A_\tau(t) \, h_\tau(t).
\end{equation}
In particular, the observable for the local occupation number $N_\gamma(t)$ is simply given by $n_\gamma(t)$. With \cref{eqn:97A2} we can write the expectation value \labelcref{eqn:1A} as
\begin{equation}\label{eqn:18YB}
	\langle A\rangle
	= \int \dif n \, \bar fAf,
\end{equation}
where $\int \dif n$ denotes the sum over local configurations according to \cref{eqn:76A}

We can generalize the notion of local observables by introducing operators $A^\prime(t)$ as matrices with elements $A^\prime_{\tau\rho}(t)$. Their expectation value is expressed by the ``quantum rule''
\begin{equation}\label{eqn:102A2}
	\langle A\rangle
	= \bar q_\tau \, A^\prime_{\tau\rho} \, \tilde q_\rho
	= \langle \bar q \, A^\prime\tilde \, q\rangle.
\end{equation}
Local observables that are constructed as sums of products of occupation numbers are represented by diagonal operators
\begin{equation}\label{eqn:19YA}
	A^\prime_{\tau\rho}
	= A_\tau\delta_{\tau\rho}.
\end{equation}
For our simple setting the operators $A^\prime$ are $4 \times 4$-matrices with elements $A^\prime_{\tau\rho}$. Occupation numbers correspond to diagonal operators
\begin{equation}\label{eqn:B5}
	N^\prime_1
	= \diag(1,0,1,0),
	\quad
	N^\prime_2
	= \diag(1,1,0,0).
\end{equation}

For diagonal operators the equivalence of the expressions in \cref{eqn:102A2} with \cref{eqn:1A,eqn:18YB} is easily verified by using the properties of the basis functions and the definition \labelcref{eqn:97A2}. The physical interpretation of observables represented by off-diagonal operators is discussed in ref.~\cite{CWQFCS}.

Our setting can be extended to an arbitrary number of species $\gamma \in \{1,\dots,M\}$. The sequences of occupation numbers $n_\gamma \in \{0,1\}$ describe now $N = 2^M$ local states. The corresponding basis functions $h_\tau$, $\tau \in \{1,\dots,2^M\}$, are defined by the possible products of $M$ factors $n_\gamma$ or $1 - n_\gamma$, in analogy to \cref{eqn:74}. If we label $\tau$ by a sequence of occupation numbers, say $(0,1,1,0,1\dots)$, the corresponding basis function $h_\tau = (1 - n_1) \, n_2 \, n_3 \, (1 - n_4) \, n_5\dots$ obtains by inserting $n$ for each value one, and $1 - n$ for each value zero. The basis functions obey 
\begin{equation}\label{eqn:B6}
	n_\gamma h_\tau
	= (n_\gamma)_\tau h_\tau,
\end{equation}
where $(n_\gamma)_\tau$ ``reads out'' the corresponding value zero or one in the sequence $\tau$. The three central relations \labelcref{eqn:76,eqn:76A,eqn:77A} remain valid for arbitrary $M$.

\subsection{Overall probability distribution}

The overall probability distribution $p\bigl(\{n\}\bigr)$ associates to each sequence of occupation numbers $\{n\} = \bigl\{n_\gamma(t)\bigr\}$ a probability. We consider a discrete set of equidistant locations $t \in \{t_\text{in},t_\text{in} + \epsilon,t_\text{in} + 2\epsilon,\dots,t_\text{f}\}$. (The continuum limit $\epsilon \to 0$ can be taken, if appropriate, at the end.) A sequence of occupation numbers specifies first all $n_\gamma(t_\text{in})$, then all $n_\gamma(t_\text{in} + \epsilon)$ and so on. If there is only one species ($M = 1$), the sequences are the configurations of an Ising chain. For an arbitrary number of species the sequences are the configurations of generalized Ising models. For the formal part of our discussion we generalize to an arbitrary weight function $w\bigl(\{n\}\bigr)$ which is not necessarily positive. For the classical statistical systems of interest $w\bigl(\{n\}\bigr)$ will then be restricted to be positive semidefinite.

A general observable depends on the sequence of occupation numbers $\bigl\{n_\gamma(t)\bigr\}$. Expectation values obey the standard statistical rule
\begin{equation}\label{eqn:12B}
	\langle A\rangle
	= \sum_{\{n\}} w[n]A[n].
\end{equation}
We switch here to the standard ``functional integral'' notation, $w[n] = w\bigl(\{n\}\bigr)$, $A[n] = A\bigl(\{n\}\bigr)$. (The naming ``functional integral'' corresponds to the continuum limit $\epsilon \to 0$ where $n_\gamma(t)$ become discrete functions of $t$. For finite $\epsilon$ the functional integral is simply a finite configuration sum which can be considered as an appropriate regularization of the functional integral.) The symbol $\sum_{\{n\}}$ denotes the sum over all configurations of occupation numbers at arbitrary $t$. We will equivalently use the symbol for functional integration,
\begin{equation}\label{eqn:79}
	\int \mathcal{D}n
	= \sum_{\{n\}}
	= \prod_t \prod_\gamma \smashoperator[r]{\sum_{n_\gamma(t)=0,1}}
	= \prod_t \int \dif n(t).
\end{equation}
\Cref{eqn:12B} assumes that the weight distribution is normalized,
\begin{equation}\label{eqn:37AA}
	\int \mathcal{D} n \, w[n]
	= 1.
\end{equation}

Boundary conditions at $t_\text{in}$ and $t_\text{f}$ can be implemented by weight distributions of the form
\begin{equation}\label{eqn:12A}
	w[n]
	= \bar f_\text{f}\bigl(n(t_\text{f})\bigr) \, K[n] \, f_\text{in}\bigl(n(t_\text{in})\bigr),
\end{equation}
where $K[n]$ is independent of the boundary conditions. The ``initial boundary term'' $f_\text{in}$ only depends on the occupation numbers at $t_\text{in}$, e.g. $n_\gamma(t_\text{in})$, while the ``final boundary term'' $\bar f_\text{f}$ involves only occupation numbers at $t_\text{f}$. In this paper we discuss quasi-local probabilities
\begin{equation}\label{eqn:21A}
	K[n]
	= \prod_t\mathcal{K}(t).
\end{equation}
We will concentrate on next-neighbor interactions where
\begin{equation}\label{eqn:80}
	K[n]
	= \prod_t\mathcal{K}\bigl(n(t + \epsilon),n(t)\bigr)
	= \prod_t\mathcal{K}(t).
\end{equation}
The product is over all time points $t_\text{in}\leq t\leq t_\text{f} - \epsilon$. The factor $\mathcal{K}(t)$ depends on two sets of neighboring occupation numbers $n_\gamma(t)$ and $n_\gamma(t + \epsilon)$. Extensions beyond next-neighbor interactions are discussed in \cref{sec:beyond next-neighbor interactions}. They do not lead to new structural elements.

Local observables $A(t)$ depend only on local occupation numbers $n_\gamma(t)$ at a given $t$. Their expectation value is given by the local probability distribution $p(t) = p\bigl(n_\gamma(t)\bigr) = p_\tau(t) \, h_\tau(t) = p_\tau(t) \, h_\tau\bigl(n(t)\bigr)$ according to \cref{eqn:1A}. With \cref{eqn:12B} the local probability distribution obtains from the overall weight distribution as a configuration sum of occupation numbers at $t^\prime$ different from $t$
\begin{equation}\label{eqn:12C}
	p(t)
	= p\bigl([n_\gamma(t)]\bigr)
	= \prod_{t^\prime \neq t} \prod_\gamma \smashoperator[r]{\sum_{n_\gamma(t^\prime)=0,1}}
	w[n].
\end{equation}

It will be our aim to express the $t$-dependence or evolution of the local probabilities $p_\tau(t)$ by an evolution law for the time dependent wave function $f(t)$ and conjugate wave function $\bar f(t)$ as given by \cref{eqn:102A1}. From the evolution of the wave functions we can compute directly the evolution of expectation values of local observables $\langle A(t)\rangle$ according to \cref{eqn:18YB}, and thereby investigate the propagation of information from the boundary into the bulk.

\subsection{Partition function and step evolution operator}

The partition function $Z$ can be written as a functional integral 
\begin{equation}\label{eqn:78}
	Z
	= \sum_{\{n\}} \bar f_\text{f}\bigl(n(t_\text{f})\bigr) \, K[n] \, f_\text{in}\bigl(n(t_\text{in})\bigr).
\end{equation}
By the definition of the overall weight distribution \labelcref{eqn:12A} it equals one,
\begin{equation}\label{eqn:14A}
	Z
	= \int\mathcal{D}n~ w[n]
	= 1.
\end{equation}
The real initial and final wave functions $\bar f_\text{f}$ and $f_\text{in}$ depend only on $n_\gamma(t_\text{f})$ and $n_\gamma(t_\text{in})$, respectively. They associate weights to different initial and final boundary conditions. Their normalization should be compatible with \cref{eqn:14A}. In \cref{eqn:78} the factor, $K[n]$ takes a real value for each configuration $\{n\}$.

Consider now the quasi-local probability distribution \labelcref{eqn:80} with next-neighbor interactions. Arbitrary $\mathcal{K}(t)$ can be expanded as
\begin{equation}\label{eqn:81}
	\mathcal{K}(t)
	= S_{\tau\rho}(t) \, h_\tau(t + \epsilon) \, h_\rho(t),
\end{equation}
with basis functions $h_\tau(t + \epsilon)$ and $h_\rho(t)$ given in terms of $n_\gamma(t + \epsilon)$ and $n_\gamma(t)$, respectively. Due to the identity \labelcref{eqn:76} \cref{eqn:81} is indeed the most general function of $n_\gamma(t + \epsilon)$ and $n_\gamma(t)$. The ``step evolution operator'' $S$, with matrix elements $S_{\tau\rho}$, is a central quantity for our discussion. We only consider invertible $S$ that can be diagonalized by complex regular matrices. The multiplicative normalization of $\mathcal{K}(t)$ is chosen such that the largest absolute value of the eigenvalues of $S$ equals one.

For $t$-independent $S$ we observe the identity 
\begin{equation}\label{eqn:BBA}
	\int \dif n(t + \epsilon) \, \mathcal{K}(t + \epsilon) \, \mathcal{K}(t)
	= h_\tau(t + 2\epsilon) \, (S^2)_{\tau\rho} \, h_\rho(t)
\end{equation}
where $S^2$ denotes the matrix multiplication
\begin{equation}\label{eqn:31A}
	(S^2)_{\tau\rho}
	= S_{\tau\sigma}S_{\sigma\rho}.
\end{equation}
This generalizes to chains of $\mathcal{K}$-factors. The matrix multiplication property allows us to write
\begin{alignedeqn}\label{eqn:31B}
	Z
	= \int \dif n(t_\text{f}) \dif n(t_\text{in}) \, &\bar f_\text{f}\bigl( n(t_\text{f})\bigr) \, h_\tau\bigl(n(t_\text{f})\bigr) \,(S^G)_{\tau\rho}\\
	&\times h_\rho\bigl(n(t_\text{in})\bigr) \, f_\text{in}\bigl(n (t_\text{in})\bigr)
\end{alignedeqn}
or
\begin{equation}\label{eqn:BBB}
	Z
	= \bar q_\tau(t_\text{f}) \, (S^G)_{\tau\rho} \, \tilde q_\rho(t_\text{in}),
\end{equation}
with $G = (t_\text{f} - t_\text{in})/\epsilon$ the number of points between $t_\text{f}$ and $t_\text{in}$ and $S^G$ the matrix product of $G$ factors $S$.

\Cref{eqn:BBB} shows the close relation of the step evolution operator $S$ with the transfer matrix. What is particular is the normalization. We employ a normalized partition function $Z = 1$. \Cref{eqn:BBB} still leaves some freedom in the relative normalization of $S$, $\tilde q(t_\text{in})$ and $\bar q(t_\text{f})$. We will choose a normalization where the largest absolute value among the eigenvalues of $S$ is normalized to one. If $S$ depends on $t$ one replaces $S^G$ by the ordered product of matrices $S(t_i)$, with $t_\text{f} - \epsilon$ to the left and $t_\text{in}$ to the right.

\subsection{Wave function from functional integral}

The wave function $f(t)$ and conjugate wave function $\bar f(t)$ correspond to partial functional integrations of the overall weight distribution $w[n]$. Here $f(t)$ involves an integration over variables $n(t^\prime)$ with $t^\prime < t$, while $\bar f(t)$ results from a similar integration of $n(t^\prime > t)$. For this purpose we split
\begin{alignedeqn}\label{eqn:82}
	K[n]
	&= K_>(t) \, K_<(t),\\
	K_<(t)
	&= \prod_{t^\prime=t_\text{in}}^{t - \epsilon} \mathcal{K}(t^\prime),\quad
	K_>(t)
	= 
	\prod_{t^\prime=t}^{t_\text{f} - \epsilon} \mathcal{K}(t^\prime).
\end{alignedeqn}
The wave functions are defined as
\begin{alignedeqn}\label{eqn:83}
	f(t)
	&= \int \mathcal{D}n(t^\prime < t) \, K_<(t) \, f_\text{in}(t_\text{in}),\\
	\bar f(t)
	&= \int \mathcal{D}n(t^\prime > t) \, \bar f_\text{f}(t_\text{f}) \, K_>(t).
\end{alignedeqn}
The integral $\int \mathcal{D}n(t^\prime < t)$ restricts the product in \cref{eqn:79} to $t_\text{in} \leq t^\prime \leq t - \epsilon$, such that $f(t)$ is a function of $n_\gamma(t)$. Similarly, $\int\mathcal{D}n(t^\prime > t)$ involves $t + \epsilon\leq t^\prime\leq t_\text{f}$ and $\bar f(t)$ also depends on $n_\gamma(t)$. These definitions allow us to express $Z$ in terms of the wave function and conjugate wave function,
\begin{equation}\label{eqn:84}
	Z
	= \int \dif n(t) \, \bar f(t) \, f(t)
	= \bar q_\tau(t) \, \tilde q_\tau(t).
\end{equation}
By suitable normalizations of $f_\text{in},\bar f_\text{f}$ and $\mathcal{K}(t)$ we can always implement $Z = 1$.

Inserting the definitions \labelcref{eqn:82,eqn:83} into \cref{eqn:12C} expresses the local probabilities in terms of the wave functions
\begin{equation}\label{eqn:SIB}
	p(t)
	= \bar f(t) \, f(t).
\end{equation}
Expanding in basis functions according to \cref{eqn:102A1},	
\begin{alignedeqn}\label{eqn:SIC}
	p(t)
	&= p_\tau(t) \, h_\tau(t)
	= \bar q_\rho(t) \, h_\rho(t) \, \tilde q_\tau(t) \, h_\tau(t)\\
	&= \sum_\tau \bar q_\tau(t) \, \tilde q_\tau(t) \, h_\tau(t),
\end{alignedeqn}
yields \cref{eqn:97A2}. This identifies the wave functions in \cref{eqn:97A2} with the ones constructed from \cref{eqn:83}. For diagonal local observables \labelcref{eqn:19YA} the expectation values therefore obey \cref{eqn:102A2}
\begin{equation}\label{eqn:SID}
	\langle A(t)\rangle
	= \bar q_\tau(t) \, A^\prime_{\tau\rho}(t) \, \tilde q_\rho(t).
\end{equation}
Knowledge of the wave functions is sufficient for the determination of $\langle A(t)\rangle$, while all additional information contained in $w[n]$ is not relevant.

\subsection{Positivity of overall probability distribution}

A local probabilistic setting \labelcref{eqn:97A2} is realized if the product $\bar q_\tau(t)\tilde q_\tau(t)$ is positive semidefinite for all $t$, and for each $\tau$ individually. While this condition is sufficient for a well defined local probability distribution, it is necessary but not sufficient for a well defined overall probability distribution. The latter requires positivity for all $w[n] = p[n]$.

The overall probability distribution \labelcref{eqn:I2a} can be represented as a product of elements of step evolution operators. We use the explicit representation
\begin{alignedeqn}\label{eqn:28A}
	w[n]
	= \bar q_\tau(t_\text{f})h_\tau(t_\text{f})~\mathcal{K}(t_\text{f} - \epsilon)\dots~\mathcal{K}(t_\text{in})\tilde q_\rho(t_\text{in})h_\rho(t_\text{in}),\\
\end{alignedeqn}
where $f_\text{in} = \tilde q_\rho(t_\text{in}) \, h_\rho(t_\text{in})$ is the initial wave function and $\bar f_\text{f} = \bar q_\tau(t_\text{f})h_\tau(t_\text{f})$ the final conjugate wave function. Insertion of \cref{eqn:81} and use of \cref{eqn:76} yields 
\begin{align}\label{eqn:28B}
	w[n]
	= \smashoperator{\sum_{\rho_1,\dots,\rho_{G + 1}}} &w_{\rho_1\rho_2\dots\rho_{G + 1}}\\
	&\times h_{\rho_1}(t_1) \, h_{\rho_2}(t_2) \dots
	h_{\rho_{G + 1}}(t_{G + 1}),\notag
\end{align}
with (no summations in the following expression)
\begin{alignedeqn}\label{eqn:28C}
	&w_{\rho_1\rho_2\dots \rho_{G + 1}}
	= \bar q_{\rho_{G + 1}}(t_{G + 1})S_{\rho_{G + 1}\rho_{G}}(t_{G})\\
	&\quad\times S_{\rho_{G}\rho_{G-1}}(t_{G-1})\dots~
	S_{\rho_3\rho_2}(t_2)S_{\rho_2\rho_1}(t_1)\tilde q_{\rho_1}(t_1).
\end{alignedeqn}
Here we have numbered the time arguments, $t_\text{in} = t_1$, $t_\text{in} + \epsilon = t_2,\dots~t_\text{f} = t_{G + 1}$.

We could specify $w[n]$ directly by the product of step evolution operators \labelcref{eqn:28B,eqn:28C}. However, arbitrary $S$ will not yield a positive overall probability distribution $p[n]$ obeying \cref{eqn:I1,eqn:I2a}. In general, \cref{eqn:28B,eqn:28C} define a weight distribution that can take negative values for certain sequences $\{n\}$. A classical probability distribution requires then that all weights are positive. In turn, this requires all coefficients $w_{\rho_1\rho_2\dots\rho_{G + 1}}$ to be positive semidefinite, since they can be identified with the probabilities to find a particular sequence of occupation numbers. If negative $w_{\rho_1\dots\rho_{G + 1}}$ occur, we deal with the weight function for some generalized statistical setting, but no longer with classical statistics. Positivity of $w$ can always be realized if all elements of the step evolution operator as well as $\tilde q_{\rho_1}(t_1)$ and $\bar q_{\rho_{G + 1}}(t_{G + 1})$ are positive or zero. This property is, however, not necessary. We do not necessarily require translation invariance such that the step evolution operators for different $t$ may differ. For example, there could be purely positive $S$ (all $S_{\rho\tau}\geq 0)$ and purely negative $S$ (all $S_{\rho\tau}\leq 0)$, with an even number of purely negative $S$.

For establishing a necessary condition for the positivity of $w$ we introduce the product (no sum over $\rho_m$)
\begin{equation}\label{eqn:28D}
	{B}_{\rho_m,\alpha\beta}(t_m)
	= S_{\alpha\rho_m}(t_m)S_{\rho_m\beta}(t_{m-1}).
\end{equation}
This factor appears in $w$, while the other factors do not depend on the index $\rho_m$ for $2\leq m\leq G$. If $B_{\rho_m,\alpha\beta}(t_m)$ has different signs for different $\rho_m$, with some given values for $\alpha,\beta$ and $t_m$, there are necessarily elements $w_{\rho_1\dots \rho_m\dots \rho_{G + 1}}$ that change sign as the index $\rho_m$ varies. Therefore some elements of $w$ must be negative and the weight factor is not a classical statistical probability distribution. A necessary condition for a classical statistical system requires that two consecutive matrix elements $S_{\alpha\rho_m}(t_m)$ and $S_{\rho_m\beta}(t_{m-1})$ either have the same sign for all $\rho_m$, or at least one element vanishes. This should hold independently of $\alpha,\beta$ and $t_m$.

For a more extended discussion of the question which type of step evolution operators are compatible with the notion of an overall classical statistical probability distribution $p[n]$ we refer to ref.~\cite{CWQFCS}. In the present paper we only discuss examples for which $S_{\tau\rho}(t)$, $\tilde q_\tau(t_\text{in})$ and $\bar q_\tau(t_\text{f})$ are all positive semidefinite.

\section{Evolution}
\label{sec:evolution}

The dependence of the local probabilities $p_\tau(t)$ on the location $t$ can be described by the $t$-dependence of the wave function $f(t)$ and the conjugate wave function $\bar f(t)$, (no sum over $\tau$ here)
\begin{equation}\label{eqn:22A}
	p_\tau(t)
	= \bar q_\tau(t) \, \tilde q_\tau(t)
	= \int \dif n(t) \, h_\tau(t) \, \bar f(t) \, f(t).
\end{equation}
We will often call the $t$-dependence of the wave function a ``time evolution'', even though it describes the dependence on some general location. The notion of evolution distinguishes the coordinate $t$, for which the evolution is considered, from other possible coordinates $\vec x$ in a multi-dimensional setting. It does not mean that $t$ and $\vec x$ are conceptually different.

\subsection{Evolution of wave functions}

The evolution of the wave function $f(t)$ follows directly from its definition \labelcref{eqn:83},
\begin{equation}\label{eqn:85}
	f(t + \epsilon)
	= \int \dif n(t)\mathcal{K}(t)f(t).
\end{equation}
Using the explicit expressions \labelcref{eqn:102A1,eqn:81} yields
\begin{alignedeqn}\label{eqn:86}
	f(t + \epsilon)
	&= \tilde q_\tau(t + \epsilon) \, h_\tau(t + \epsilon)\\
	&= \int \dif n(t) \, S_{\tau\rho}(t) \, h_\tau(t + \epsilon) \, h_\rho(t) \, \tilde q_\sigma(t) \, h_\sigma(t)\\
	&= \sum_{\tau,\rho} S_{\tau\rho}(t) \, \tilde q_\rho(t) \, h_\tau(t + \epsilon)\int \dif n(t)h_\rho(t)\\
	&= S_{\tau\rho}(t) \, \tilde q_\rho(t) \, h_\tau(t + \epsilon).
\end{alignedeqn}
Here we have used the relations \labelcref{eqn:76,eqn:76A}. \Cref{eqn:86} establishes the linear evolution law
\begin{equation}\label{eqn:87}
	\tilde q_\tau(t + \epsilon)
	= S_{\tau\rho}(t)\tilde q_\rho(t).
\end{equation}

Using analogous steps one obtains for the conjugate wave function the evolution 
\begin{equation}\label{eqn:89}
	\bar q_\tau(t)
	= \bar q_\rho((t + \epsilon))S_{\rho\tau}(t),
\end{equation}
or, for invertible matrices $S$,
\begin{equation}\label{eqn:90}
	\bar q_\tau(t + \epsilon)
	= \bar q_\rho(t)(S^{-1}(t))_{\rho\tau}
	= \bigl(\tran{S}(t)\bigr)^{-1}_{\tau\rho} \, \bar q_\rho(t).
\end{equation}
The time evolution of the classical wave function $\tilde q$ and conjugate wave function $\bar q$ is therefore described by the step evolution operator $S$, as encoded in the overall probability distribution by \cref{eqn:80,eqn:81}. The naming ``step evolution operator'' reflects the role of $S$ for the evolution in a minimal discrete time step.

\subsection{Classical Ising-spin systems}

For a classical Ising-spin system the factor $\mathcal{K}(t)$ can be expressed by an exponential
\begin{equation}\label{eqn:128B1}
	\mathcal{K}(t)
	= \exp\{-\mathcal{L}(t)\},
\end{equation}
where $\mathcal{L}(t)$ depends on $n_\gamma(t)$ and $n_\gamma(t + \epsilon)$. For 
\begin{equation}\label{eqn:35A}
	\bar f_\text{f}\bigl(n(t_\text{f})\bigr) \, f_\text{in}\bigl(n(t_\text{in})\bigr)\geq 0
\end{equation}
the product $\bar f_\text{f} K f_\text{in}$ is positive semidefinite. Therefore
\begin{equation}\label{eqn:35B}
	p[n]
	= \bar Z^{-1} \, \bar f_\text{f} \, K \, f_\text{in}
\end{equation}
is a normalized probability distribution. For general $\mathcal{L}(t)$ the partition function and the classical wave functions are not yet normalized. We therefore take $\bar Z > 0$ arbitrary, while we will later achieve $Z = 1$ by a suitable additive constant in $\mathcal{L}(t)$ and normalization of the boundary factors.

For the Ising model with general $\bar Z > 0$ one has the standard functional integral expression for expectation values of local observables
\begin{equation}\label{eqn:LP2}
	\langle A(t)\rangle = \bar Z^{-1} \int \mathcal{D}n \, \bar f_\text{f}(n(t_\text{f})) \, e^{-S_\text{cl}[n]} \, A(t) \, f_\text{in}\bigl(n(t_\text{in})\bigr),
\end{equation}
where we employ in \cref{eqn:78} the classical action $S_\text{cl}[n]$,
\begin{equation}\label{eqn:LP3}
	K[n] = \exp\{-S_\text{cl}[n]\}~,\quad~ S_\text{cl}[n]
	= \sum_t\mathcal{L}(t).
\end{equation}
(Recall that a local observable $A(t)$ is an arbitrary function of occupation numbers $n(t) = \bigl( s(t) + 1\bigr)/2$ at a given $t$.) Using the factorization \labelcref{eqn:82} one arrives at
\begin{equation}\label{eqn:LP4}
	\langle A(t)\rangle = \bar Z^{-1} \int \dif n(t) \bar f(t) A(t) f(t),
\end{equation}
with $\bar Z$ given by \cref{eqn:84}. We can express $A(t)$ as a linear combination of basis observables $h_\tau(t)$ according to \cref{eqn:18YA}. The expansion \labelcref{eqn:102A1} of $f(t)$ and $\bar f(t)$ in terms of basis functions defines the local probabilities as (sum over $\tau$ only if indicated)
\begin{equation}\label{eqn:LP5}
	p_\tau (t)
	= \bar Z^{-1} \bar q_\tau (t) \, \tilde q_\tau (t),
	\quad
	\bar Z = \sum_\tau \bar q_\tau(t) \, \tilde q_\tau (t),
\end{equation}
such that \cref{eqn:LP4} is consistent with \cref{eqn:18YB}.

We want to work with a normalization of the wave function where \cref{eqn:97A2} holds. By a suitable additive constant in $\mathcal{L}(t)$ and/or suitable multiplicative constants in $f_\text{in} \text{ and } \bar f_\text{f} $ we can always achieve $Z = 1$, such that $p_\tau = \bar q_\tau\tilde q_\tau$. We will adjust the additive constant in $\mathcal{L}(t)$ such that the largest eigenvalue of $S$ obeys $|\lambda| = 1$. Then only $f_\text{in}\bar f_\text{f}$ has to be normalized accordingly.

The most general action involving next-neighbor interactions and local terms can be written as
\begin{equation}\label{eqn:S10}
	\mathcal{L}(t) = M_{\tau\rho}(t) \, h_\tau(t + \epsilon) \, h_\rho(t).
\end{equation}
By use of \cref{eqn:76,eqn:77A} one finds
\begin{equation}\label{eqn:S11}
	\exp\{-\mathcal{L}(t) \} = \sum_{\tau,\rho} \exp\bigl(-M_{\tau\rho}(t)\bigr) \, h_\tau (t + \epsilon) \, h_\rho(t),
\end{equation}
and therefore the matrix elements of the step evolution operator 
\begin{equation}\label{eqn:S12}
	S_{\tau\rho} (t)
	= \exp\bigl(-M_{\tau\rho}(t)\bigr).
\end{equation}
The step evolution operator $S$ has only positive elements, $S_{\tau\rho}\geq 0$. In this case we call $S$ a ``positive matrix''. (In the present paper a positive matrix means that all matrix elements are positive or zero, in distinction to the positivity of all eigenvalues.)
The trivial evolution $S_{\tau\rho} = \delta_{\tau\rho}$ is found if the diagonal elements of $M$ vanish, $M_{\tau\tau} = 0$, while all off-diagonal elements diverge, $M_{\tau\rho} \to \infty$ for $\tau \neq \rho$. Our choice of basis functions allows for a straightforward and efficient computation of the step evolution operator or transfer matrix by \cref{eqn:S10,eqn:S12}.

All these statements apply equally for multicomponent Ising spins or occupation numbers $n_\gamma(t)$. The standard Ising-type models assume translation symmetry, with $M_{\tau\rho}$ and $S_{\tau\rho}$ independent of $t$. As an explicit example we next describe the one-dimensional Ising model in our formalism. A few generalized Ising-type models are discussed in \cref{app:generalized Ising models}.

\subsection{One-dimensional Ising model}

The one-dimensional Ising model in a homogeneous magnetic field is one of the best known exact solutions in statistical physics. It is therefore a good example to demonstrate our formalism based on classical wave functions explicitly. The known properties of the exact solution are recovered from the solution of the evolution equation for the wave function. Furthermore, the use of the occupation number basis for the computation of the transfer matrix and step evolution operator, as well as the solution of the model based on the behavior of classical wave functions, can be generalized to many other spin models.

The one-dimensional Ising model with next-neighbor interaction is described by a single spin $s(t) = \pm 1$ for each site $t$. The factor $\mathcal{K}(t)$ in \cref{eqn:81,eqn:128B1} reads
\begin{alignedeqn}\label{eqn:S1}
	\mathcal{K}(t) 
	&= \exp\Bigl\{\beta s(t + \epsilon) s(t) +\frac\gamma2\bigl( s(t + \epsilon) + s(t)\bigr)\Bigr\}\\
	&= \exp\Bigl\{\beta(2 n(t + \epsilon) - 1) (2 n(t) - 1)\Big \}\\
	&\hphantom{{}=}\times \exp\Bigl\{\gamma\bigl( n(t + \epsilon) + n(t) - 1\bigr)\Bigr\}\\
	&= \exp\Bigl\{\beta \bigl[h_1(t + \epsilon) - h_2(t + \epsilon)\bigr] \bigl[h_1(t) - h_2(t)\bigr]\Bigr\}\\
	&\times \exp\Bigl\{\tfrac\gamma2\bigl(h_1(t + \epsilon) - h_2(t + \epsilon) + h_1(t) - h_2(t)\bigr)\Bigr\}.
\end{alignedeqn}
Here we use a single occupation number $n(t) = (1 + s(t))/2$ and basis functions $h_1 = n$, $h_2 = 1 - n$. The constants $\beta$ and $\gamma$ are related to the next-neighbor coupling $J$ and the magnetic field $H$ by $\beta = J/(k_BT)$, $\gamma = H/(k_BT)$. We consider $\beta > 0$.

From the relation \labelcref{eqn:76} one infers
\begin{equation}\label{eqn:167A}
	(h_1 - h_2)^2
	= h_1 + h_2
	= 1
\end{equation}
and computes
\begin{equation}\label{eqn:63A}x
	\mathcal{K}(t)
	= \mathcal{K}_\beta(t) \, \mathcal{K}_\gamma(t),
\end{equation}
with
\begin{alignedeqn}\label{eqn:63B}
	\mathcal{K}_\beta (t) 
	&= \cosh\beta\Bigl(\bigl( h_1(t + \epsilon) + h_2(t + \epsilon)) (h_1(t) + h_2(t)\bigr)\Bigr)\\
	&+ \sinh\beta\Bigl(\bigl( h_1(t + \epsilon) - h_2(t + \epsilon)) (h_1(t) - h_2(t)\bigr)\Bigr),\\
\end{alignedeqn}
and
\begin{alignedeqn}\label{eqn:S2}
	\mathcal{K}_\gamma(t)
	&= 1 + \frac{1}{2}(\cosh\gamma - 1)
	\Bigl(1 +\\
	& + \bigl( h_1(t + \epsilon) - h_2(t + \epsilon)\bigr)\bigl( h_1(t) - h_2(t)\bigr)\Bigr)
	\\
	& +\frac{1}{2}\sinh\gamma\bigl( h_1(t + \epsilon) - h_2(t + \epsilon) + h_1(t) - h_2(t)\bigr).
\end{alignedeqn}
This expresses $\mathcal{K}(t)$ in terms of the basis functions
\begin{alignedeqn}\label{eqn:63C}
	\mathcal{K}(t)
	&= e^{\beta + \gamma}h_1(t + \epsilon)h_1(t)
	 + e^{\beta - \gamma}h_2(t + \epsilon)h_2(t)\\
	& + e^{-\beta}\bigl( h_1(t + \epsilon)h_2(t) + h_2(t + \epsilon)h_1(t)\bigr),
\end{alignedeqn}
and one infers the well known result for the transfer matrix
\begin{equation}\label{eqn:S3}
	\bar S
	= \begin{pmatrix}
		e^{\beta + \gamma} & e^{-\beta}\\
		e^{-\beta} & e^{\beta - \gamma}
	\end{pmatrix}.
\end{equation}

At this point the probability distribution is not normalized and one may compute $\bar Z(\beta,\gamma)$ from \cref{eqn:BBB}. The normalization can be achieved by a multiplicative constant
\begin{equation}\label{eqn:168A}
	S
	= e^{-\varphi}\bar S.
\end{equation}
Employing the same factor for the boundary term $\bar q(t_\text{f})q(t_\text{in})$ we obtain from \cref{eqn:BBB}
\begin{equation}\label{eqn:168B}
	Z
	= e^{-(G + 1)\varphi}\bar Z
	= 1.
\end{equation}
We can identify $\varphi$ with the free energy per degree of freedom
\begin{equation}\label{eqn:168C}
	\varphi(\beta,\gamma)
	= \frac{1}{G + 1}\ln \bar Z(\beta,\gamma).
\end{equation}

\paragraph{Free energy} The association of the free energy density $\varphi$ with the normalization of the step evolution operator in \cref{eqn:168A} permits us to compute $\varphi(\beta,\gamma)$ from the evolution of the wave function. For this purpose we recall the evolution of the wave function
\begin{equation}\label{eqn:EA}
	\tilde q_\tau(t_\text{in} + n\epsilon)
	= (S^n)_{\tau\rho}\tilde q_\rho(t_\text{in}).
\end{equation}
Since $S$ is a real symmetric matrix it can be diagonalized by orthogonal transformations. The eigenvalues $\lambda_j$ are real. Eigenvectors with $\lambda_j > 1$ will grow, while eigenvectors with $\lambda_j < 1$ decrease
\begin{equation}\label{eqn:EB}
	\tilde q^{(j)}_\tau(t + n\epsilon)
	= \lambda^n_j\tilde q^{(j)}_\tau(t_\text{in}).
\end{equation}

Consider now the final ``time'' $t_\text{f}$,
\begin{equation}\label{eqn:EC}
	\tilde q_\tau(f_\text{f})
	= (S^G)_{\tau\rho}\tilde q_\rho(t_\text{in}).
\end{equation}
With 
\begin{equation}\label{eqn:ED}
	Z
	= \bar q_\tau(t_\text{f})\tilde q_\tau(t_\text{f})
	= 1
\end{equation}
we conclude that for any finite $\bar q(t_\text{f})$ \cref{eqn:ED} can hold only if $\tilde q(t_\text{f})$ is finite and different from zero. (We exclude here particularly ``fine tuned'' final states $\bar q(t_\text{f})$ that are orthogonal to a diverging part of $\tilde q(t_\text{f})$). For $G \to \infty$ this requires that one eigenvalue of $S$ equals one, while the other eigenvalue is smaller than one. This condition fixes the multiplicative factor in the step evolution operator or normalized transfer matrix
\begin{equation}\label{eqn:EE}
	S
	= \begin{pmatrix}
		e^{\beta + \gamma - \varphi} & e^{-\beta - \varphi}\\
		e^{-\beta - \varphi} & e^{\beta - \gamma - \varphi}
	\end{pmatrix}.
\end{equation}
The largest eigenvalue of $S$ should be equal to one. This in turn determines $\varphi$ as a function of $\beta$ and $\gamma$. In consequence, the thermodynamics can be extracted from the normalization of the step evolution operator.

From the eigenvalue condition $\det(S - 1) = 0$ one infers
\begin{equation}\label{eqn:EG}
	e^{-\varphi}
	= \frac{2}{\sinh 2\beta} \biggl(e^\beta\cosh \gamma \mp \csqrt{e^{-2 \beta} + \sinh^2\gamma e^{2\beta}}\biggr),
\end{equation}
or
\begin{equation}\label{eqn:EH}
	\varphi(\beta,\gamma)
	= \ln\biggl[e^\beta\cosh \gamma \pm \csqrt{e^{-2 \beta} + \sinh^2\gamma e^{2 \beta}}\biggr].
\end{equation}
For the appropriate sign this is the standard exact solution for the free energy of the one-dimensional Ising model. The largest eigenvalue of $\bar S$ matters, corresponding to the plus sign in \cref{eqn:EH} or the minus sign in \cref{eqn:EG}. We observe that $e^\varphi$ is an eigenvalue of the transfer matrix $\bar S$ in \cref{eqn:S3}, as expected. From the free energy $F = (G + 1)\varphi$ the magnetization and thermodynamic quantities of the Ising model can be computed. In other words, the requirement that the wave function $\tilde q(t)$ does not diverge or vanish for $(t_\text{f} - t_\text{in})/\epsilon \to \infty$ can be used for the computation of the thermodynamic equilibrium properties.

For more general models with not too large $N = 2^M$ this property may be useful for a determination of equilibrium properties by a numerical solution of the discrete evolution equation \labelcref{eqn:87}.

If the largest eigenvalue of $S$ equals one, and all other eigenvalues are smaller than one, the product $S^n$ will converge for $n \to \infty$ to a scaling form $S_\ast$. This implies the matrix identity
\begin{equation}\label{eqn:EI}
	S \, S_\ast
	= S_\ast,
	\quad
	S_\ast
	= \lim_{n \to \infty}S^n.
\end{equation}
In App. C we use this relation in order to determine $\varphi$ and $S_\ast$. For the one-dimensional Ising model the explicit scaling form $S_\ast$ is found as
\begin{equation}\label{eqn:EW}
	S_\ast
	= \frac{1}{2} \begin{pmatrix}
		1 + \sinh \gamma \, e^{2\beta } g & g\\
		g & 1 - \sinh\gamma \, e^{2\beta} g
	\end{pmatrix},
\end{equation}
with
\begin{equation}\label{eqn:EX}
	g
	= \frac{1}{\csqrt{1 + \sinh^2\gamma \, e^{4\beta}}}.
\end{equation}
One verifies $\det S_\ast = 0$, corresponding to one eigenvalue one and the other zero.

The explicit form of the step evolution operator reads
\begin{equation}\label{eqn:EY}
	S
	= \frac{1}{\cosh \gamma \, e^{2\beta} + g^{-1}} \begin{pmatrix}
		e^{2 \beta + \gamma} & 1\\
		1 & e^{2 \beta - \gamma}
	\end{pmatrix}.
\end{equation}
For $\gamma = 0$ one has the particular simple form $g = 1$ and 
\begin{equation}\label{eqn:EZ}
	S_\ast
	= \frac{1}{2} \begin{pmatrix}
		1 & 1\\
		1 & 1
	\end{pmatrix},
	\quad
	S
	= \frac{1}{2\cosh\beta} 	\begin{pmatrix}
		e^\beta & e^{-\beta}\\
		e^{-\beta} & e^\beta
	\end{pmatrix}.
\end{equation}

Knowing $S_\ast$ we can fix (for $G \to \infty$) the normalization of the initial and final factors, with \cref{eqn:BBB} reading 
\begin{equation}\label{eqn:FA}
	\bar q_\tau(t_\text{f})(S_\ast)_{\tau\rho}\tilde q_\rho(t_\text{in})
	= Z_b.
\end{equation}
We can decompose $\tilde q(t_\text{in})$ and $\bar q(t_\text{f})$ into eigenvectors of $S_\ast$ with eigenvalues one or zero. Only the eigenvectors with eigenvalue one contribute to $Z_b$. They can be normalized by multiplication with $1/\csqrt{Z_b}$ in order to achieve $Z = 1$.

\paragraph{Magnetization} We can compute the magnetization in the bulk directly from the asymptotic behavior of the wave function. Exploiting \cref{eqn:EA} for large $n$ only the largest eigenvalue of $S$ will contribute. Thus the wave function $\tilde q(t)$ converges to an equilibrium value $\tilde q_\ast$ inside the bulk, given by
\begin{equation}\label{eqn:FB}
	S\tilde q_\ast
	= \tilde q_\ast,
	\quad
	S_\ast \, \tilde q_\ast
	= \tilde q_\ast.
\end{equation}
This fixes the wave function up to a multiplicative constant
\begin{equation}\label{eqn:FC}
	\tilde q_{\ast2}
	= \bigl(1/g - \sinh\gamma e^{2\beta}\bigr) \, \tilde q_{\ast1}.
\end{equation}
A similar argument for the conjugate wave function yields for large $m$ (using $S = \tran{S}$ in \cref{eqn:89})
\begin{equation}\label{eqn:FD}
	\bar q(t_\text{f} - m\epsilon)
	= S^m\bar q(t_\text{f})
	= S^\ast\bar q(t_\text{f})
	= \bar q_\ast,
\end{equation}
implying for the equilibrium conjugate wave function
\begin{equation}\label{eqn:FE}
	\bar q_{\ast2}
	= \bigl(1/g - \sinh\gamma e^{2\beta}\bigr) \, \bar q_{\ast1}.
\end{equation}
For properly normalized boundary conditions ($Z_b = 1$) we can employ the normalization \labelcref{eqn:84} 
\begin{equation}\label{eqn:FF}
	\bar q_{\ast1}\tilde q_{\ast1} + \bar q_{\ast2}\tilde q_{\ast2}
	= 1
\end{equation}
which implies
\begin{alignedeqn}\label{eqn:FG}
	\bar q_{\ast1}\tilde q_{\ast1}
	&= \frac{1}{2N_1},\\
	N_1
	&= 1 + \sinh^2\gamma e^{4\beta} - \frac1g\sinh\gamma e^{2\beta}.
\end{alignedeqn}

Knowing the bulk wave functions explicitly we can compute expectation values in the bulk from \cref{eqn:SID}
\begin{equation}\label{eqn:FH}
	\langle A\rangle
	= \tran{\bar{q}}_*A^\prime\tilde q_\ast.
\end{equation}
For example, the occupation number $N^\prime = \diag(1,0)$ has in the bulk the expectation value
\begin{equation}\label{eqn:FI}
	n_\ast
	= \langle n\rangle
	= \bar q_{\ast1}\tilde q_{\ast1}
	= 
	\frac{1}{2N_1}.
\end{equation}
This translates to the average spin
\begin{equation}\label{eqn:FJ}
	s_\ast
	= 2n_\ast - 1
	= \frac{1}{N_1} - 1
	= \frac{\Delta}{1 + \Delta},
\end{equation}
where
\begin{equation}\label{eqn:FK}
	\Delta
	= \sinh \gamma e^{2\beta} \bigl(1/g + \sinh\gamma e^{2\beta}\bigr).
\end{equation}

We observe that $s_\ast$ only depends on the parameter combination
\begin{equation}\label{eqn:FL}
	\delta
	= \sinh \gamma e^{2\beta},
\end{equation}
with 
\begin{equation}\label{eqn:FM}
	\Delta
	= \delta(\sqrt{1 + \delta^2} + \delta),
	\quad
	s_\ast
	= \frac{\delta}{\csqrt{1 + \delta^2}}.
\end{equation}
For $\delta \to \infty$ one finds the usual saturation
\begin{equation}\label{eqn:FN}
	s_\ast
	= 1 - \frac{1}{2\delta^2}.
\end{equation}
As expected, $s_\ast$ coincides with the mean spin as computed from the free energy
\begin{equation}\label{eqn:FO}
	\frac{\partial \varphi}{\partial\gamma}
	= s_\ast.
\end{equation}

\paragraph{Effect of boundary conditions} Besides the asymptotic behavior in the bulk the explicit knowledge of the evolution of the wave functions permits a detailed description of the influence of boundary conditions on the expectation values of local spins. For the boundary problem we split the initial wave function $\tilde q(t_\text{in})$, which fixes the boundary term, into a part proportional to $\tilde q_\ast$ and a part proportional to the eigenvector of $S$ with eigenvalue $\lambda_- < 1$,
\begin{equation}\label{eqn:FP}
	\tilde q(t_\text{in})
	= c_1\bigl(\tilde q_\ast + \delta \tilde q(t_\text{in})\bigr),
\end{equation}
where 
\begin{equation}\label{eqn:FQ}
	S\delta \tilde q
	= \lambda_ - \delta \tilde q.
\end{equation}
The evolution equation \labelcref{eqn:EA} yields
\begin{equation}\label{eqn:FR}
	\tilde q(t_\text{in} + n\epsilon)
	= c_1\bigl(\tilde q_\ast + \lambda^n_-\delta \tilde q(t_\text{in})\bigr).
\end{equation}
Arbitrary initial wave functions approach $c_1\tilde q_\ast$ as $n$ increases, according to 
\begin{equation}\label{eqn:FS}
	\tilde q(t)
	= c_1
	\Bigl(\tilde q_\ast + \exp\Bigl(-\tfrac{t - t_\text{in}}{\xi}\Bigr) \, \delta \tilde q(t_\text{in})\Bigr).
\end{equation}
The correlation length $\xi$ is related to the second eigenvalue of $S$,
\begin{equation}\label{eqn:FT}
	\xi
	= \frac{\epsilon}{\ln (1/\lambda_-)}.
\end{equation}

For a quantitative estimate of the memory of the boundary conditions after $n$ steps into the bulk we need the second eigenvalue $\lambda_-$ of $S$. Since one eigenvalue of $S$ equals one we compute $\lambda_-$ from 
\begin{equation}\label{eqn:FU}
	\lambda_-
	= \det S
	= \frac{e^{4\beta} - 1}{\bigl(1/g + \cosh\gamma e^{2\beta}\bigr)^2}.
\end{equation}
Restricting the discussion to $\gamma = 0$ one finds
\begin{equation}\label{eqn:FV}
	\lambda_-
	= 
	\frac{e^{2\beta} - 1}{e^{2\beta} + 1}
	= \tanh \beta.
\end{equation}
The correlation length $\xi$ diverges for $\beta \to \infty$
\begin{equation}\label{eqn:FX}
	\frac{\xi}{\epsilon}
	= \frac{1}{\ln \cosh \beta}
	\to	\frac{e^{2\beta}}{2}.
\end{equation}
It is the same correlation length as the one defined by the decay of the correlation function for the known exact solution of the Ising model.

The conjugate wave function for $t$ in the vicinity of $t_\text{in}$, with large $G \to \infty$, can be approximated by the equilibrium value
\begin{equation}\label{eqn:FY}
	\bar q(t)
	= \frac{\bar q_\ast}{c_1}
	= \frac{\bar q_{\ast1}}{c_1} \begin{pmatrix}
		1\\
		\csqrt{1 + \delta^2} - \delta
	\end{pmatrix},
\end{equation}
where the normalization is chosen such that $Z = 1$. The expectation value $\langle n(t)\rangle$ follows as
\begin{alignedeqn}\label{eqn:FZ}
	\langle n(t)\rangle
	&= \bar q_1(t)\tilde q_1(t)
	= \bar q_{\ast1}\bigl(\tilde q_{\ast1} + \delta \tilde q_1(t)\bigr)\\
	&= n_\ast + \bar q_{\ast1}\delta \tilde q_1(t)
	= n_\ast + \bar q_{\ast1}\delta \tilde q_1(t_\text{in})
	\frac{\delta \tilde q_1(t)}{\delta\tilde q_1(t_\text{in})}.
\end{alignedeqn}
With 
\begin{equation}\label{eqn:GA}
	n_\ast + \bar q_{\ast1}\delta \tilde q_1(t_\text{in})
	= \langle n(t_\text{in})\rangle
\end{equation}
\cref{eqn:FS} yields the simple final result
\begin{equation}\label{eqn:GB}
	\langle n(t)\rangle
	= n_\ast + \bigl(n(t_\text{in}) - n_\ast\bigr) \exp\Bigl(-\tfrac{t - t_\text{in}}{\xi}\Bigr).
\end{equation}

We observe that the orthogonality of eigenvectors to different eigenvalues implies
\begin{equation}\label{eqn:GC}
	\bar q_{\ast1}\delta \tilde q_1 + \bar q_{\ast2}\delta \tilde q_2
	= 0.
\end{equation}
The bounds $0\leq \langle n(t_\text{in})\rangle\leq 1$ limit the allowed values of $\delta \tilde q (t_\text{in})$, while $\tilde q(t_\text{in})$ can be arbitrary by use of the constant $c_1$.

We conclude that the classical wave function $\tilde q(t)$, together with the conjugate wave function $\bar q(t)$, solve the issue of the influence of boundary conditions for the one-dimensional Ising model completely. This method is closely analogous to the use of the transfer matrix. The simple concept of wave functions allows a straightforward use of methods known from quantum mechanics. It is suitable for more complex models and the discussion of general properties.

\subsection{Different types of evolution}

Many classical statistical systems share the qualitative properties oft the one-dimensional Ising model. For this class of systems the transfer matrix $\bar S$ has positive real eigenvalues. The step evolution operator $S$ is then related to $\bar S$ by a multiplicative constant, chosen such that the largest eigenvalue of $S$ equals one. If the second largest eigenvalue $\lambda_-$ is separated from the largest eigenvalue, it defines a finite correlation length $\xi$ according to \cref{eqn:FT}. The memory of boundary terms is exponentially damped, similar to \cref{eqn:GB}. For some subset of initial conditions the damping even occurs faster, according to other eigenvalues of $S$ smaller than $\lambda_-$. This scenario with a separated largest eigenvalue of the transfer matrix is realized if the dimension of $\bar S$ is finite and all elements are strictly positive, $\bar S_{\tau\rho} > 0$.

This loss of boundary information is, however, not the most general behavior. Interesting situations arise if the largest eigenvalue of $S$ becomes degenerate. In this limit the correlation length diverges. This situation is typically realized for critical phenomena. For practical purposes it is sufficient that $\xi$ exceeds the size of the system $t_\text{f} - t_\text{in}$, as realized for the strongly coupled or low temperature Ising model for large $\beta$.

One may also encounter unstable situations where the transfer matrix $\bar S$ has negative eigenvalues. For the one-dimensional Ising model this is realized for $\beta < 0$. Typically, the equilibrium state $\tilde q_\ast$ is not invariant under translations by $\epsilon$ in this case. The equilibrium state may still have translation symmetry with respect to translations by $2\epsilon$, as for the antiferromagnetic Ising model for $\beta < 0$.

It is possible that $S$ has complex eigenvalues. This can occur if $S$ is not symmetric. In this case one expects oscillatory behavior. An example is the four-state oscillator chain, as realized by the four by four matrix 
\begin{equation}\label{eqn:GD}
	S
	= \begin{pmatrix}
		1 - \eta & 0 & \eta & 0\\
		\eta & 1 - \eta & 0 & 0\\
		0 & 0 & 1 - \eta & \eta\\
		0 & \eta & 0 & 1 - \eta
	\end{pmatrix}.
\end{equation}
The four eigenvalues are
\begin{equation}\label{eqn:GE}
	\lambda_0
	= 1,
	\quad
	\lambda_1\pm
	= 1 - \eta \pm i \eta,
	\quad
	\lambda_2
	= 1 - 2 \eta.
\end{equation}
For $0\leq \eta\leq 1$ all entries of $S$ are positive and the model can be realized with Ising spins. Since some elements of $S$ vanish it is a type of ``constrained Ising model''. Its detailed realization is discussed in \cref{app:generalized Ising models}. We may avoid negative eigenvalues by requiring $\eta < 1/2$.

The (unnormalized) eigenvectors to the eigenvalues $\lambda_0,\lambda_{1 + },\lambda_{1-},\lambda_2$ are given by 
\begin{equation}\label{eqn:GF}
	\begin{pmatrix} 1\\1\\1\\1 \end{pmatrix},
	\quad
	\begin{pmatrix} 	1\\-i\\i\\-1 \end{pmatrix},
	\quad
	\begin{pmatrix} 1\\i\\-i\\-1 \end{pmatrix},
	\quad
	\begin{pmatrix} 1\\-1\\-1\\1 \end{pmatrix}.
\end{equation}
Since the real part of $\lambda_{1\pm}$ is smaller than one, one encounters damped oscillations of the wave function if the initial wave function contains components $\sim v_{1\pm}$.

Whenever a unique largest eigenvalue one of $S$ is separated from the other eigenvalues $\lambda_i$, with $|\lambda_i| < 1$, one may suspect a loss of memory of boundary conditions with a finite correlation length. The situation is subtle, however. The reason is that an exponentially decaying $\tilde q - \tilde q_\ast$ may be compensated by an exponentially increasing $\bar q - \bar q_\ast$. This could lead to undamped oscillations of the local probabilities. In the next section we will define a density matrix as a bilinear in $\tilde{q}$ and $\bar{q}$. The general solution of the evolution equation \labelcref{eqn:I8} admits indeed undamped oscillatory behavior \cite{CWQFCS}. One then has to investigate if such density matrices can be realized by suitable boundary conditions. The answer is negative, such that the evolution is indeed described by damped oscillations \cite{CWQFCS}.

One may ask if there are simple exact models where the loss of boundary information by the damping towards the equilibrium state is absent. This is indeed the case. A sufficient condition for keeping the memory of boundary conditions at $t_\text{in}$ far inside the bulk, and even at the opposite boundary at $t_\text{f}$, is the presence of more than one largest eigenvalue of $S$ with $|\lambda| = 1$. (Largest is here in the sense of largest $|\lambda|$.) The simplest and rather trivial example is $S = 1$. In this case the expectation value of a local observable $A(t)$ is simply given by
\begin{equation}\label{eqn:GG}
	\langle A(t)\rangle
	= \bar q_\tau(t_\text{f})A^\prime_{\tau\rho}(t)\tilde q_\rho(t_\text{in}).
\end{equation}
This holds independently of $t$, including $t = t_\text{f}$. The expectation value obviously depends on the initial wave function $q(t_\text{in})$.

A somewhat less trivial example is 
\begin{equation}\label{eqn:GH}
	S
	= \begin{pmatrix}
		1 & 0 & 0 & 0\\
		0 & 0 & 1 & 0\\
		0 & 1 & 0 & 0\\
		0 & 0 & 0 & 1
	\end{pmatrix}.
\end{equation}
In the notation of \cref{eqn:74}, with two species of occupation numbers interpreted as fermionic particles at two different sites, this corresponds to one particle jumping from one site to the other, while states with no particles or two particles, one at each site, are $t$-independent. The matrix $S$ has three eigenvalues $ + 1$, and one eigenvalue $-1$. Since $S^2 = 1$ one finds
\begin{equation}\label{eqn:GI}
	\langle A(t)\rangle
	= \tran{\bar{q}}(t_\text{f})S^{m(t)} A^\prime(t)S^{n(t)}\tilde q(t_\text{in}),
\end{equation}
with 
\begin{alignedeqn}\label{eqn:GJ}
	m(t) & = \begin{cases*}
	0 & for $(t_\text{f} - t)/\epsilon$ even,\\
	1 & for $(t_\text{f} - t)/\epsilon$ odd.
	\end{cases*}\\
	n(t) & =
	\begin{cases*}
	0 & for $(t - t_\text{in})/\epsilon$ even,\\
	1 & for $(t - t_\text{in})/\epsilon$ odd.
	\end{cases*}
\end{alignedeqn}
For even $G = (t_\text{f}-t_\text{in})/\epsilon$ one has $m(t) = n(t)$, while $m$ and $n$ differ for $G$ odd. In particular, one finds at $t_\text{f}$ 
\begin{align}\label{eqn:GHa}
	\langle A(t_\text{f})\rangle
	= 
	\begin{cases*}
	\tran{\bar{q}}(t_\text{f})A^\prime(t_\text{f})\tilde q(t_\text{in}) & for $G$ even\\
	\tran{\bar{q}}(t_\text{f})A^\prime(t_\text{f})S\tilde q(t_\text{in}) & for $G$ odd
	\end{cases*}.
\end{align}
In both cases the information on the boundary at $t_\text{in}$ is transported completely to the other boundary at $t_\text{f}$.

Let us restrict the discussion for simplicity to $G$ even and to $t$-independent local observables, represented by the same operator $\hat A$ for each $t$. Operators commuting with $S$ have static expectation values
\begin{equation}\label{eqn:GHb}
	\langle A(t)\rangle
	= \tran{\bar{q}}(t_\text{f})A^\prime\tilde q(t_\text{in}).
\end{equation}
On the other hand, if $A^\prime \, S = -S \, A^\prime$, the expectation value oscillates with period $2\epsilon$
\begin{equation}\label{eqn:GHc}
	\langle A(t + \epsilon)\rangle
	= \langle -A(t)\rangle.
\end{equation}
An example is
\begin{equation}\label{eqn:GK}
	A^\prime
	= \begin{pmatrix}
		0 & 0 & 0 & 0\\
		0 & 1 & 0 & 0\\
		0 & 0 & -1 & 0\\
		0 & 0 & 0 & 0 
	\end{pmatrix},
\end{equation}
which measures the difference $n_2-n_1$. Obviously, this example realizes oscillating local probabilities $p_\tau(t)$, with period $2\epsilon$. Indeed, the wave function obeys an oscillating behavior
\begin{equation}\label{eqn:GL}
	\tilde q_2(t + \epsilon)
	= \tilde q_3(t),
	\quad\tilde q_3(t + \epsilon)
	= \tilde q_2(t),
\end{equation}
while $q_1$ and $q_4$ are static. The same holds for the conjugate wave function, such that 
\begin{equation}\label{eqn:GM}
	p_2(t + \epsilon)
	= p_3(t),
	\quad
	p_3(t + \epsilon)
	= p_2(t),
\end{equation}
with static $p_1$ and $p_4$.

A further example of complete information transport is given by \cref{eqn:GD}, with $\eta = 1$. It corresponds to $\tilde q_2(t + \epsilon) = \tilde q_1(t)$, $\tilde q_4(t + \epsilon) = \tilde q_2(t)$, $\tilde q_3(t + \epsilon) = \tilde q_4(t)$, $\tilde q_1(t + \epsilon) = \tilde q_3(t)$. This evolution is periodic with period $4\epsilon$. Correspondingly the eigenvalues of $S$ are $\pm1,\pm i$. With $\tran{S} \, S = 1$ the step evolution operator
\begin{equation}\label{eqn:GN}
	S
	= 
	\begin{pmatrix}
		0 & 0 & 1 & 0\\
		1 & 0 & 0 & 0\\
		0 & 0 & 0 & 1\\
		0 & 1 & 0 & 0
	\end{pmatrix}
\end{equation}
describes a rotation. The conjugate wave function $\bar q(t)$ follows the same evolution as $\tilde q(t)$ according to \cref{eqn:90}. Therefore the local probabilities $p(t)$ oscillate with period $4\epsilon$, according to $p_2(t + \epsilon) = p_1(t)$, $p_4(t + \epsilon) = p_2(t)$, $p_3(t + \epsilon) = p_4(t)$, $p_1(t + \epsilon) = p_3(t)$.

These simple examples demonstrate already that the issue of information transport is expected to differ substantially from a simple approach to equilibrium in the bulk whenever more than one eigenfunction to eigenvalues of $S$ with $|\lambda| = 1$ exists. In this case there is no unique equilibrium wave function $\tilde q_\ast$ as in the Ising model, cf. \cref{eqn:FC}. Memory of boundary conditions is expected to occur due to a ``degeneracy'' of generalized equilibrium wave functions.

\subsection{Quantum mechanics}

Let us investigate the particular case where $S = R$ is a rotation matrix. Then $f(t)$ and $\bar f(t)$ obey the same evolution law. If they are equal for one particular time point $\bar t$ they will remain equal for all $t$. The wave function $f(t)$ depends on the initial value $f_\text{in}(t_\text{in})$, while $\bar f(t)$ depends on $\bar f_\text{f}(t_\text{f})$. One can always choose $\bar f_\text{f}(t_\text{f})$ such that $\bar f(t) = f(t)$.
(This is formally achieved by solving the evolution law \labelcref{eqn:85} or \labelcref{eqn:87} in order to compute $f(t_\text{f})$, and then to choose $\bar f_\text{f}(t_\text{f}) = f(t_\text{f})$.) We will assume here the boundary condition $\bar f(t_\text{f}) = f(t_\text{f})$, while more general choices of $\bar f_\text{f}$ are discussed in ref.~\cite{CWQFCS}.

A \textit{normalized} classical wave function $q(t)$ is defined \cite{CWCW} by the square root of the local probability up to a sign function $\hat s_\tau(t) = \pm 1$,
\begin{equation}\label{eqn:CW1}
	q_\tau(t)
	= \csqrt{p_\tau(t)} \, \hat s_\tau(t),
	\quad
	q^2_\tau(t)
	= p_\tau(t).
\end{equation}
According to \cref{eqn:22A} it is realized if
\begin{equation}\label{eqn:CW2}
	\bar q(t)
	= \tilde q(t)
	= q(t).
\end{equation}
For $\bar f(t) = f(t)$ we can indeed replace both $\tilde q$ and $\bar q$ by a common normalized classical wave function $q$.

For $\bar f(t) = f(t)$ one has, from \cref{eqn:84}
\begin{alignedeqn}\label{eqn:91}
	Z
	&= \int \dif n(t)f^2(t)\\
	&= \int \dif n(t) \, q_\tau(t) \, q_\rho(t) \, h_\tau(t) \, h_\rho(t)
	= \sum_\tau q^2_\tau(t).
\end{alignedeqn}
A normalized wave function obeys 
\begin{equation}\label{eqn:92}
	\sum_\tau q^2_\tau
	= 1,
\end{equation}
and this normalization is preserved by the evolution if $S$ describes a rotation. In this case it is sufficient that the initial wave function $f_\text{in}$ is normalized. For a normalized wave function also $Z$ is normalized, $Z = 1$. An evolution that preserves the length of the vector $q_\tau(t)$ will be called unitary. (In the present case of real $q_\tau$ the unitary transformations are rotations.)

We can now identify $p_\tau(t) = q^2_\tau(t)$ with the local probability for finding at time $t$ an appropriate combination of occupation numbers. For $M$ species of occupation numbers this is precisely the setting of quantum mechanics for the special case of a real $2^M$-component wave function. If the rotation matrix $R_{\tau\rho}$ is compatible with a complex structure we can order the $2^M$ real components into a complex $2^{M-1}$-component wave function that obeys a unitary evolution, as we will see in the next section.

Expectation values of occupation numbers at a given $t$ can be computed from the normalized wave function as
\begin{alignedeqn}\label{eqn:94}
	\langle n_\gamma(t)\rangle
	&= \int \dif n(t)f^2(t)n_\gamma(t)\\
	&= \sum_\tau p_\tau(t)n_{\gamma,\tau}(t),
\end{alignedeqn}
using $n_\gamma(t) \, h_\tau(t) = n_{\gamma,\tau}(t) \, h_\tau(t)$ according to \cref{eqn:75}. Employing the diagonal operators $N^\prime_\gamma$ defined in \cref{eqn:B5} we also have the quantum rule for expectation values
\begin{equation}\label{eqn:95}
	\langle n_\gamma(t)\rangle
	= q_\tau(t)(N^\prime_\gamma)_{\tau\rho}q_\rho(t).
\end{equation}
For a normalized wave function and $\bar f = f$ the basic definition of these expectation values is given by the functional integral 
\begin{equation}\label{eqn:96}
	\langle n_\gamma(t)\rangle
	= \int \mathcal{D}n\bar f_\text{f} K n_\gamma(t)f_\text{in}.
\end{equation}
In accordance with the previous general discussion, \cref{eqn:94} follows from \cref{eqn:96} by using the split \labelcref{eqn:82} and the definition of the wave functions \labelcref{eqn:83}.

The question arises under which circumstances the step evolution operator can be a rotation if $p[n]$ is a classical probability distribution. Acceptable local probabilities are found for arbitrary rotation matrices $S = R$, since $q^2_\tau\geq 0$. In contrast, a positive overall probability distribution is, in general, not realized. For arbitrary rotations the weight distribution $w[n]$, defined by \cref{eqn:28B,eqn:28C}, will not be positive semidefinite. For rotations there are typically negative matrix elements $S_{\tau\rho}$. This holds, in particular, for all (non-trivial) infinitesimal rotations. Exceptions of finite rotations with only positive matrix elements of $S$ exist, however, as demonstrated by the examples \labelcref{eqn:GH,eqn:GN}.

In case of translation symmetry and for finite $M$ it is easy to convince oneself that step evolution operators corresponding to infinitesimal rotations cannot be realized by a positive $p[n]$. Indeed, for infinitesimal rotations the factor $B_{\rho_m,\alpha\beta}(t_m)$ in \cref{eqn:28D} exhibits both positive and negative values as $\rho_m$ is varied. We conclude that in case of translation symmetry and finite $M$ the realization of a quantum evolution by step evolution operators that are rotation matrices differing only infinitesimally from one is not compatible with an overall classical probability distribution $p[n]$. Nevertheless, one can find interesting examples of quantum evolution. They can be realized by extensions of the examples \labelcref{eqn:GH,eqn:GN} or if translation symmetry is abandoned. It is also possible that a subsystem follows a quantum evolution even though the total system does not. Furthermore, for $M \to \infty$ new possibilities open up. The notion of infinitesimal may now refer to the neighborhood in a dense space of states.

In any case, particular conditions are needed for the whole system to follow a quantum evolution. Generic generalized Ising-type models \labelcref{eqn:128B1,eqn:S10} do not follow a unitary evolution. Since all matrix elements $S_{\tau\rho}$ are positive semidefinite, the matrix $S$ is a rotation matrix only for special cases. Such special cases are static memory materials and will be discussed in \crefrange{sec:simple static memory materials}{sec:general static memory materials}. While the generic Ising-spin systems do not represent a quantum system of the type discussed above, we will argue that the evolution of the local probabilities admits undamped oscillatory behavior under a rather wide range of circumstances. For these cases a suitable subsystem follows a unitary time evolution. We briefly discuss in \cref{app:non-linear unitary evolution} that the generic evolution can be viewed as a non-linear unitary evolution of a suitably defined normalized wave function. The quantum subsystems are then subsystems that follow a linear unitary evolution.

\section{Generalized Schrödinger equation}
\label{sec:generalized schrödinger equation}

This section addresses the continuum limit of the evolution \labelcref{eqn:87,eqn:90}. In the continuum limit the wave functions obey linear differential equations - the generalized Schrödinger equation. If the evolution admits a complex structure the generalized Schrödinger equation takes the usual complex form. Only the generalization of the Hamiltonian operator is, in general, not hermitian for classical statistical systems. Similar to quantum mechanics one can define a density matrix, which evolves according to a generalized von Neumann equation.

\subsection{Continuous evolution limit}

A continuous time evolution can be realized if the change of the wave function after one or a few time steps is in some sense small. For small changes after two time steps we define
\begin{equation}\label{eqn:FF1}
	W(t)
	= \frac{1}{2\epsilon}\bigl(S(t) - S^{-1}(t - \epsilon)\bigr),
\end{equation}
such that 
\begin{equation}\label{eqn:FF2}
	\partial_t \tilde q(t)
	= \frac{1}{2 \epsilon}\bigl( \tilde q(t + \epsilon) - \tilde q(t - \epsilon)\bigr)
	= W(t)\tilde q(t).
\end{equation}
If the limit $\epsilon \to 0$ can be taken, $\tilde q(t)$ is a differentiable function and $\partial_t \tilde q$ becomes the standard derivative. For the particular case where $S$ is independent of $t$ and deviates from the unit matrix only by elements $\sim \epsilon$ one has
\begin{equation}\label{eqn:128B2}
	S_{\tau\rho}
	= \delta_{\tau\rho} + \epsilon W_{\tau\rho},
	\quad
	S
	= 1 + \epsilon W.
\end{equation}
This is, however, not the only possibility how a continuum limit can be realized.

The evolution equation \labelcref{eqn:FF2} is a real equation for a real wave function. We can formally write it as a complex equation which makes the difference to the Schrödinger equation in quantum mechanics apparent. We split $W$ into a hermitian (symmetric) and antihermitian (antisymmetric) part and write without loss of generality
\begin{equation}\label{eqn:FF3}
	W
	= J - i H,
\end{equation}
with $H$ and $J$ hermitian matrices. For real wave functions $\tilde q$ and real $W$ the Hamilton operator $H$ is antisymmetric and purely imaginary, while $J$ is real and symmetric. This yields a generalized Schrödinger equation 
\begin{equation}\label{eqn:46K}
	\partial_t \tilde q = -i H \, \tilde q + J \, \tilde q.
\end{equation}
For quantum systems with $t$-invariant orthogonal $S$ the matrix $J$ vanishes.

For the conjugate wave function \cref{eqn:90} implies 
\begin{alignedeqn}\label{eqn:NXY1}
	\partial_t\bar q(t)
	&= -\tran{\tilde{W}}(t) \, \bar q(t),\\
	\tilde W(t)
	&= \frac{1}{2\epsilon}\bigl[S(t - \epsilon) - S^{-1}(t)\bigr].
\end{alignedeqn}
For $t$-independent $S$ one has $\tilde W = W$ and therefore
\begin{equation}\label{eqn:FF4}
	\partial_t\bar q
	= -\tran{W} \, \bar q
	= -i H \, \bar q - J \, \bar q.
\end{equation}
For quantum systems $\tilde q$ and $\bar q$ obey the same evolution equation which amounts to the Schrödinger equation for the normalized wave function $q$. For $J \neq 0$ the evolution of $\tilde q$ and $\bar q$ differs.

We may discuss simple examples. For the one-dimensional Ising model with $\gamma = 0$ the continuous evolution is realized for $\beta \to \infty$. In this case one has
\begin{alignedeqn}\label{eqn:HA}
	S
	= \begin{pmatrix}
		1 - e^{-2 \beta} & e^{-2 \beta}\\
		e^{-2 \beta} & 1 - e^{-2 \beta} 
	\end{pmatrix},
	\quad
	W
	= J
	= 
	\begin{pmatrix}
		-\omega & \omega\\
		\omega & -\omega 
	\end{pmatrix},\\
\end{alignedeqn}
with 
\begin{equation}\label{eqn:HB}
	\omega
	= \frac{e^{-2 \beta}}{\epsilon}.
\end{equation}
The equilibrium solution 
\begin{equation}\label{eqn:HC}
	\tilde q_\ast
	= \frac{1}{\csqrt{2}} \begin{pmatrix}
		1\\
		1
	\end{pmatrix}
\end{equation}
corresponds to the zero eigenvalue of $W$. The general solution of the generalized Schrödinger equation \labelcref{eqn:46K},
\begin{equation}\label{eqn:HD}
	\tilde q(t)
	= c_1\tilde q_\ast + c_2e^{-2 \omega (t - t_\text{in})} \begin{pmatrix}
		1\\
		-1
	\end{pmatrix},
\end{equation}
is the equivalent of \cref{eqn:FS}, with $\xi^{-1} = 2\omega$.

Another example is the four-state oscillator chain with step evolution operator \labelcref{eqn:GD}. For small $\eta$, we define $\omega = \eta/\epsilon$ and extract
\begin{alignedeqn}\label{eqn:HE}
	J
	&= \frac{1}{2}
	\begin{pmatrix}
		-2\omega & \omega & \omega & 0\\
		\omega & -2\omega & 0 & \omega\\
		\omega & 0 & -2\omega & \omega\\
		0 & \omega & \omega & -2\omega
	\end{pmatrix}
	,\\
	H
	&= \frac{i}{2}
	\begin{pmatrix}
		0 & -\omega & \omega & 0\\
		\omega & 0 & 0 & -\omega\\
		-\omega & 0 & 0 & \omega\\
		0 & \omega & -\omega & 0
	\end{pmatrix}.
\end{alignedeqn}
The general solution of the evolution equation \labelcref{eqn:46K} shows damped oscillations, as expected.

\subsection{Complex wave function and generalized Schrödinger equation}

In quantum mechanics we are used to employ complex wave functions. Rather trivially we can write every complex wave function as a real wave function with twice the number of components. A hermitian Hamiltonian becomes in this real language an antisymmetric purely imaginary matrix, such that $-i H$ is real, as in \cref{eqn:46K}. In the opposite direction a real wave function can be combined to a complex wave function with half the number of components if a suitable complex structure exists. This complex structure is very useful in quantum mechanics. One often encounters a complex structure also for the evolution in classical statistics described in this paper. A complex structure does not need a unitary evolution.

For many interesting cases there exists a basis for which the $N \times N$-matrix $W$ can be written in terms of $N/2 \times N/2$-matrices $W_1$ and $W_2$ in the form
\begin{equation}\label{eqn:CS1}
	\begin{pmatrix*}[r]
		W_1 & W_2\\
		-W_2 & W_1
	\end{pmatrix*}.
\end{equation}
We can then write the evolution equation in the form of a complex generalized Schrödinger equation,
\begin{equation}\label{eqn:CS2}
	i\partial_t\psi
	= G\psi,
	\quad
	 G
	= \hat H + i\hat J.
\end{equation}
For the generalized Schrödinger equation $G$ is not hermitian. The hermitian and antihermitian parts are associated to 
$\hat H = \hat H^\dagger$ and $\hat J = \hat J^\dagger$, respectively, with
\begin{equation}\label{eqn:CS3}
	\hat H= W_{2S} + iW_{1A},
	\quad
	\hat J
	= W_{1S} - iW_{2A},
\end{equation}
and $W_{iS},W_{iA}$ the symmetric and antisymmetric parts of $W_i$.

According to the block structure \labelcref{eqn:CS1} we group the components of $\tilde q$ as
\begin{equation}\label{eqn:CS4}
	\tilde q= \begin{pmatrix}
		\tilde q_R\\
		\tilde q_I
	\end{pmatrix}
\end{equation}
and define the complex wave function as
\begin{equation}\label{eqn:CS5}
	\psi
	= \tilde q_R + i\tilde q_I.
\end{equation}
The equivalence of \cref{eqn:CS2} with \cref{eqn:46K} is established by insertion of \cref{eqn:CS5}. The complex $N/2 \times N/2$-matrices $\hat H$ and $\hat J$ correspond to the antisymmetric and symmetric parts of $W$, respectively
\begin{alignedeqn}\label{eqn:CS6}
	W_A
	= -i H
	= \begin{pmatrix*}[r]
		W_{1A} & W_{2S}\\
		-W_{2S} & W_{1A}
	\end{pmatrix*},\\
	W_S
	= J
	= \begin{pmatrix*}[r]
		W_{1S} & W_{2A}\\
		-W_{2A} & W_{1S}
	\end{pmatrix*}.
\end{alignedeqn}

For the conjugate wave function we employ
\begin{equation}\label{eqn:CS7}
	\bar\psi
	= \bar q_R - i\bar q_I.
\end{equation}
For $\tilde W = W$, \cref{eqn:FF4} is transformed to
\begin{equation}\label{eqn:CS8}
	-i\partial_t\bar\psi
	= \tran{G} \, \bar\psi
	= (\hat H^\ast + i\hat J^\ast)\bar\psi.
\end{equation}
For quantum systems with $\bar q = \tilde q$ one has $\bar\psi = \psi^\ast$. For $\hat J = 0$, \cref{eqn:CS8} is the complex conjugate of \cref{eqn:CS2}. Thus for antisymmetric $W$ and $\bar{q} = \tilde{q}$ we recover the standard complex Schrödinger equation of quantum mechanics if a complex structure \labelcref{eqn:CS1} exists. With \cref{eqn:CS7} and $\bar\psi = \psi^\ast$ it is easy to verify that \cref{eqn:SID} becomes in the complex formulation the standard expression of quantum mechanics for the expectation value of observables.

For a generic evolution with $J \neq 0$ one observes
\begin{equation}\label{eqn:CS9}
	\partial_t (\tran{\bar\psi} \, \psi)
	= 0,
\end{equation}
in accordance with \cref{eqn:84}. On the other hand, one finds
\begin{equation}\label{eqn:CS10}
	\partial_t(\psi^\dagger\psi)
	= 2\psi^\dagger\hat J\psi.
\end{equation}
The form of \cref{eqn:CS10} shows that the antihermitian part of $G$ acts as a generalized damping term that can change the norm $|\psi|$.

\subsection{Density matrix}

For ``pure states'' the real classical density matrix $\rho^\prime_{\tau\rho}(t)$ obtains by multiplication of the wave function with its conjugate
\begin{equation}\label{eqn:D1a}
	\rho^\prime_{\tau\rho}(t)
	= \tilde q_\tau(t)\bar q_\rho(t).
\end{equation}
(Primes are used here in order to make the distinction to an equivalent complex formulation more easy to follow.) The diagonal elements are the local probabilities (no sum here)
\begin{equation}\label{eqn:D2a}
	p_\tau (t)
	= \rho^\prime_{\tau\tau}(t).
\end{equation}
This holds by virtue of \cref{eqn:97A2}, and no particular conditions on $\tilde q$ or $\bar q$ need to be imposed except for the normalization $Z = 1$.

For pure states the quantum expression for expectation values of local observables follows directly from \cref{eqn:SID},
\begin{equation}\label{eqn:174A}
	\langle A(t)\rangle
	= A^\prime_{\rho\tau} \, \rho^\prime_{\tau\rho}(t)
	= \Tr(A^\prime \, \rho^\prime).
\end{equation}
Mixed states can be obtained for the general boundary conditions \labelcref{eqn:I2}. The expression \labelcref{eqn:174A} continues to hold.

The time evolution of the density matrix follows directly from \cref{eqn:87,eqn:90}
\begin{equation}\label{eqn:D9a}
	\rho^\prime_{\tau\rho}(t + \epsilon)
	= S_{\tau\alpha}(t)\rho^\prime_{\alpha\beta}(t)S^{-1}_{\beta\rho}(t).
\end{equation}
In the continuum limit, $\epsilon \to 0$, one finds with \cref{eqn:FF1,eqn:FF2,eqn:FF4} and for sufficiently smooth $q(t)\approx \bigl( q(t + \epsilon) + q(t - \epsilon)\bigr) /2$,
\begin{alignedeqn}\label{eqn:D11a}
	\partial_t\rho^\prime_{\tau\rho}
	= W_{\tau\alpha}(t)\rho^\prime_{\alpha\rho}(t) - \rho^\prime_{\tau\alpha}(t)\tilde W_{\alpha\rho}(t).
\end{alignedeqn}
For $S$ independent of $t$ one has $W = \tilde W$. More generally, we concentrate in the following on the case $\tilde W(t) = W(t)$, for which
\begin{alignedeqn}\label{eqn:231}
	\partial_t\rho^\prime
	= [W,\rho^\prime].
\end{alignedeqn}

If $W$ admits a complex structure according to \cref{eqn:CS1} the pure state density matrix obtains from the complex wave function $\psi$ and conjugate wave function $\bar\psi$ as 
\begin{equation}\label{eqn:163AA}
	\rho_{\lambda\sigma}(t)
	= \psi_\lambda(t)\bar\psi_\sigma(t).
\end{equation}
This generalizes to mixed states according to the generalized boundary condition \labelcref{eqn:I2},
\begin{equation}\label{eqn:164AB}
	\rho_{\lambda\sigma}(t)
	= \sum_\alpha w_\alpha\psi^{(\alpha)}_\lambda(t) \bar\psi^{(\alpha)}_\sigma(t).
\end{equation}
The complex evolution equation is the generalization of the von Neumann equation to non-hermitian $G$,
\begin{equation}\label{eqn:164AC}
	i\partial_t\rho
	= [G,\rho]
	= [\hat H,\rho] + i[\hat J,\rho].
\end{equation}

We conclude that the quantities carrying the local information, such as the wave functions $\tilde q$, $\bar q$ or $\psi$, $\bar\psi$ or the density matrices $\rho^\prime$ or $\rho$, all obey linear evolution equations. The superposition principle holds. Memory materials can be realized if $J$ or $\hat J$ vanish on a subspace with more than one (real) dimension. (The one-dimensional case corresponds to a unique equilibrium state.) Oscillating local probabilities further require $H \neq 0$ or $\hat H \neq 0$.

Similar to quantum mechanics, the density matrix is a convenient concept for the description of subsystems. ``Reduced'' or ``coarse grained'' density matrices for subsystems can be obtained by taking a partial trace. A necessary condition for a unitary evolution of the subsystem is the vanishing of $J$ or $\hat{J}$ for the reduced evolution equation. In other words, the symmetric part of $W$ or the antihermitian part of $G$ have to commute with the reduced $\rho$. This condition is, however, not sufficient to guarantee a non-trivial behavior of a subsystem as undamped oscillations. An additional condition is the realization of the potentially oscillating behavior by appropriate boundary conditions.

A quantum behavior for subsystems typically holds for idealized isolated subsystems. In many circumstances the isolation may only be approximate, resulting in decoherence \cite{ZE,JZ,ZU} or syncoherence \cite{CWQM} for the subsystem. Decoherence or syncoherence in subsystems can be described by additional terms in the evolution equation. In their presence a pure state density matrix can evolve into a mixed state density matrix (decoherence), or a mixed state can evolve into a pure state (syncoherence).

\section{Basic concepts and quantum formalism}
\label{sec:basic concepts}

Before proceeding to detailed examples of static memory materials it may be useful to summarize the main features of our formalism of classical wave functions and density matrices. For readers less interested in the formal aspects of this paper we have written this section to be self-contained, necessarily involving some repetition of material of the previous sections. Our formalism is based on the notion of a ``classical wave function'' $\tilde q(t)$ and the conjugate wave function $\bar q(t)$ \cite{CWCW}. Here $t$ denotes the location of a hypersurface in the bulk, with boundary conditions set at $t_\text{in}$ and $t_\text{f}$. The expectation values of local observables can be computed from $\tilde q(t)$ and $\bar q (t)$. It is therefore sufficient to understand the $t$-dependence or ``evolution'' of these wave functions.

It is remarkable that the formalism for information transport in classical statistical systems turns out to be conceptually close to quantum mechanics. It resembles Feynman's derivation of the wave function from the path integral for quantum mechanics \cite{FEY}. In particular, the change of local probabilities between different positions of hypersurfaces in the bulk is described by a linear evolution equation for a classical density matrix, rather than by the local probabilities alone. Indeed, we can construct from $\tilde q(t)$ and $\bar q(t)$ a ``classical density matrix'' $\rho^\prime(t)$ at a given location $t$. It is bilinear in $\tilde q(t)$ and $\bar q(t)$, $\rho^\prime_{\tau\rho} = \tilde q_\tau\bar q_\rho$, and permits to compute expectation values by the standard quantum formula
\begin{equation}\label{eqn:I7}
	\langle A(t)\rangle = \Tr \bigl\{A^\prime(t) \, \rho^\prime(t)\bigr\}.
\end{equation}
Similar to quantum mechanics $A^\prime(t)$ is an operator associated to the local observable $A(t)$. The local probabilities are the diagonal elements of the density matrix $\rho^\prime(t)$. The evolution law for the density matrix is linear, while any formulation in terms of the local probability distribution alone would result in a complicated non-linear equation. It is the presence of additional local information in the off-diagonal elements of the density matrix that renders the evolution simple and allows for the superposition principle for solutions.

The central issue for information transport is the evolution of the wave function between two neighboring points or hypersurfaces $t$ and $t + \epsilon$. The evolution law for the classical wave function is linear
\begin{equation}\label{eqn:8FA}
	\tilde q(t + \epsilon)
	= S\tilde \, q(t),
\end{equation}
%(9)
with $S$ the ``step evolution operator''. The step evolution operator is related to the transfer matrix \cite{TM,MS,FU} by a suitable multiplicative renormalization. For the step evolution operator $S$ all eigenvalues obey $|\lambda|\leq 1$, with a set of ``largest eigenvalues'' $|\lambda| = 1$.

One often can define a continuum limit $\epsilon \to 0$, which reads for $t$-independent $S$ 
\begin{equation}\label{eqn:8FB}
	\partial_t\tilde q
	= \frac{\dif \tilde q}{\dif t}
	= W\tilde q,
\end{equation}
where 
\begin{equation}\label{eqn:8FC}
	W
	= \frac{1}{2\epsilon} (S - S^{-1}).
\end{equation}
\Cref{eqn:8FB} is a linear differential equation for the wave function and constitutes the generalized Schrödinger equation.

The usual complex Schrödinger equation can always be written in the real form \labelcref{eqn:8FB} by splitting a complex wave function into real and imaginary parts, $\psi = \tilde q_R + i\tilde q_I$, such that it becomes a real linear first order differential equation for $\tilde q = (\tilde q_R,\tilde q_I)$. Inversely, if $W$ has appropriate properties for the introduction of a complex structure (cf. \cref{sec:generalized schrödinger equation}) we can write \cref{eqn:8FB} in the familiar complex form
\begin{equation}\label{eqn:11AA}
	i\partial_t\psi
	= G\psi.
\end{equation}

It is a key difference between the evolution of the wave function for classical statistical systems and the Schrödinger equation for quantum mechanics that $G$ is, in general, not hermitian, or $W$ not antisymmetric. Decomposing $G$ into its hermitian and antihermitian parts
\begin{equation}\label{eqn:11AB}
	G
	= \hat H + i\hat J,
	\qquad
	\hat H^\dagger
	= \hat H,
	\qquad
	\hat J^\dagger
	= \hat J,
\end{equation}
one finds that $\hat J$ is responsible for the loss of information as $t$ increases from some boundary or initial value $t_\text{in}$ into the bulk. For memory materials $\hat J\psi$ vanishes in the bulk. The evolution in the bulk obeys then the Schrödinger equation and is unitary. Static memory materials are therefore quantum simulators for which the dependence of observables on the location $t$ traces the time evolution of observables in a quantum system with the same $\hat H$.

Similarly, the dependence of the density matrix on $t$ obeys an evolution equation that is a generalization of the von Neumann equation in quantum mechanics
\begin{equation}\label{eqn:I8}
	\partial_t\rho^\prime(t)
	= [W,\rho^\prime(t)].
\end{equation}
In particular, this equation describes the $t$-dependence of the local probabilities, which are given by the diagonal elements of $\rho^\prime$, $p_\tau(t) = \rho^\prime_{\tau\tau}(t)$. In the presence of a complex structure \cref{eqn:I8} translates to a generalized von Neumann equation for the complex density matrix $\rho$,
\begin{equation}\label{eqn:11AC}
	i\partial_t\rho
	= [G,\rho].
\end{equation}
One recovers the unitary evolution according to the von Neumann equation for $\hat J = 0$, or more generally, for $[\hat J,\rho] = 0$. For $\hat J \neq 0$ \cref{eqn:11AC} one finds a modification of the von Neumann equation. Also in classical statistics a pure state density matrix remains pure state in the course of the evolution. For decoherence in subsystems one expects additional terms similar to the Lindblad equation \cite{KO,LI,ZO}.

It is remarkable that the simple linear time evolution \labelcref{eqn:I8,eqn:11AC} is formulated for a density matrix. No such simple evolution law can be formulated in terms of the local probabilities $p_\tau(t) = \rho^\prime_{\tau\tau}(t)$ alone. The density matrix arises as the natural object for the description of information transport in a completely classical statistical context. It is not an object specific to quantum systems. The off-diagonal elements of the density matrix should be considered as a genuine part of the local information at a given $t$. Only with this information, which goes beyond the local probabilities, a simple evolution law can be formulated.

The linearity of the evolution equation implies the superposition principle. Again, this central principle for the wave interpretation is not a specific quantum feature. It characterizes the classical statistical information transport as well. Particle-wave duality appears in classical statistics. The discrete ``particle properties'' are associated to the discrete values of observables, while the continuous wave aspect arises from the continuous probabilistic description in terms of classical wave functions or the density matrix.

If the step evolution operator $S$ admits a unique largest eigenvalue $|\lambda| = 1$, the evolution is characterized by the approach to a unique equilibrium state, which is the eigenstate to this largest eigenvalue. In this case the memory of boundary conditions is lost in the bulk, with a typical behavior as in \cref{eqn:I8-1}. Indeed, the eigenstates to eigenvalues $|\lambda| < 1$ are damped towards zero by multiple repetition of \cref{eqn:8FA}. Static memory materials can be realized if the set of largest eigenvalues of $S$ is not unique. In particular for complex largest eigenvalues, $\lambda = e^{i\alpha}$, $\alpha \neq 0,\pi$, one finds an oscillatory behavior similar to \cref{eqn:I6}. Such undamped oscillations occur in the subsector of wave functions that are eigenstates to eigenvalues $|\lambda| = 1$. The boundary information concerning this subsector is transported inside the bulk without loss of memory.

The hermitian and antihermitian parts of $G$ (or antisymmetric and symmetric parts of W) have a direct relation to the eigenvectors of the step evolution operator. For the subspace of eigenvectors to eigenvalues $|\lambda| = 1$ the length of the vector $\tilde q$ is conserved. The evolution within this subspace is unitary, with antisymmetric $W$ or hermitian $G$. The symmetric part of $W$ or antihermitian part of $G$ acts only on the subspace corresponding to eigenvalues $|\lambda| < 1$. These parts of the wave function go to zero for large $\Delta t = t - t_\text{in}$, typically exponentially with a correlation length $\xi$. For the asymptotic behavior inside the bulk the components of $\tilde q$ corresponding to $|\lambda| < 1$ can be neglected. The asymptotic evolution inside the bulk is therefore always the unitary ``quantum evolution''.

For systems with a unique equilibrium state (unique eigenvalue $|\lambda| = 1$) the ``quantum evolution'' is trivial, however, corresponding to $\hat H = 0$. A non-vanishing Hamiltonian $\hat H$, with the typical associated oscillatory behavior of the wave function, can only be realized if the step evolution operator has more than one eigenvalue $|\lambda| = 1$. (Memory of boundary conditions can also be conserved for $\hat H = 0$, provided the equilibrium is not unique. In this case the memory corresponds to $t$-independent expectation values of local observables that depend on the boundary conditions.)

The asymmetric diagonal Ising model \labelcref{eqn:6XA} with $\beta \to \infty$ is a memory material for which all eigenvalues of the step evolution operator obey $|\lambda| = 1$. The complete boundary information is transported into the bulk. Correspondingly, the classical wave function obeys a real Schrödinger equation \labelcref{eqn:8FB} with antisymmetric $W$. We will find that this model actually describes the quantum field theory for free massless relativistic fermions in two-dimensional Minkowski space. This suggests that the association of the variable $t$ with an emergent effective time may have a deeper meaning.

In a complex formulation the Schrödinger equation for a one-particle excitation involves the momentum operator $\hat P$,
\begin{equation}\label{eqn:NF1}
	i\partial_t\psi
	= \hat H\psi
	= \hat P\psi,
	\quad
	\hat P
	= -i\partial_x,
\end{equation}
with generalizations to multi-particle states. The information in the initial wave function $\psi(t_\text{in},x)$ is transported completely to the bulk,
\begin{equation}\label{eqn:NF2}
	\psi(t,x)
	= \psi(t_\text{in},x - t + t_\text{in}).
\end{equation}
Initial oscillations in $x$ translate to oscillations in $t$ for fixed $x$, e.g.
\begin{equation}\label{eqn:NF3}
	\psi(t,x)
	= c_\mathcal{N}\bigl(\cos\bigl[\omega(x - t + t_\text{in}) + \alpha\bigr] + 1\bigr).
\end{equation}
Many other initial conditions are possible. Distributions that are located around $x_0$ at $t_\text{in}$ will propagate into the bulk as wave packets, localized around $x_0 + t - t_\text{in}$.

The realization of such memory materials could offer new possibilities for information processing. A large amount of information could be transported. In the ideal case these are complete sequences of bits. Even if individual bits can not be controlled separately, any probabilistic distribution of initial bits will be transported to the bulk. Furthermore, the initial information is available at every $t$. The shift of characteristic features in $x$ for different $t$ could also be exploited.

\section{Simple static memory materials}
\label{sec:simple static memory materials}

In \crefrange{sec:simple static memory materials}{sec:general static memory materials} we discuss detailed examples and general features of static memory materials. Static classical statistical systems with the property that the information about the boundary conditions at $t_\text{in}$ is not completely lost at the other boundary at $t_\text{f}$, or inside the bulk at arbitrary $t$, may be called ``static memory materials''. In short, the material keeps memory of its boundary condition. Memory materials can be realized in the presence of more than one eigenvalue $\lambda$ of the step evolution operator $S$ with $|\lambda| = 1$. If all eigenvalues of $S$ obey $|\lambda| = 1$ the information is completely transmitted. If there is only a subset of eigenvalues $|\lambda_i| = 1$, information relating to a subsystem corresponding to this subset is transmitted, while other parts of the boundary information can be lost by exponential damping. The limits between the different cases get somewhat washed out in the presence of eigenvalues with $|\lambda|$ only slightly smaller than one. If some of the eigenvalues $\lambda_i$ with $|\lambda_i| = 1$ have an imaginary part the local probability distribution will typically show oscillations. In this section we discuss materials for which all eigenvalues of the step evolution operator obey $|\lambda_i| = 1$. In \cref{sec:general static memory materials} we investigate systems where only a subset of $\lambda_i$ has unit absolute value.

\subsection{Unique jump chains}

One of the simplest examples for a memory material is the ``unique jump chain''. It is a one dimensional chain, labeled by positions $t_i$, on which an excitation, defect or impurity, called ``particle'' for our purpose, can propagate or ``jump'' from a given property at $t_\text{in}$ to another one at the next neighbor at $t_i + \epsilon$. Let the number of particles first be conserved, e.g. the same for all $t_i$. The chain may have at each $t_i$ several sites labeled by $\gamma$. Alternatively, $\gamma$ may label different internal properties of the particle. For a unique jump chain there is a unique possibility for every particle configuration at $t$ to jump to a particle configuration at $t + \epsilon$. All elements of the step evolution operator $S$ are one or zero, with a single one in each column. This property should hold for jumps in both directions, e.g. from $t$ to $t + \epsilon$ or from $t + \epsilon$ to $t$, such that $S$ is invertible, with a single one in each row. Thus $S$ is a rotation matrix and unique jump chains are quantum systems. If we take boundary conditions such that $\bar q = \tilde q$ the evolution can be described by a normalized classical wave function $q(t)$.

The realization by Ising spins involves constrained Ising models, as discussed in \cref{app:generalized Ising models}. The elements $M_{\tau\rho}$ in \cref{eqn:S10} vanish or diverge according to \cref{eqn:S12}. For our particular example neighboring configurations with a different number of particles are forbidden. For neighboring configurations with the same particle number all possibilities except one are forbidden as well. For the simplest case with two species, $\gamma = 1,2$, there are two possibilities for neighboring one particle states
\begin{equation}\label{eqn:M1}
	\begin{array}{rcc}
		t + \epsilon: & 01~10 & 01~10\\
		t: & \underbrace{01~10}_{(a)} & \underbrace{10~01}_{(b)} 
	\end{array}.
\end{equation}
While $(a)$ corresponds to $S = 1$, $(b)$ is realized for $S$ given by \cref{eqn:GH}.

The limitation to a conserved particle number is instructive, but not necessary. What is needed is only the uniqueness of the jump. The step evolution operator \labelcref{eqn:GN} can be seen as particle number non conserving jumps. Alternatively, this can be viewed as a single particle with four different internal properties labeled by $\tau$. The period of oscillations is $4\epsilon$ in this case.

For $M$ difference species of occupation numbers $n_\gamma(t),\gamma = 1\dots M$, $\tau = 1\dots 2^M$, a suitable choice of $S$ can achieve oscillations with a maximum period of $\tau_p = N\epsilon$, $N = 2^M$. Smaller periods are also possible, by closing circles of jumps after $n$ steps. Thus possible periods are 
\begin{equation}\label{eqn:M2}
	\tau_p
	= n \, \epsilon,
	\quad
	2 \leq n \leq N.
\end{equation}
There is no need that $t_\text{f} - t_\text{in}$ precisely corresponds to the completion of some oscillation. Different subsystems may also exhibit different periods. We conclude that for large $G = (t_\text{f} - t_\text{in})/\epsilon$ and large $M$ rather arbitrary time evolutions of expectation values $\langle A(t)\rangle$ can be realized without loss of memory. (A minimal period of $2\epsilon$ is simply a consequence of discrete time steps.)

The way to realize unique jump chains is to forbid certain types of neighboring configurations, implementing zero elements $S_{\tau\rho}$. Characterizing Ising models by $M_{\tau\rho}$ in \cref{eqn:S10,eqn:S12}, with step evolution operators $S_{\tau\rho} = \exp(-M_{\tau\rho})$, this can be achieved for classical Ising spin systems by letting $M_{\tau\rho} \to \infty$ for the forbidden combinations of sets of occupation numbers. If the unique allowed finite $M_{\tau\rho}$ have all the same value, the normalization of the probability distribution by an additive constant in $\mathcal{L}$ will lead for the finite values to $M_{\tau\rho} = 0$, $S_{\tau\rho} = 1$. As an alternative we can achieve the unique jumps by letting $M_{\tau\rho} \to -\infty$ for the allowed transition, keeping finite $M_{\tau\rho}$ for the forbidden transitions. If all $M_{\tau\rho}$ are equal for the allowed jumps, the normalization of the step evolution operator will again result in $S_{\tau\rho} = 1$ or $0$ for the allowed or forbidden jumps.

More generally, it is sufficient that the elements $M_{\tau\rho}$ belong to two classes with small and large values. All elements of the class with small $M_{\tau\rho}$ should be equal. It is then sufficient that the difference between the large $M_{\tau\rho}$ and the small ones diverges. After subtraction of the additive normalization constant the small elements all equal zero, while all large elements diverge. The infinite difference between the large and small elements of $M$ can be achieved by the ``zero temperature limit'' $\beta \to \infty$ similar to the Ising model.

Unique jumps chains bare resemblance to deterministic cellular automata \cite{CA,ICJ,TH,TH2,EL}. For a given initial sequence of occupation numbers at $t_\text{in}$ the sequence at every later time step is uniquely fixed. For every time step from $t$ to $t + \epsilon$ the sequence changes in a deterministic way. The probabilistic element concerns then only the probabilities of initial sequences at $t_\text{in}$. This is determined by both the boundary factors at $t_\text{in}$ \textit{and} $t_\text{f}$.

\subsection{Single-particle propagation in two dimensions}

Let us next describe a model for the propagation of a single particle, realized by a single impurity, excitation or defect, in two dimensions. We use the $N = 2^M$ values of $\tau$ in order to label the coordinate $x$ of a second dimension. Correspondingly, the wave function $f_\tau(t)$ can be written as $f(t,x)$. Neighboring $\tau$ correspond to neighboring $x$, with $\tau = (x - x_\text{in} + \epsilon)/\epsilon$. For finite $M$ and finite $\epsilon$ the coordinate $x$ extends from $x_\text{in}$ to $x_\text{f} = x_\text{in} + (N - 1)\epsilon$. We may adopt for simplicity periodic boundary conditions with $x_\text{f} + \epsilon$ identified with $x_\text{in}$, or 
\begin{equation}\label{eqn:M2A}
	x_\text{in} + N \epsilon
	= x_\text{in}.
\end{equation}

The step evolution operator $S_{\tau\rho}$ is assumed to be independent of $t$ and can be written as $S(x,y)$. Let us consider the particular unique jump chain
\begin{equation}\label{eqn:M3}
	S(x,y)
	= \delta(x - \epsilon,y),
\end{equation}
with discrete $\delta$-function $\delta(x,y) = 1$ for $x = y$ and $\delta(x,y) = 0$ otherwise. This implies the evolution of the wave function
\begin{equation}\label{eqn:M4}
	q(t + \epsilon,x)
	= q(t,x - \epsilon).
\end{equation}
The matrix $S$ has ones just above the diagonal. For $N = 4$ and periodic boundary conditions it reads
\begin{equation}\label{eqn:M5}
	S
	= \begin{pmatrix}
	0 & 1 & 0 & 0\\
	0 & 0 & 1 & 0\\
	0 & 0 & 0 & 1\\
	1 & 0 & 0 & 0
	\end{pmatrix},
\end{equation}
describing an evolution of the same type as for \cref{eqn:GN}, with period $4\epsilon$.

For large $N$ we may consider the jump from $x$ to $x + \epsilon$ as a small change and use the approximation of a continuous evolution equation. From 
\begin{equation}\label{eqn:M6}
	q(t + \epsilon,x) - q(t - \epsilon,x)
	= q(t,x - \epsilon) - q(t,x + \epsilon),
\end{equation}
we obtain, by dividing both sides by $2\epsilon$,
\begin{equation}\label{eqn:M7}
	\partial_t q(t,x)
	= -\partial_x q(t,x).
\end{equation}
This describes the propagation of a particle moving to the positive $x$-direction as $t$ increases. The general solution reads
\begin{equation}\label{eqn:M8}
	q(t,x)
	= f(t - x).
\end{equation}
Multiplying \cref{eqn:M7} formally by $i$ yields a Schrödinger equation, with hermitian Hamiltonian $H = P = -i\partial_x$. The solutions of this quantum system admits indeed real wave functions. With 
\begin{equation}\label{eqn:M11}
	S^{-1}(x,y)
	= \tran{S}(x,y)
	= S(y,x)
	= \delta(x + \epsilon,y),
\end{equation}
the conjugate wave function shows the same evolution as $q$, and the local probabilities obey $p(t + \epsilon,x) = p(t,x - \epsilon)$.

Formally, the insertion of \cref{eqn:M3} into the definition \labelcref{eqn:FF1} yields indeed an antisymmetric matrix $W$,
\begin{equation}\label{eqn:M9A}
	W(x,y)
	= \frac{1}{2\epsilon}\bigl(\delta(x - \epsilon,y) - \delta(x,y - \epsilon)\bigr),
\end{equation}
where we use
\begin{equation}\label{eqn:M9B}
	S^{-1}(x,y)
	= \delta(x,y - \epsilon).
\end{equation}
For periodic boundary conditions in $x$, \cref{eqn:M2A}, the time-period is $N\epsilon$. The different powers $S^n$ can be seen as elements of the abelian rotation group $SO(2)$, more precisely the subgroup $Z_N$. With $S^N = 1$ and $S^n \neq 1$ for $1\leq n < N$ the $N$ eigenvalues of $S$ are given by 
\begin{equation}\label{eqn:M10}
	\lambda_m
	= \exp\biggl(\frac{2\pi im}{N}\biggr),
	\quad
	|\lambda_m|
	= 1,
	\quad
	1 \leq m \leq N.
\end{equation}
For $N \to \infty$ we can interpret $S$ as an infinitesimal $SO(2)$-rotation.

The Schrödinger equation \labelcref{eqn:M7} describes the propagation of a single Majorana-Weyl fermion \cite{CWMW} in one time and one space dimension. The occupation number is one or zero, as appropriate for a fermion. The Weyl condition implies that the particle moves only in one direction, towards increasing $x$ in our case. The Majorana condition permits a real one-component wave function $q(t,x)$. Majorana-Weyl spinors do not admit a mass term, implying the energy-momentum relation $E = |p|$. It is remarkable that a quantum system with time evolution arises from a classical statistical setting that has not introduced time as an external parameter. The coordinate $t$ has been introduced just as a position label on a one-dimensional chain. Nevertheless, all properties of the quantum theory for a Majorana-Weyl fermion in two-dimensional Minkowski space are represented by this classical statistical system.

\section{Ising models for massless relativistic fermions}
\label{sec:ising models}

In this section we describe a class of asymmetric diagonal Ising models for which the ``zero temperature state'' for $\beta \to \infty$ is exactly solvable. They are perfect memory materials for which the boundary information is completely transmitted to the bulk. The solutions of the evolution equation can be interpreted as the propagation of an arbitrary number of free massless fermions in two-dimensional Minkowski space.

\subsection{Multi-fermion systems}

In the previous example we have treated $t$ and $x$ in a different fashion. Whereas $t$ labels different occupation numbers, $x$ has labeled different sequences of $M$ occupation numbers with values one or zero. We next discuss memory materials where $t$ and $x$ are treated equally as locations on a two-dimensional lattice. Instead of $n_\gamma(t)$ we now discuss occupation numbers $n(t,x)$, with $x$ corresponding to $\gamma$. For periodic boundary conditions in $x$ we have now $M$ different values for $x$, $x = x_\text{in} + (m - 1)\epsilon$, $1\leq m\leq M$, $x_\text{f} = x_\text{in} + (M - 1)\epsilon$, $x_\text{in} + M\epsilon = x_\text{in}$.

A static memory material can be realized as an Ising type model with interactions only among diagonal neighbors in one direction
\begin{alignedeqn}\label{eqn:M12}
	\mathcal{L}(t)
	&= -\beta\sum_x\Bigl\{n(t + \epsilon,x + \epsilon)n(t,x)\\
	& + \bigl( n(t + \epsilon,x + \epsilon) - 1\bigr)\bigl( n(t,x) - 1\bigr) - 1\Bigr\}.
\end{alignedeqn}
We consider an attractive interaction, $\beta > 0$, and the limit $\beta \to \infty$. Equivalently, we can write the classical action $S_\text{cl} = \sum_t\mathcal{L}(t)$ in terms of Ising spins $s(t,x)$,
\begin{alignedeqn}\label{eqn:M13}
	\mathcal{L}(t)
	= -\frac{\beta}{2}\sum_x\bigl\{s(t + \epsilon,x + \epsilon)s(t,x) - 1\bigr\}.
\end{alignedeqn}
This ``asymmetric diagonal Ising model'' can be solved exactly in terms of the free propagation of an arbitrary number of Weyl fermions.

This solution is easily understood by considering two neighboring sequences of $x$-occupation numbers, e.g. $\bigl\{n(t + \epsilon,x)\bigr\}$ and $\bigl\{n(t,x)\bigr\}$. Whenever the occupation numbers $n(t + \epsilon,x + \epsilon)$ and $n(t,x)$ are the same, either both one or both zero, the factor $\mathcal{K}(t)$ in \cref{eqn:128B1} receives a factor one. If they are different, $n(t + \epsilon,x + \epsilon) \neq n(t,x)$, a factor $e^{-\beta}$ occurs. The leading term in $\mathcal{K}(t)$ thus copies the sequence $n(t,x)$ from $t$ to $t + \epsilon$, now translated by one unit in the positive $x$-direction. For this configuration of two neighboring displaced sequences $\mathcal{K}(t)$ assumes the value $\mathcal{K}(t) = 1$. In other words, the leading term copies a bit sequence $\bigl\{n(x)\bigr\}$ at $t$ to $t + \epsilon$, displacing all bits by one unit in $x$. A copy error in one bit reduces $\mathcal{K}(t)$ by a factor $e^{-\beta}$, and errors in $k$-bits produce a penalty $e^{-k\beta}$. In the limit $\beta \to \infty$ the weight of configurations with copy errors goes to zero. This realizes a unique jump chain, where every sequence $\bigl\{n(x)\bigr\}$ at $t$ is mapped uniquely to a sequence $\bigl\{n^\prime(x)\bigr\}$ at $t + \epsilon$. This model is a perfect copy-machine of any initial sequence $\bigl\{n(x)\bigr\}$ at $t_\text{in}$ to later times, with an $\epsilon$-displacement in the $x$-direction for every $\epsilon$-advance in $t$.

The wave function $f(t)$ is a function of the configuration $\bigl[n(x)\bigr]$ of occupation numbers $n(t,x)$, $f(t) = f\bigl( t;\bigl[n(x)\bigr]\bigr)$. In this language the general exact solution of the evolution equation takes the form 
\begin{equation}\label{eqn:188A}
	f\bigl( t + \epsilon; \bigl[n(x)\bigr]\bigr)
	= f\bigl( t;\bigl[\tilde n(x)\bigr]\bigr),
\end{equation}
where $[\tilde n(x)]$ obtains from $[n(x)]$ by shifting all zeros and ones of the sequence by one place to the left. In other words, at $t + \epsilon$ the value of $f$ for a given configuration $\bigl[n(t + \epsilon,x)\bigr]$ is the same as the value of $f$ at $t$ for a configuration $\bigl[\tilde n(t,x)\bigr]$.

We may phrase this situation in terms of the transfer matrix $\bar S$. A given sequence $\bigl[n(x)\bigr]$ corresponds to a given basis function $h_\tau$. The sequence displaced by one unit, e.g. $n^\prime(x) = n(x + \epsilon)$, corresponds to a different basis function $h_{\alpha(\tau)}$. (The map $\tau \to \alpha(\tau)$ depends on the particular ordering in $\tau$.) The transfer matrix $\bar S_{\rho\tau}$ equals one whenever $\rho = \alpha(\tau)$, while all other elements are suppressed by factors $e^{-k\beta}$, with $k$ the ``number of errors''. The relative size of the suppression remains the same if we normalize $\bar S$ by a multiplicative factor in order to obtain the step evolution operator $S$. For $\beta \to \infty$ one has $\bar S = S$, where $S_{\rho\tau} = 1$ for $\rho = \alpha(\tau)$, and $S_{\rho\tau} = 0$ for $\rho \neq \alpha(\tau)$. We recognize a unique jump chain. The evolution of the wave function is given by 
\begin{equation}\label{eqn:M14}
	q_{\alpha(\tau)}(t + \epsilon)
	= q_\tau(t),
	\quad
	q_\tau(t + \epsilon)
	= q_{\alpha^{-1}(\tau)}(t).
\end{equation}

The displacement by one $x$-unit of the copied $\bigl[n(x)\bigr]$-sequence does not change the number of ones in this sequence. If we associate each $n(x) = 1$ with a particle present at $x$ (and $n(x) = 0$ with no particle present), the total number of particles is the same at $t$ and $t + \epsilon$. The step evolution operator conserves the particle number. We can therefore decompose the wave function $q_\tau(t)$ into sectors with different particle numbers $F$. They do not mix by the evolution and can be treated separately.

The sectors with $F = 0$ and $F = M$ are static. The sector with $F = 1$ describes the propagation of a single particle. It is characterized by a single-particle wave function $q^{(1)}(t,x)$, where $x$ denotes the location of the single particle or the position of the unique one in the sequence $\bigl[n(x)\bigr]$. Its evolution obeys \cref{eqn:M4}, and the previous discussion of the propagation of a single particle can be taken over completely for the wave function in this subsection. A similar discussion holds for $F = M - 1$, where $x$ denotes now the position of a hole.

The subsector with $F = 2$ is characterized by the positions $x$ and $y$ of the two particles. There is no distinction which particle sits at $x$ and which one at $y$, and we may use an antisymmetric wave function
­\begin{equation}\label{eqn:M15}
q(t;x,y)
	= -q(t;y,x).
\end{equation}
Using variables
\begin{equation}\label{eqn:M16}
	s
	= \frac{x + y}{2},
	\quad
	r
	= x - y
\end{equation}
the evolution obeys
\begin{equation}\label{eqn:M17}
	q(t + \epsilon;s,r)
	= q(t;s - \epsilon,r).
\end{equation}
\noindent
The distance $r$ between the occupied sites remains the same for all $t$. The evolution equation
\begin{equation}\label{eqn:M18}
	\partial_tq
	= -\partial_s q
	= -(\partial_x + \partial_y)q
\end{equation}
has the same structure as \cref{eqn:M7}, with inter-particle distance $r$ an additional label not affected by the evolution. The case $F = M - 2$ can be treated in the same way, now with two ``holes'' (sites with $n = 0$ in even environment of $n = 1$) playing the role of particles. This setting is easily generalized to sectors with arbitrary $F$, all inter-particle distances being conserved.

\subsection{Single-particle wave function}

For a single particle $(F = 1)$ the wave function $q(t,x)$ depends on the position $x$ of the particle or the unique Ising spin up. In the continuum limit the evolution reads
\begin{equation}\label{eqn:CF1}
	\partial_t q(x)
	= \int_y W(x,y)q(y),
\end{equation}
with $W(x,y)$ a derivative operator
\begin{equation}\label{eqn:CF2}
	W(x,y)
	= \frac{1}{2\epsilon}\bigl(\delta(x,y + \epsilon) - \delta(x,y - \epsilon)\bigr)
	= -\delta(x - y)\partial_y.
\end{equation}

The two-dimensional lattice with points
\begin{equation}\label{eqn:CF3}
	(t,x)
	= (t_\text{in} + n^\prime\epsilon,x_\text{in} + m^\prime\epsilon)
\end{equation}
can be composed of an ``even sublattice'' with $n^\prime + m^\prime$ even and an ``odd sublattice'' where $n^\prime + m^\prime$ is odd. The propagation of a particle on a diagonal does not mix the points of the even and odd sublattice. We can employ this observation for the introduction of a simple complex structure. The complex conjugation corresponds to an involution $K$ acting on $q(t,x) = q(n^\prime,m^\prime)$ by reversing the sign on the odd sublattice,
\begin{equation}\label{eqn:CF4}
	Kq(n^\prime,m^\prime)
	= (-1)^{n^\prime + m^\prime}q(n^\prime,m^\prime).
\end{equation}
For even $t$ we may restrict the positions to even $x$ by denoting
\begin{equation}\label{eqn:CF5}
	q(t,x)
	= q_R(t,x),
	\qquad
	q(t,x + \epsilon)
	= q_I(t,x),
\end{equation}
while for odd $t$ we take odd $x - \epsilon$ with 
\begin{equation}\label{eqn:CF6}
	q(t,x - \epsilon)
	= q_R(t,x - \epsilon),
	\quad
	q(t,x)
	= q_I(t,x - \epsilon).
\end{equation}
The complex wave function is defined by 
\begin{equation}\label{eqn:CF7}
	\psi(t,x)
	= q_R(t,x) + iq_I(t,x).
\end{equation}
The map $K$ translates to the complex conjugation of $\psi$.

The matrix $W$ does not mix $q_R$ and $q_I$. It is antisymmetric and acts in the same way on the two blocks $q_R$ and $q_I$, such that in the language of \cref{eqn:CS6} one has $W_{1S} = W_{2S} = W_{2A} = 0$. In the complex formulation $W$ is represented by the hermitian Hamiltonian
\begin{equation}\label{eqn:CF8}
	\hat H
	= \hat P,
	\qquad
	\hat P(x,y)
	= -i\delta (x - y)\partial_y.
\end{equation}
We recognize the momentum operator $\hat P$ and the standard form of the Schrödinger equation for a simple free massless Weyl fermion in two dimensions,
\begin{equation}\label{eqn:CF9}
	i \partial_t \psi
	= \hat P \, \psi.
\end{equation}

As an example for a specific initial condition we may consider 
\begin{alignedeqn}\label{eqn:CF10}
	q(t_\text{in},x)
	&= q_\text{in}(x)
	= c \cos^2\biggl(\frac{\omega x}{2} + \alpha\biggr),\\
	\psi(t_\text{in},x)
	&= \frac{1 + i}{\csqrt{2}}q_\text{in}(x),
\end{alignedeqn}
with normalization $c$ chosen such that 
\begin{equation}\label{eqn:CF11}
	\int_x q^2_\text{in}(x)
	= 1.
\end{equation}
With this initial condition the solution of the evolution equation \labelcref{eqn:CF9} reads
\begin{equation}\label{eqn:CF12}
	\psi(t,x)
	= \frac{c(1 + i)}{2\csqrt{2}} \Bigl\{\cos\bigl[\omega(x - t + t_\text{in}) + 2\alpha\bigr] + 1\Bigr\}.
\end{equation}
For fixed $x$ the solution oscillates with $t$, with period $2\pi/\omega$. This example provides for a simple existence proof of static memory materials with oscillating local probabilities in the bulk.

Depending on initial conditions the Schrödinger equation \labelcref{eqn:CF9} has many other solutions as, for example, propagating wave packets. The expectation value of the momentum operator $\hat P$ can be computed from the usual rule of quantum mechanics and is conserved. We also stress that a ``particle'' is an excitation with respect to a given ``ground state''. For the discussion above we have assumed the ground state to be the unique configuration with $F = 0$, e.g. all spins down. Many other ground states are possible, for example with half filling. The criterion for a possible ground state is its independence on $t$. All probability amplitudes $f(t_\text{in},x)$ that are invariant under translations in $x$ lead to such a ``ground state''. Local excitations (as compared to a given ground state) will be described by classical wave functions that obey \cref{eqn:CF1} -- see ref.~\cite{CWQFT} for a related discussion.

\subsection{Memory materials for Dirac, Weyl, Majorana and Majorana-Weyl fermions in two dimensions}

The Ising type model \labelcref{eqn:M13} describes a quantum field theory for free Weyl fermions in two-dimensional Minkowski space. (For a more detailed discussion of wave functions and the realization of Lorentz-symmetry cf. refs.~\cite{CWCMW,CWQFT}.) The extension to free Dirac fermions is straightforward. Dropping the Weyl constraint one has ``left movers'' as well as ``right movers''. A free massless Dirac fermion can be viewed as two Weyl fermions, one moving to the left and the other to the right. Correspondingly, we introduce two species of occupation numbers $n_\alpha(t,x)$ at each site of the lattice, $\alpha = 1,2$, $\gamma = (\alpha,x)$. The weight factor in the partition function is now specified by $(\beta \to \infty)$
\begin{alignedeqn}\label{eqn:M19}
	\mathcal{L}(t)
	= -\smash{\frac\beta2 \sum_x} \bigl\{&s_1(t + \epsilon,x + \epsilon)s_1(t,x)\\
	& + s_2(t + \epsilon,x - \epsilon)s_2(t,x) - 2\bigr\}.
\end{alignedeqn}
While the species $\alpha = 1$ describes the right movers, the species $\alpha = 2$ accounts for the left movers.

Each species $\alpha = 1,2$ can actually be viewed as describing two distinct Majorana-Weyl spinors. They correspond to particles on the even or odd sublattice. Altogether we distinguish four sorts of particles that propagate independently. They correspond to occupation numbers $n_\alpha^{(\text{even})}$ and $n_\alpha^{(\text{odd})}$, or $n_{\alpha R}$ and $ n_{\alpha I}$. The total number of particles for each sort is conserved separately. We therefore could introduce separate wave functions
\begin{equation}\label{eqn:M20}
	q^{F_{\alpha\delta}}_\sigma(t),
	\quad\alpha
	= 1,2,
	\quad\delta
	= R,I,
\end{equation}
with $F_{\alpha\delta}$ the number of particles of sort $(\alpha,\delta)$, and $\sigma$ denoting the locations of these particles at different $x$. They will not be mixed by the evolution.

A Majorana constraint can be imposed by eliminating one of the sublattices, keeping, for example, only the odd sublattice. The odd sublattice is again a quadratic lattice, now with lattice distance $\epsilon^\prime = \csqrt{2} \, \epsilon$. The diagonal neighbors on the original lattice become next neighbors on the sublattice. Majorana-Weyl or Majorana spinors can therefore be realized by Ising models with next-neighbor interaction. What is important is that species $\alpha = 1$ has only interactions in one direction, say the $z_1$-direction, while species $\alpha = 2$ has only interactions in the orthogonal $z_2$-direction. A memory material for Majorana spinors is realized for 
\begin{align}\label{eqn:M21}
	\smash{\sum_t\mathcal{L}(t)
	= -\frac\beta2 \sum_{z_1,z_2}}
	\bigl\{&s_1(z_1 + \epsilon^\prime,z_2)
	s_1(z_1,z_2)\\
	&+ s_2(z_1,z_2 + \epsilon^\prime)s_2(z_1,z_2) - 2\bigr\},\notag
\end{align}
with $\epsilon^\prime = \csqrt{2} \, \epsilon$. The coordinates $t$ and $x$ are given by
\begin{equation}\label{eqn:M22}
	t
	= \tfrac{1}{\csqrt{2}}(z_1 + z_2),
	\qquad
	x
	= \tfrac{1}{\csqrt{2}}(z_1 - z_2).
\end{equation}	
Initial boundaries and final boundaries at constant $t$ correspond now to diagonals on the $(z_1,z_2)$-lattice. Also the evolution described by the transfer matrix is in a diagonal direction.

We have described here two-dimensional fermions by an ``Euclidean quantum field theory'' for Ising spins. The fermionic character is expressed by the property that occupation numbers take values one or zero. We have not put much weight on the issue of statistics. Usually, quantum field theories for fermions are described by Grassmann functional integrals, which make the fermionic statistics manifest by the anticommuting properties of the Grassmann variables. In ref.~\cite{CWFGI} we construct an explicit map between the generalized Ising model \labelcref{eqn:M19} and an equivalent Grassmann functional integral.

\subsection{Experimental and computer realization of two-dimensional static memory materials}

One may wonder if static memory materials exist in nature for certain solids, if they can be technically produced, or if they can be experimentally simulated by ultracold atoms on a lattice. If one of the memory materials describing the propagation of free fermions could be realized for practical use in a computer, it would provide new algorithmic possibilities. Not only the information stored in whole bit sequences could be substantial. The realization of oscillating information or the transport of information along diagonal directions may open new aspects. The close analogy to quantum evolution may suggest some analogies to quantum computers.

A ``Majorana-Weyl material'' corresponds to \cref{eqn:M21} with $s_2$ left out. The crucial ingredient is the absence of the interaction in one of the directions and the limit of very large $\beta$. The limit $\beta \to \infty$ can be approached by lowering the temperature towards the ground state. The absence of other interactions is more difficult to realize. Consider a weight factor $e^{-S_\text{cl}}$
\begin{alignedeqn}\label{eqn:M23}
	S_\text{cl}
	&= -\frac\beta2 \sum_{z_1,z_2}\bigl\{s(z_1 + \epsilon^\prime,z_2)s(z_1,z_2)\\
	& + \sigma s(z_1,z_2 + \epsilon^\prime)s(z_1,z_2) - c\bigr\}.
\end{alignedeqn}
Depending on the sign of $\sigma$ the interaction in the $z_2$-direction will favor an alignment of spins in this direction, or alternating signs. Without imposing boundary conditions the ground state has then uniform signs or alternating stripes. In both cases it is unique up to an overall flip of signs for all spins. One may therefore expect that a large part of the information is damped out for any $\sigma \neq 0$. In the presence of fixed values of the spins at the initial boundary the aligned state involves errors in the transmission of boundary information. For small enough $|\sigma|$ there will be a competition between the ordering tendency due to $\sigma$ and the memory of the initial spin sequence on a boundary.

For $\beta \to \infty$ the question if memory of the initial boundary is transmitted or not amounts to the determination of the minimum of $S_\text{cl}$ in the presence of fixed values of boundary spins. Consider first $\sigma > 0$ with a tendency to alignment of all spins. For simplicity we begin with the case where boundary conditions are given for fixed $z_1 = \tilde t$. One may compute the difference of the action $\Delta S_\text{cl}$ between a state where the initial information is transmitted and a state where it is lost. For the first state the spins at $\tilde x = z_2$ have for all $\tilde t$ the same value as at $\tilde t_\text{in}$, while for the second state they are all positive or all negative for all $\tilde t\geq\tilde t_\text{in} +\epsilon$. Consider an initial state where all $s(\tilde t_\text{in},\tilde x)$ are positive except for an interval $[\tilde x_-,\tilde x_+]$ with $D$ sites where they are all negative. The difference in action amounts to 
\begin{equation}\label{eqn:198A}
	\Delta S_\text{cl}
	= \beta \, \sigma \, G - \frac\beta2 \, D,
\end{equation}
since the ``memory state'' has $2G$ spin flips on the sides of the domains with negative spin, while the ``no-memory state'' has $D$ errors in the transition from $t_\text{in}$ to $t_\text{in} + \epsilon$. The boundary information is transmitted for $D > 2\sigma G$, and lost for $D < 2\sigma G$. For small $D$ and large $G$ very small $\sigma $ are required for memory transmission. Otherwise memory is transmitted only partially. For a given $\sigma$ and $G$ the transmitted part of the memory is the one stored in large enough domains, $D > 2\sigma G$. The situation for $\sigma < 0$ is similar. (An exception is the case of periodicity in $\tilde x$ with an odd number of $\tilde x$-points. This necessitates a defect in the stripe sequence, and the information of the location of the defect is transmitted for arbitrary values of $\sigma$.) Using similar arguments one finds that imposing boundary conditions at fixed $t$ leads to the same result as for boundary conditions at fixed $\tilde t$.

For transmitting a maximum amount of information one would attempt to realize $|\sigma|\ll1$. Since interactions in one of the directions can often not be eliminated completely a good strategy may be to realize some random distribution of $\sigma$ that is tunable such that $\sigma$ vanishes in the average. Another possibility is the use of symmetry. For $\sigma = 0$ the weight factor \labelcref{eqn:M23} is invariant under a reversal of the sign of all $s$ for $z_2$ odd, while $\sigma \neq 0$ breaks this symmetry.

Beyond the preparation of the material one needs the ability to impose an initial wave function on some boundary, for example by imposing a fixed sequence of spins on this boundary. Similarly, a read out mechanism has to measure the expectation values of spins on some ``final boundary''. If the boundaries are at fixed $z_1$ the transfer of information is in the $z_1$-direction. This realizes the mathematically trivial case $S = 1$. For a realization of the fermion system described above the boundaries should be at fixed $t$ and therefore on diagonals in the $(z_1,z_2)$-lattice.

We finally observe that the periodic boundary conditions in the $x$-direction introduced before are of a purely technical nature and not needed for real systems. The information propagation is a local phenomenon. For the information at a given $(t,x)$ only the ``past light cone'' at $\bigl( t_\text{in},x_\text{in} = x\pm(t - t_\text{in} )\bigr)$ plays a role. If the range in $x$ is large enough this light cone does not feel the presence of periodic boundary conditions or not. Our memory material realizes in a simple way the causal structure of a quantum field theory, with a maximal velocity of propagation of information. (In our units the ``speed of information'' or speed of light is set to one.)

While a material realization of static memory materials is an experimental challenge, their numerical realization on a computer seems rather straightforward. The probability distribution $p[n]$ can be sampled by a suitable Monte-Carlo algorithm. There is no problem to implement only diagonal interactions in one direction and to investigate the limit $\beta \to \infty$. Finite large $\beta$ can be used to investigate how the loss of memory sets in. Suitable boundary conditions can be imposed without particular problems. Expectation values of local observables can be ``measured'' for all $t$. Such a ``computer experiment'' should reveal the predicted oscillatory patterns.

\subsection{Oscillating local probabilities}

The asymmetric diagonal Ising model \labelcref{eqn:M13} with $\beta \to \infty$ provides for a simple example of a classical statistical system with oscillating local probabilities. The probability distribution
\begin{equation}\label{eqn:193XA}
	p[n]
	= p[s]
	= \exp\biggl\{\frac\beta2\sum_{t,x} \bigl[s(t + \epsilon,x + \epsilon) \, s(t,x) - 1\bigr]\biggr\} f_\text{in}
\end{equation}
is considered here for simplicity on a torus in the $x$-direction. We have taken $\bar f_\text{f} = 1$ while $f_\text{in} = f_\text{in}\bigl(s(t_\text{in},x)\bigr)$ specifies boundary values at $t_\text{in}$.The conjugate wave function $\bar q(t_\text{f})$ corresponding to $\bar f_\text{f} = 1$ reads $\bar q_\tau(t_\text{f}) = 1$. It is invariant under the evolution, resulting for all $t$ and $\tau$ in 
\begin{equation}\label{eqn:193XB}
	\bar q_\tau(t)
	= 1.
\end{equation}
For the particular conjugate wave function \labelcref{eqn:193XB} the local probability distribution is given by the wave function,
\begin{equation}\label{eqn:193XC}
	p_\tau(t)
	= \tilde q_\tau(t).
\end{equation}

Consider now a boundary term
\begin{equation}\label{eqn:193XD}
	f_\text{in}
	= f(t_\text{in})
	= \exp\biggl\{-\sum_x \tilde{h}(x) \, s(t_\text{in},x) + c\biggr\},
\end{equation}
with periodic $\tilde{h}(x),\tilde n \in \mathbb{Z}$
\begin{equation}\label{eqn:193XE}
	\tilde{h}\Bigl(x + \tfrac{2\pi\tilde n}{\omega}\Bigr)
	= \tilde{h}(x).
\end{equation}
Since $f_\text{in}$ is positive for all values of the initial spins $s(t_\text{in},x) = 2n(t_\text{in},x) - 1$, the distribution $p[n]$ in \cref{eqn:193XA} defines indeed a probability distribution (for an appropriate normalization constant $c$). The boundary term determines the expectation values of the spins or occupation numbers at the initial boundary 
\begin{equation}\label{eqn:193XF}
	\bigl\langle s(t_\text{in},x)\bigr\rangle
	= \int \dif n(t_\text{in})\bigl(2n(t_\text{in},x) - 1\bigr) f_\text{in}
	= g(x)
\end{equation}
The periodicity of $\tilde{h}(x)$ results in periodic $g(x)$ 
\begin{equation}\label{eqn:193XG}
	g\Bigl(x + \tfrac{2\pi\tilde n}{\omega}\Bigr)
	= g(x).
\end{equation}

From the general solution of the evolution equation for the wave function \labelcref{eqn:188A} we can compute the local probabilities for all $t$, cf. \cref{eqn:22A} for $\bar f(t) = 1$,
\begin{alignedeqn}\label{eqn:193XH}
	p(t)
	&= f(t),\\
	p_\tau(t)
	&= \int \dif n(t) \, h_\tau(t)f(t).
\end{alignedeqn}
(There should be no confusion between the ``initial magnetic field'' $\tilde{h}(x)$ and the basis functions $h_\tau$.) According to
\begin{alignedeqn}\label{eqn:193XI}
	f(t_\text{in} + \epsilon)
	= \exp\Bigl\{-\sum_x \tilde{h}(x - \epsilon)s(t_\text{in} + \epsilon,x) + c\Bigr\},\\
	f(t)
	= \exp\Bigl\{-\sum_x \tilde{h}(x - t + t_\text{in})s(t,x) + c\Bigr\},
\end{alignedeqn}
we conclude from the periodicity of $\tilde{h}$ the periodicity of $f$,
\begin{equation}\label{eqn:193XJ}
	f\Bigl(t + \tfrac{2 \pi \tilde n}{\omega}\Bigr)
	= f(t).
\end{equation}
The local probabilities $p_\tau(t)$ are periodic functions of $t$. Correspondingly, the expectation values of spins at a given position $x$ are periodic in time
\begin{equation}\label{eqn:193XK}
	\langle s(t,x)\rangle
	= g(x - t + t_\text{in}).
\end{equation}

The initial wave function \labelcref{eqn:193XD} could be realized by a periodic magnetic field only present at the initial boundary $t_\text{in}$. In terms of fermion numbers this initial state contains contributions from all sectors with fixed particle numbers. If we want to realize a one fermion state the corresponding boundary wave function $f_\text{in}$ should vanish for all configurations that contain more than one or zero particles. An example for a positive wave function and positive $p[n]$ is 
\begin{equation}\label{eqn:193XL}
	f_\text{in}
	= c \sum_x \cos^2\Bigl(\tfrac{\omega x}{2} + \alpha\Bigr) \, n(t_\text{in},x) \prod_{x^\prime \neq x} \bigl(1 - n(t_\text{in},x^\prime)\bigr).
\end{equation}
Local probabilities and expectation values are again periodic, obeying \cref{eqn:193XJ,eqn:193XK}. We have discussed the corresponding one-particle wave function above.

\section{General static memory materials}
\label{sec:general static memory materials}

Based on the examples discussed so far one may ask what are the general conditions for memory materials. For low $N$ we can explicitly classify the positive matrices that lead to several ``largest eigenvalues'' $|\lambda| = 1$. This is done in \cref{app:step evolution operators} for $N = 2$ and $N = 4$. For large $N$ we rather may concentrate on general properties, as the absence of a unique equilibrium state or the presence of conserved quantum numbers.

\subsection{General conditions for static memory materials}

An important ingredient for the realization of static memory materials is a degeneracy of the equilibrium state. By equilibrium state we mean here the bulk, with boundaries moved far away. If the equilibrium state is unique there will always be a tendency to approach this state when $t$ starts to deviate from the boundary at $t_\text{in}$. Uniqueness of the equilibrium state comes in pair with a loss of memory of boundary conditions. In case of degeneracy, however, there is an ambiguity which may prevent the complete loss of memory. Typically, the information that can be transmitted by a memory material reflects a degeneracy of the equilibrium state. For the Weyl material given by \cref{eqn:M13}, $\beta \to \infty$, the degeneracy concerns an arbitrary sequence of $\bigl\{s(x)\bigr\}$ at a given $t$. For every such sequence there exists an associated equilibrium state. A particular equilibrium state is realized whenever all spins on a given diagonal have the same sign -- which sign does not matter. More generally, the non-uniqueness of the equilibrium state is a necessary condition for a memory material. In the other direction the preservation of boundary information in a memory material implies that the equilibrium state cannot be unique.

Let us next assume that the different degenerate equilibrium states can be labeled by the expectation values $\langle A_i(\bar t)\rangle$ of local observables at some given $\bar t$. If those are sufficient to map out completely the degeneracy of the equilibrium state, the expectation values $\langle A_i(t)\rangle$ are fixed in terms of $\langle A_i(\bar t)\rangle$ for all $t$. This can describe a non-trivial evolution of $\langle A_i(t)\rangle$, rather than an approach to a unique value in case of a unique equilibrium state. A memory material is then realized if one can connect $\langle A_i(\bar t)\rangle$ to some appropriate initial boundary conditions.

\subsection{Spontaneous symmetry breaking}

A rather common reason for a degenerate equilibrium state is spontaneous symmetry breaking. Consider a two-dimensional Ising model with $Z_2$-symmetry at low enough temperature, such that the $Z_2$-symmetry is spontaneously broken. The equilibrium state has either $\langle s\rangle = m$ or $\langle s\rangle = -m$. Depending on the boundary condition at $t_\text{in}$ the bulk will be in one of the two equilibrium states. This information can be transmitted to another boundary at $t_\text{f}$. Even though being a rather trivial example, with two eigenvalues of $S$ equal to one, and all other $|\lambda| < 1$, it demonstrates that static memory materials exist in nature. For models with spontaneous symmetry breaking in the bulk the selection of the equilibrium state depends on the boundary condition.

We can understand the fermion models of the preceding section in terms of a symmetry which multiplies for a given $t$ each $s(x)$ with $\delta(x) = \pm 1$. If at $t + \epsilon$ one multiples $s(x + \epsilon)$ with $\delta(x)$, the expression $\mathcal{L}(t)$ in \cref{eqn:M13} is invariant. This extends to other $\mathcal{L}(t^\prime)$ by multiplying appropriately shifted variables $s(x + k\epsilon)$ with $\delta(x)$. As a result, all spins on a given diagonal can be multiplied by the same sign, without changing $S_\text{cl} = \sum_t\mathcal{L}(t)$. This model can be decomposed into independent one-dimensional Ising models. For one-dimensional Ising models spontaneous symmetry breaking is possible in the zero temperature limit $\beta \to \infty$. In this case the boundary conditions decide which one of the two ground states of the one-dimensional Ising model is realized, all spins up or all down. This is done for every diagonal independently. For finite $\beta$ the one-dimensional Ising model has a unique ground state and no longer features spontaneous symmetry breaking. The Ising model \labelcref{eqn:M13} for finite $\beta$ is no longer a memory material.

An example for a memory material with finite $\beta$ is a three dimensional generalized Ising model on a lattice with points $(t,x,y)$, given by
\begin{align}\label{eqn:231A}
	\mathcal{L}(t)
	= -\frac\beta 2 \smash{\sum_{x,y}}
	\bigl\{&s(t + \epsilon,x + \epsilon,y)s(t,x,y)\\
	& + s(t,x,y + \epsilon)s(t,x,y) - c(\beta)\bigr\}.\notag
\end{align}
This model is composed of independent two-dimensional Ising models on diagonal hypersurfaces. In the infinite volume limit (infinitely many points in the $t$-, $x$- and $y$-directions) spontaneous symmetry breaking occurs for $\beta > \beta_c$. The boundary information at $t_\text{in}$ decides for every surface separately which one of the two ground states is selected. This constitutes an example for a memory material that is not a unique jump chain. Local probabilities and observables at fixed $x$ and $y$ may oscillate with $t$ for suitable boundary conditions. Much more complicated and interesting situations of information transport can be conceived in the presence of spontaneous symmetry breaking. Still rather simple, but somewhat less trivial examples concern the propagation of topological defects.

\subsection{Conserved quantities}

A simple obstacle for the existence of a unique equilibrium state are conserved quantities. Examples are conserved particle numbers or a conserved number of topological defects. If the evolution conserves in each step a certain quantity as, for example, the total number of spins up, this quantity is the same at every $t$ as at $t_\text{in}$. Different values of the conserved quantity specify different equilibrium states or families of equilibrium states, such that a unique equilibrium state is excluded. The memory of the conserved quantity is transmitted and a static memory material can be realized. In app. B we explicitly discuss some simple models with conserved quantities.

In case of conserved quantities the step evolution operator becomes 
block diagonal, with different sectors corresponding to the different possible values $c_k$ of the conserved quantity $C$. The expectation value of $C$ at $t_\text{in}$ is given by the probabilities $p_k(t_\text{in})$ that $C$ takes the value $c_k$ at $t_\text{in}$,
\begin{equation}\label{eqn:CQ1}
	\langle C(t_\text{in})\rangle
	= \sum_k p_k(t_\text{in}) \, c_k.
\end{equation}
By a conserved quantity we understand that $\langle C(t)\rangle$ is independent of $t$. Typically this is realized by $t$-independent probabilities $p_k(t)$. In this case the relative weight of the different sectors is not changed by the evolution. Therefore, each block of $S$ corresponding to a given sector must have at least one eigenvalue with $|\lambda| = 1$.

If each sector has precisely one eigenvalue $\lambda = 1$, and all other eigenvalues $|\lambda| < 1$, the memory is preserved in a rather simple fashion. Each sector approaches its own separate equilibrium state. The total ``equilibrium state'' is the weighted sum of the individual equilibrium states, and the transmitted information concerns the probabilities $p_k$. Our example of oscillating local probabilities for generalized Ising models with conserved particle numbers demonstrates that more complex memory transport is possible.

For observables that are expressed by $t$-independent operators $A^\prime$, and for $t$-independent step evolution operators $S$, the expectation value $\langle A\rangle$ is independent of $t$ if $A^\prime$ commutes with $S$,
\begin{equation}\label{eqn:232A}
	[A^\prime,S]
	= 0.
\end{equation}
This follows directly from \cref{eqn:SID,eqn:87,eqn:90},
\begin{alignedeqn}\label{eqn:232B}
	\langle A(t + \epsilon)\rangle
	&= \langle\bar q(t + \epsilon)A^\prime\tilde q(t + \epsilon)\rangle\\
	&= \langle \bar q(t) S^{-1}A^\prime \, S\tilde q(t) \rangle
	= \langle \bar q(t) A^\prime\tilde q(t)\rangle\\
	&= \langle A(t)\rangle.
\end{alignedeqn}
In the other direction, a quantity is conserved for a given state if $\langle S^{-1} \, A^\prime \, S\rangle = \langle A^\prime\rangle$. If we require this condition for all possible states one finds the condition \labelcref{eqn:232A}. For every $t$ the expectation value of conserved quantities is given by the boundary conditions
\begin{equation}\label{eqn:232C}
	\langle A(t)\rangle
	= \langle A(t_\text{in})\rangle
	= \langle \bar q(t_\text{in})A^\prime\tilde q(t_\text{in})\rangle.
\end{equation}

Despite the possible presence of conserved quantities a unique equilibrium state cannot preserve memory of boundary conditions. In case of a unique equilibrium state in the bulk and large enough $t_\text{f} - t_\text{in}$ we can replace $\bar q(t_\text{in})$ by the equilibrium conjugate wave function $\bar q_\ast$,
\begin{alignedeqn}\label{eqn:232D}
	\langle A(t_\text{in})\rangle
	&= \langle \bar q_\ast A^\prime\tilde q(t_\text{in})\rangle
	= \langle \bar q_\ast S^{-1}_\ast A^\prime\tilde q(t_\text{in})\rangle\\
	&= \langle \bar q_\ast A^\prime S_\ast\tilde q(t_\text{in})\rangle
	= \langle \bar q_\ast A^\prime\tilde q_\ast\rangle.
\end{alignedeqn}
For $A^\prime$ commuting with $S$ the initial value $\langle A(t_\text{in})\rangle$ is uniquely given by \cref{eqn:232D}. It cannot be chosen freely and no information associated to $\langle A(t)\rangle$ can be transported into the bulk. (In particular, this holds for $A^\prime = S$.)

The scaling step operator $S_\ast$, defined by \cref{eqn:EI}, is a projector,
\begin{equation}\label{eqn:232E}
	S^2_\ast
	= S_\ast.
\end{equation}
Its eigenvalues are one or zero. This limit matrix $S_\ast$ always exists for a unique equilibrium state where only one eigenvalue equals one. For degenerate bulk states $S_\ast$ may no longer exists. There are no longer unique equilibrium wave functions $\tilde q_\ast$, $\bar q_\ast$, and non-trivial information can now be encoded in $\langle A(t_\text{in})\rangle$. If $S_\ast$ exists, more than one eigenvalue differs from zero for degenerate bulk states.

\subsection{Memory materials as asymptotic states}

The length of eigenvectors to eigenvalues of $S$ with $|\lambda| < 1$ decreases as $t$ increases. For the asymptotic behavior of the wave function for large $t - t_\text{in}$ (typically $t - t_\text{in}\gg\xi$) all eigenvectors to eigenvalues $|\lambda| < 1$ have become very close to zero. Neglecting them one is still left with the subspace spanned by the eigenvectors to eigenvalues $|\lambda| = 1$. For wave functions $\tilde q$ belonging to this subspace the length $|\tilde q|
$ no longer changes as $t$ increases. The evolution far inside the bulk always becomes the unitary evolution of quantum mechanics.

The unitary evolution can be trivial. If the step evolution operator projected on the ``bulk subspace'' is the unit matrix, one has $H = 0$ and the wave function $\tilde q_\ast$ becomes independent of $t$. This is automatically the case if the equilibrium state is unique -- cf. the discussion of the Ising model in \cref{sec:evolution}. Such a behavior may also occur in the presence of conserved quantities or for degenerate equilibrium states. In this case memory of boundary conditions is preserved in the bulk, while local probabilities are independent of $t$. There are, however, also models for which the evolution equation in the bulk follows a non-trivial quantum mechanical evolution with $H \neq 0$. A very simple example with $M = 4$ is displayed in \cref{app:step evolution operators}, where we discuss explicitly how the appropriate two-component subsystem is constructed. Other examples are the two-dimensional fermion models discussed in \cref{sec:ising models}.

In summary, the quantum evolution with orthogonal step evolution operators for suitable subsystems is the generic case for large $t - t_\text{in}$ (or $t_\text{f} - t)$ far inside the bulk. If this evolution is not trivial, the generalized Ising model can be used as a quantum simulator of a quantum system with the same $H$. The space-dependence of observables in the generalized Ising model can be used to ``measure'' the time dependence of observables in quantum mechanics.

The class of quantum systems that can be realized as appropriate memory materials is restricted by the spectrum of the step evolution operator, more precisely by the set of eigenvalues with $|\lambda| = 1$. If the (sub)space of eigenvalues with $|\lambda| = 1$ has finite dimension $M^\prime$, one can have at most $M^\prime$ different eigenvalues. This can describe discrete quantum mechanics, where the continuous time evolution is replaced by the evolution operator for discrete time steps, $U(t + \epsilon,t)$. The time evolution of continuous quantum mechanics requires $M^\prime \to \infty$. This is realized for the continuum limit of the two-dimensional fermion models discussed before.

\subsection{Quantum evolution in generalized Ising models and unique jump chains}

For generalized Ising models with a positive step evolution operator, $S_{\tau\rho}\geq 0$, one may investigate the conditions for a ``unitary evolution'' corresponding to quantum mechanics. We want to classify positive orthogonal $N \times N$-matrices, $\tran{S} \, S = 1$, for finite $N$. The unique possibilities are unique jump operators. One concludes that a quantum evolution for the whole system (not subsystems), that does not correspond to the almost deterministic unique jump chains, requires an infinite number of degrees of freedom $M$. In practice, this condition is not very strong since $N$ increases very rapidly with $M$.

The orthogonality condition 
\begin{equation}\label{eqn:OC1}
	\sum_j S_{ij}S_{kj}
	= \sum_j S_{ji}S_{jk}
	= \delta_{ik}
\end{equation}
requires for $S_{ij}\geq 0$ that all terms in the sums vanish separately for $i\not= k$
\begin{equation}\label{eqn:OC2}
	S_{ij}S_{kj}
	= 0, S_{ji} S_{jk}
	= 0\quad \text{for}~ i \neq k.
\end{equation}
Furthermore, invertibility of $S$ requires that each column and each row has at least one non-zero element. We want to show that two or more non-vanishing elements in a given row or column lead to a contradiction to the property \labelcref{eqn:OC2}. Then positive orthogonal matrices have in each column and each row precisely one positive non-zero element. The normalization condition \labelcref{eqn:OC1} for $i = k$ implies that all these elements equal one and $S$ is therefore a ``unique jump operator''.

In order to establish the contradiction we assume that for a given $i$ there are two non-zero elements $S_{ij_1}$ and $S_{ij_2}$. Then \cref{eqn:OC2} implies $S_{kj_1} = S_{kj_2} = 0$ for all $k\not= i$. More generally, each $S_{i\bar j}\not= 0$ ``blocks'' the row $\bar j$ for all other $k$, in the sense $S_{k\bar j} = 0$. Once we have distributed $N$ non-zero values $S_{ij}$ (for different $i$) all columns are blocked. If for a given row $i$ we have already blocked two columns $j_1$ and $j_2$, at most $N - 2$ other rows can admit a non-zero element before all columns are blocked. There remains then at least one row without any non-zero element, which contradicts the invertibility of $S$.

For finite $N$ we conclude that orthogonal step evolution operators that are not unique jump operators require some of the elements $S_{\tau\rho}$ to be negative. This cannot be realized by a classical Ising spin system with positive step evolution operator \labelcref{eqn:S12}. We recall, however, that negative elements $S_{\tau\rho}$ can still be compatible with a positive weight distribution $w[n]$. Also subsystems can follow a quantum evolution even for positive $S$ that are not unique jump operators. We present a simple explicit example in \cref{app:step evolution operators}. For the limit $N \to \infty$ the simple counting argument above is not necessarily valid. For the action \labelcref{eqn:231A} we have realized an Ising type model that admits spontaneous symmetry breaking for finite $\beta$ in the ``infinite volume limit'' corresponding to $N \to \infty$. This constitutes an example for a quantum evolution of a subsystem that is realized by step evolution operators that are not unique jump operators.

\subsection{Extension to quantum statistics}

Even though the main emphasis of the present paper is on information transport in classical statistical systems, many aspects of the formalism can be taken over to information transport in quantum statistical systems in thermal equilibrium. For this purpose we employ the Matsubara formalism for quantum statistics which is based on a ``functional integral'' where Euclidean time is compactified on a torus with circumference $1/T$. In our two-dimensional setting of \cref{sec:ising models} the periodic variable $x$ corresponds then to (discretized) Euclidean time, while $t$ plays the role of a space coordinate. The wave function and density matrix describe the evolution of the quantum statistical equilibrium system in space and can therefore be used for an investigation of information transport in space. The two-dimensional models describe the one-dimensional quantum statistical models, with appropriate generalization to higher dimensions.

\section{Conclusions and discussion}
\label{sec:conclusions}

We have developed a formalism for understanding the influence of boundary conditions on observables in the bulk for static classical statistical systems. Key elements are local probabilities at particular points $t$ of a wire or, more generally, on particular hypersurfaces of a multi-dimensional system. Local probabilities permit the computation of expectation values of local observables. They obtain from the overall probability distribution by integrating out all degrees of freedom except those on the given hypersurface.

The transport of information between neighboring hypersurfaces is, however, not described in terms of the local probabilities alone. It rather can be encoded in the evolution equation for classical wave functions. More generally, the evolution of the local probabilities is described by a density matrix which obeys a generalization of the von Neumann equation. The local probabilities are the diagonal elements of this density matrix. It is remarkable that the evolution of the local probabilities is not described by a linear first order differential equation involving the probabilities alone. The appropriate linear first order differential equation also needs the off-diagonal elements of the density matrix as necessary local information for the evolution. A structure well known from quantum mechanics arises naturally in the context of classical statistics.

The analogy to quantum mechanics becomes even more striking if the density matrix of the system (or an appropriate subsystem) obeys a unitary evolution. In this case the evolution equation for the density matrix is given precisely by the von Neumann equation with an appropriate hermitian Hamiltonian. We have presented several explicit examples for static classical statistical systems with this property. For these systems we have solved the boundary problem exactly by use of the quantum formalism.

For a non-unitary evolution the generalized von Neumann equation can describe the approach to ``equilibrium states'' in the bulk. The only difference between the evolution of the classical wave function and the Schrödinger equation for the quantum wave function concerns the lack of hermiticity of the evolution operator $G$ in \cref{eqn:CS2} - and similarly for the evolution of the density matrix \labelcref{eqn:164AC}. The antihermitian part of $G$ drives the wave function or density matrix towards generalized equilibrium states. This family of states is reached asymptotically in the bulk. On these asymptotic states $G$ acts as a hermitian Hamiltonian operator and the evolution becomes unitary.

One may view the family of asymptotic states in the bulk as a subsystem obeying the unitary ``quantum evolution''. Even for pure classical states of the overall system the subsystem may be described by a mixed system in terms of a ``coarse grained'' density matrix. Projecting the overall evolution equation onto the subsystem typically generates additional terms, similar to the Lindblad equation \cite{KO,LI,ZO}. They account for decoherence \cite{ZE,JZ,ZU} or syncoherence \cite{CWQM}.

In case of a unique equilibrium state the unitary evolution of the subsystem is a trivial $t$-independent behavior, while oscillations may be found if no unique equilibrium state exists. If the asymptotic bulk state has more than one independent component of the wave function $\tilde q$, the system constitutes a static memory material. For such memory materials the properties of a boundary remain relevant for observables in the bulk. Boundary information is transported from the boundary to the bulk, and also to some other boundary. The usual complete loss of memory of boundary conditions in the approach to an equilibrium bulk state does not occur. Very generally, static memory materials become possible if there is no unique equilibrium state in the bulk. If one encounters with comparable probability a family of ``equilibrium states'' in the bulk, the boundary information can be transported as selection and evolution within this family.

For static memory materials one often encounters oscillating local probabilities and expectation values. This occurs whenever the Hamiltonian $\hat H$ acting on the asymptotic states differs from zero. We have discussed the possible experimental realization of static memory materials, as well as their numerical simulation. The very particular properties of static memory materials may find applications in information processing or elsewhere. Boundary information becomes available ``simultaneously'' at all local $t$-hypersurfaces.

Realizing a static memory material in experiment or by numerical simulation constitutes a ``quantum simulator''. The observed oscillations in space as a function of $t$ are mapped one to one to the oscillations in time of the corresponding quantum system described by the same von Neumann equation. The time dependence of quantum systems can be made visible by the space dependence in static memory materials.

The close correspondence between the evolution in space of local probabilistic information with the time evolution in quantum mechanics suggests that quantum systems can be understood as appropriate subsystems of classical statistical systems, as proposed earlier \cite{CWEQM,CWQM}. In the present paper we have not addressed the issue of non-commuting observables. This is related to observables represented by off-diagonal operators. A typical off diagonal operator appearing in our setting is the Hamiltonian. Also derivatives of local observables with respect to $t$ can be represented by off diagonal operators \cite{CWPT}. A more extensive discussion of the question which type of observables are described by non-diagonal operators can be found in refs.~\cite{CWFGI,CWQFCS}.

The particular ``quantum properties'' related to non-commuting operators, as the uncertainty relation, are rooted in ``incomplete statistics'' \cite{CWIS}. Incomplete statistics emerges naturally in our setting where the local probabilistic information is given by the density matrix. The density matrix contains statistical information beyond the local probabilities. This information is sufficient for the computation of expectation values of additional observables - namely the ones represented by off-diagonal operators. It is incomplete in the sense that the probabilities for finding simultaneously the values of two non-commuting observables are not part of the local information contained in the density matrix. Typically, the overall probability distribution may include the information about joint probabilities, but the latter is lost in the reduction to the local density matrix. Quite generally, incomplete statistics characterizes subsystems. In this case the appropriate subsystem is associated to the local probabilistic information, as given by the density matrix.

A second quantum issue that we have not addressed here concerns correlations. Different types of sequences of measurements and associated conditional probabilities are described by different correlation functions \cite{CWQM}. Ideal classical sequences of measurements are described by the classical correlation function which involves joint probabilities. While computable from the overall probability distribution, the classical correlation needs information beyond the local subsystem characterized by the density matrix. Ideal quantum measurements are described by quantum correlations that are computable from the information of the subsystem. The quantum correlations involve non-commuting operator products and violate \cite{CWQM} Bell's inequalities \cite{BE}.

We can associate the sequence of hypersurfaces with a concept of probabilistic time \cite{CWPT}. In this view time emerges as an ordering concept of general statistical systems. In our case it orders the hypersurfaces on which ``local physics'' is defined. Remarkably, time and quantum mechanics emerge together in our formalism. (The emergent time should not be confounded with some ``external time'' in dynamical systems. With respect to external dynamics our systems are static.) In two - or higher - dimensional classical statistical systems there is no a priori difference between time and space directions. The time direction is simply the one in which the transport of statistical information is studied. As one of our examples we have studied asymmetric Ising models in two dimensions. For these models there is no difference between the time direction $(t)$ and the space direction $(x)$. Nevertheless, the system describes the propagation of free fermions in Minkowski space, with the associated Lorentz-symmetry. The asymmetric signature of the Minkowski metric arises as a result of the evolution in a particular $t$-direction. This picture describes the world by an overall probability distribution, covering past, present and future. Quantum mechanics arises by ``integrating out'' the past and future, concentrating on the probabilistic information of the ``present'' local subsystem.

The present paper has developed and employed the quantum formalism for the description of static memory materials or quantum simulators. We have presented a few explicit examples for static classical overall probability distributions that realize a unitary quantum evolution and induce oscillating local probabilities in the bulk. Nevertheless, our examples still describe rather simple physical situations as the propagation of non-interacting massless fermions in two-dimensional spacetime. It will be interesting to see if more complex situations can be realized by classical overall probability distributions, as massive particles in a potential with typical quantum effects as interference and tunneling.

Despite its present limitations this paper contributes a possible answer to the question why our probabilistic description of the world uses probability amplitudes or wave functions rather than directly the probabilities. Probability amplitudes arise as the natural concept if one deals with local observables defined on a hypersurface at the location $t$ for static systems or time in quantum mechanics. Local probabilities obtain by integrating out the ``past'' and the ``future''. Each of these integrations separately produces a probability amplitude, with local probabilities bilinear in these amplitudes. Typical evolution equations are linear in the amplitudes, implementing the superposition principle fort their possible solutions. Corresponding evolution laws formulated only in terms of local probabilities are complicated non-linear equations. It is the simplicity of the linear evolution law for the wave functions or the density matrix that singles out these ``quantum concepts'' for the description of the local probabilistic information.

If one accepts the view that time and evolution are ordering concepts in a classical statistical description of our world, our approach also answers the question why evolution is described by quantum mechanics. Integrating out the past and future, with a number of time steps going to infinity, all information is lost except the one for subsystems described by quantum mechanics. Moving boundaries to infinity, any remaining possible non-trivial evolution is described by quantum mechanics.

\paragraph{Acknowledgment} This work is part of and supported by the DFG Collaborative Research Centre ``SFB 1225 (ISOQUANT)''

\begin{appendices}
\crefalias{section}{appendix}
\renewcommand{\theequation}{\Alph{section}.\arabic{equation}}
\numberwithin{equation}{section}

\section{Beyond next-neighbor interactions}
\label{sec:beyond next-neighbor interactions}

In the main text we have concentrated on quasi-local probability distributions \labelcref{eqn:I2a} for which $\mathcal{K}(t)$ involves only spins or occupation numbers on two neighboring $t$-layers. In this appendix we consider more general quasi-local probability distributions and show that the discussion in the main text also covers these more general settings. The representation of $K[n]$ as a product of factors $\mathcal{K}(t)$ involving each only occupation numbers on two neighboring layers can be extended to products of factors involving three or more neighboring layers. We will show that this extended setting can be represented in terms of interactions on a block lattice, with suitable block-occupation numbers on two neighboring layers of the block lattice. In this sense our discussion of next-neighbor interactions covers these more extended settings as well.

As an example we consider a product of building blocks $\mathcal{G}(t)$ containing each three neighboring occupation numbers,
\begin{alignedeqn}\label{eqn:E1}
	K[n]
	&= \prod\nolimits_t^\prime \mathcal{G}(t),\\
	\mathcal{G}(t)
	&= G_{\tau\rho\sigma} \, h_\tau(t + \epsilon) \, h_\rho(t) \, h_\sigma(t-\epsilon).
\end{alignedeqn}
Here the product $\prod^\prime_t$ extends from $t = t_\text{in} + \epsilon$ to $t = t_\text{f} - \epsilon$. The initial and final factors involve now two neighboring occupation numbers
\begin{equation}\label{eqn:E2}
	Z
	= \int \mathcal{D} n \, \bar f_\text{f} \bigl(n(t_\text{f}),n(t_\text{f} - \epsilon)\bigr) \, K[n] \, f_\text{in}\bigl(n(t_\text{in} + \epsilon),n(t_\text{in})\bigr).
\end{equation}
(Factors involving only $n(t_\text{in})$ or $n(t_\text{f})$ are a special case of \cref{eqn:E2}.) For simplicity we assume the number of lattice points to be even.

We can group two neighboring lattice points into points of a block lattice with sites at $t + \epsilon/2$. For the blocks the values of $t$ are given by $t = t_\text{in} + 2m\epsilon$, such that the block sites are at $t_\text{in} + \epsilon/2,t_\text{in} + 5\epsilon/2$ and so on. We employ extended wave functions for the occupation numbers in the blocks 
\begin{equation}\label{eqn:E3}
	f\Bigl(t + \tfrac{\epsilon}{2}\Bigr)
	= q_{\rho\tau} \Bigl(t + \tfrac{\epsilon}{2}\Bigr) h_\rho(t + \epsilon) \, h_\tau(t).
\end{equation}
For $M$ different occupation numbers on a given site the number of occupation numbers on the block sites is $2M$. Thus $N = 2^M$ states on a single site, labeled by $\tau$, imply $N^2 = 2^{2M}$ states for each site of the block lattice, labeled now by the double index $(\tau,\rho)$. Correspondingly, the wave functions $\tilde q_{\rho\tau}\bigl(t + \frac\epsilon2\bigr)$ can be viewed as $N^2$-component vectors or $N \times N$-matrices. The initial and final factors in \cref{eqn:E2} can directly be expressed in terms of the block-wave functions,
\begin{alignedeqn}\label{eqn:E4}
	f_\text{in}
	&= \tilde q_{in,\rho\tau}\Bigl(t_\text{in} + \tfrac\epsilon2\Bigr)
	 \, h_\rho(t_\text{in} + \epsilon) \, h_\tau(t_\text{in}),\\
	\bar f_\text{f}
	&= \bar q_{f,\rho\tau}\Bigl(t_\text{f} - \tfrac\epsilon2\Bigr) \, h_\rho(t_\text{f}) \, h_\tau(t_\text{f} - \epsilon).
\end{alignedeqn}

The factors $\mathcal{G}(t + \epsilon)$ and $\mathcal{G}(t + 2\epsilon)$ involve only the occupation numbers belonging to the blocks at $t + \epsilon/2$ and $t + 5\epsilon/2$. We group them together to a ``block evolution factor'' involving only two neighboring blocks
\begin{equation}\label{eqn:E5}
	\bar{\mathcal{K}}\bigl(t + \tfrac{\epsilon}{2}\bigr)
	= \mathcal{G}(t + 2\epsilon) \, \mathcal{G}(t + \epsilon),
\end{equation}
or 
\begin{alignedeqn}\label{eqn:E6}
	\bar{\mathcal{K}}\bigl(t + \tfrac{\epsilon}{2}\bigr)
	= \bar S_{\tau\rho,\sigma\lambda}
	\bigl(t + \tfrac{\epsilon}{2}\bigr) \, h_\tau
	(t + 3\epsilon) \, h_\rho(t + 2\epsilon)\\
	\times h_\sigma(t + \epsilon) \, h_\lambda(t),
\end{alignedeqn}
where (no sums here)
\begin{equation}\label{eqn:E7}
	\bar S_{\tau\rho,\sigma\lambda}
	\bigl(t + \tfrac{\epsilon}{2}\bigr)
	= G_{\tau\rho\sigma}	(t + 2\epsilon) \, G_{\rho\sigma\lambda}(t + \epsilon).
\end{equation}
With 
\begin{equation}\label{eqn:E8}
	K[n]
	= \smashoperator{\prod_{t\in \text{block}}} \bar{\mathcal{K}}\bigl(t + \tfrac{\epsilon}{2}\bigr),
\end{equation}
we recover the structure of next-neighbor interactions between blocks. The only change is the extended wave function \labelcref{eqn:E3} which involves now an extended number of states, with index $\tilde \rho = (\rho,\tau)$ taking $N^2$ values. This is easily generalized: Factors with up to four neighboring occupation numbers involve the $N^3$ different states in blocks of three sites, and so on.

The setting with a block lattice can also be employed in case of next-neighbor interactions \labelcref{eqn:80,eqn:81}. In this case one has
\begin{equation}\label{eqn:E9}
	\bar{\mathcal{K}}\Bigl(t + \tfrac\epsilon2\Bigr)
	= \mathcal{K}(t + 2\epsilon) \, \mathcal{K}(t + \epsilon) \, \mathcal{K}(t).
\end{equation}
The ``double site'' wave function $\tilde q_{\rho\tau}\bigl(t + \frac\epsilon2\bigr)$ transports now additional information beyond the one contained in the ``simple site'' wave function $\tilde q_\tau(t)$.

\section{Generalized Ising models}
\label{app:generalized Ising models}

In this appendix we display several simple generalized Ising models that can be solved exactly. They demonstrate as explicit examples several features of the discussion in the main text. In particular, constraints can be used in order to formulate situations with conserved quantities.

\subsection{Constrained Ising models}

Ising spins can be associated with occupation numbers, $n_\gamma = 2s_\gamma - 1$, for particles being present ($n_\gamma = 1$, $s_\gamma = 1$) or absent ($n_\gamma = 0$, $s_\gamma = -1$). Two Ising spins, $\gamma = 1,2$, correspond to two species of particles. We can construct models for which the total particle number $N_\text{tot} = n_1 + n_2$ is preserved by the evolution. A conserved particle number enhances the symmetry of the model. For example, if the particle number is conserved modulo two one has an additional $Z_2$-symmetry. With respect to this symmetry the configurations with $N_\text{tot} = 0,2$ are considered as even, while those with $N_\text{tot} = 1$ are odd. The step evolution operator conserves the $Z_2$-symmetry if it does not mix sectors with even or odd $N_\text{tot}$.

In the notation of \cref{sec:functional integral} with basis functions $h_\tau$, $\tau = 1\dots 4$, this implies for the step evolution operator $S$ that all elements $S_{\tau\rho}$ must vanish for which $\tau$ refers to an even basis element and $\rho$ to an odd one, or $\tau$ to an odd element and $\rho$ to an even one. In the occupation number basis even (odd) elements have an even (odd) number of occupied states $n_\gamma = 1$. With the definitions \labelcref{eqn:74} the basis elements $h_1$ and $h_4$ are even, while $h_2$ and $h_3$ are odd. The $Z_2$ symmetry maps $h_{2,3} \to -h_{2,3}$, with $h_{1,4}$ invariant. A $Z_2$-symmetric evolution is therefore realized by $S_{12} = S_{13} = S_{42} = S_{43} = 0$, and the same for the transposed elements $S_{21} = S_{31} = S_{24} = S_{34} = 0$. With \cref{eqn:S12} this implies that the corresponding elements $M_{\tau\rho}$ must diverge.

We want to understand the evolution of the wave function,
\begin{equation}\label{eqn:S13}
	q_\tau(t + \epsilon)
	= \sum_\rho \exp(-M_{\tau\rho}) \, q_\rho(t),
\end{equation}
in some more detail. Diverging elements $M_{\tau\rho}$ can be used to divide $S_{\tau\rho}$ into blocks such that the system has conserved quantities. For example, for 
\begin{alignedeqn}\label{eqn:141A}
	M_{12}
	&= M_{13}
	= M_{14}
	= M_{21}
	= M_{31}
	= M_{41}\\
	&= M_{24}
	= M_{34}
	= M_{42}
	= M_{43}
	= \kappa,
\end{alignedeqn}
and $\kappa \to \infty$, the matrix $S$ has non-vanishing elements only for $S_{11}$, $S_{44}$, $S_{22}$, $S_{23}$, $S_{32}$ and $S_{33}$. This guarantees that $\langle N_1 + N_2 \rangle$ is conserved, $\langle N_1(t) + N_2(t) \rangle = \text{const}$.

The limit $\kappa \to \infty$ can be interpreted as a constraint on the system. For example, for $M_{12} \to \infty$ all configurations with $h_1(t + \epsilon) h_2(t) \neq 0$ would lead to $\exp\{-\mathcal{L}(t)\} = 0$ and therefore not contribute to the functional integral. This implies effectively the constraint $h_1(t + \epsilon) h_2(t) = 0$. The three constraints $h_1(t + \epsilon) h_2(t) = 0,\; h_1(t + \epsilon) h_3(t) = 0,\;h_1(t + \epsilon) h_4(t) = 0$ require for the configurations that can contribute to the functional integral the selection rule $n_1(t) = n_2(t) = n_1(t + \epsilon) = n_2(t + \epsilon) = 1$ or $n_1(t + \epsilon) = 0$ or $n_2(t + \epsilon) = 0$. Similarly, the constraints $h_2(t + \epsilon) h_1(t) = 0$, $h_3(t + \epsilon) h_1(t) = 0$ and $h_4(t + \epsilon) h_1(t) = 0$ admit non-zero contributions only for $n_1(t + \epsilon) = n_2(t + \epsilon) = n_1(t) = n_2(t)$ or $n_1(t) = 0$ or $n_2(t) = 0$. Combining the two sets of three constraints leaves for the contributing combinations $(n_1(t), n_2(t), n_1(t + \epsilon), n_2(t + \epsilon))$ only the possibilities $(1,1,1,1)$, $(0,0,0,0)$, $(0,1,0,1)$, $(1,0,1,0)$, $(0,1,1,0)$, $(1,0,0,1)$, $(1,0,0,0)$, $(0,1,0,0)$, $(0,0,1,0)$ and $(0,0,0,1)$. If we further impose the constraints $h_2(t + \epsilon) h_4(t) = 0$, $h_3(t + \epsilon) h_4(t) = 0$, $h_4(t + \epsilon) h_2(t) = 0$, $h_4(t + \epsilon) h_3(t) = 0$ the last four possibilities are excluded as well. The remaining six contributions obey all $n_1(t + \epsilon) + n_2(t + \epsilon) = n_1(t) + n_2(t)$, while the configurations that violate this selection rule do not contribute to the functional integral.

Conserved quantities and symmetries as the $Z_2$-symmetry are actually often more easily expressed as properties of the step evolution operator $S$, where they can be realized by vanishing elements. For a more direct comparison with Ising models we will continue here the formulation as a standard action, characterized by $M_{\tau\rho}$. We may consider a class of models with part of the elements of $M$ given by \cref{eqn:141A}. The limit $\kappa \to \infty$ of our example therefore imposes effectively the constraint
\begin{equation}\label{eqn:S14}
	n_1(t + \epsilon) + n_2(t + \epsilon) = n_1(t) + n_2(t).
\end{equation}
The model is therefore a ``constrained Ising model'' where the total number of spins up is required to be the same for two neighboring sites. This constraint is implemented effectively by a term in the action
\begin{align}\label{eqn:S15}
	\mathcal{L}_{\kappa} 
	= \kappa \Bigl\{&n^\prime_1 \, n^\prime_2(-3 n_1 \, n_2 + 3 n_1 + 3 n_2 - 1)\\
	&+ (n^\prime_1 + n^\prime_2)(1 - n_1 - n_2)+ (n \leftrightarrow n^\prime)\Bigr\},\notag
\end{align}
with $n^\prime_\gamma = n_\gamma (t + \epsilon)$, $n_\gamma = n_\gamma(t)$, taking the limit $\kappa \to \infty$. In terms of Ising spins this reads
\begin{equation}\label{eqn:DB}
	\mathcal{L}_{\kappa}
	= \frac{\kappa}{8}\bigl\{5 - (s^\prime_1 + s^\prime_2) (s_1 + s_2) + s^\prime_1 \, s^\prime_2 + s_1 \, s_2 - 3 s^\prime_1 \, s^\prime_2 \, s_1s_2\bigr\}.
\end{equation}

For many models considered in this paper certain entries of the step evolution operator vanish. This can be seen as a constraint that certain transitions between configurations at $t$ and configurations at $t + \epsilon$ are not allowed. Typically, vanishing elements of the step evolution operator $S$ can be implemented by ``constraint terms'' of the type of \cref{eqn:S15} with $\kappa \to \infty$.

\subsection{Four-step oscillator chain}

As an example for a constrained Ising model we may take the four-step oscillator chain as specified by the evolution operator \labelcref{eqn:GD}. It can be realized by an Ising model with 
\begin{alignedeqn}\label{eqn:B.6}
	\mathcal{L}
	&= -\ln (1 - \eta)
	(h^\prime_1h_1 + h^\prime_2h_2 + h^\prime_3h_3 + h^\prime_4h_4)\\
	&\hphantom{{}=} -\ln \eta(h^\prime_2h_1 + h^\prime_1h_3 + h^\prime_3h_4 + h^\prime_4h_2)\\
	&\hphantom{{}=} + \kappa
	(h^\prime_1h_2 + h^\prime_1h_4 + h^\prime_2h_3 + h^\prime_2h_4\\
	&\hphantom{{}=} + h^\prime_3h_1 + h^\prime_3h_2 + h^\prime_4h_1 + h^\prime_4h_3),
\end{alignedeqn}
with $\kappa \to \infty$. Here we use the shorthands $h^\prime_\tau = h_\tau(t + \epsilon)$, $h_\tau = h_\tau(t)$. We can write this in terms of Ising spins using \cref{eqn:74}
\begin{alignedeqn}\label{eqn:B.7}
	h_1
	&= \frac14(1 + s_1 + s_2 + s_1 \, s_2)\\
	h_2
	&= \frac14(1 - s_1 + s_2 - s_1 \, s_2)\\
	h_3
	&= \frac14(1 + s_1 - s_2 - s_1 \, s_2)\\
	h_4
	&= \frac14(1 - s_1 - s_2 + s_1 \, s_2).
\end{alignedeqn}
This model has interactions involving up to four spins. Its purpose is not a particular interest for its realization, but rather a simple explicit demonstration of oscillating local probabilities.

\section{Scaling form of step evolution operator for Ising model}

The relation \labelcref{eqn:EI} for the scaling form of the evolution operator $S_\ast$, e.g. $SS_\ast = S_\ast$, can be employed for the determination of $S_\ast$ as well as for the free energy density $\varphi$. We perform the computation here for the Ising model in view of possible applications to more complex settings. The form of a matrix equation \labelcref{eqn:EI} can exploit directly possible symmetries of $S$.

Let us write $S$ in terms of the Pauli matrices, $\tau_0 = 1$, $\mu = 0,1,2,3$,
\begin{equation}\label{eqn:EJ}
	S
	= s_\mu\tau_\mu,
\end{equation}
with 
\begin{alignedeqn}\label{eqn:EK}
	s_0
	&= e^{\beta - \varphi}\cosh\gamma,
	\quad
	s_1
	= e^{-\beta - \varphi}\\
	s_2
	&= 0,
	\quad
	s_3
	= e^{\beta - \varphi}\sinh\gamma.
\end{alignedeqn}
With 
\begin{equation}\label{eqn:EL}
	S_\ast
	= a_\mu\tau_\mu
\end{equation}
one has
\begin{align}\label{eqn:EM}
	SS_\ast
	&= e^{\beta - \varphi} (\cosh \gamma \, a_0 + \sinh\gamma \, a_3){ + e^{-\beta - \varphi}a_1}\\
	& + \Bigl[e^{\beta - \varphi}(\cosh\gamma \,  a_3 + \sinh\gamma \,  a_0){ + ie^{-\beta - \varphi}a_2}\Bigr]{\tau_3}\notag\\
	& + \Bigl[e^{\beta - \varphi}\cosh \gamma \,  a_1 - ie^{\beta - \varphi}\sinh \gamma \,  a_2{ + e^{-\beta - \varphi}a_0}\Bigr]\tau_1\notag\\
	& + \Bigl[e^{\beta - \varphi}\cosh \gamma \,  a_2 + ie^{\beta - \varphi}\sinh\gamma \,  a_1 - ie^{-\beta - \varphi}a_3\Bigr]\tau_2.\notag
\end{align}
Since $S$ is symmetric, also $S^\ast$ is symmetric and the coefficient in front of $\tau_2$ has to vanish,
\begin{equation}\label{eqn:EN}
	a_2
	= 0,
	\quad
	e^{\beta - \varphi}\sinh \gamma \,  a_1
	= e^{-\beta - \varphi}a_3,
\end{equation}
such that 
\begin{equation}\label{eqn:EO}
	a_3
	= e^{2\beta}\sinh \gamma \,  a_1.
\end{equation}
Comparing in \cref{eqn:EI} the coefficients of $\tau_1$ implies 
\begin{equation}\label{eqn:EP}
	e^{\beta - \varphi}\cosh \gamma \,  a_1 + e^{-\beta - \varphi} a_0
	= a_1
\end{equation}
or
\begin{equation}\label{eqn:EQ}
	a_0
	= (e^{\beta + \varphi} - e^{2\beta}\cosh \gamma) a_1.
\end{equation}
Similarly, one finds for the coefficient of $\tau_3$
\begin{alignedeqn}\label{eqn:ER}
	\sinh \gamma \, a_0
	&= (e^{\varphi - \beta} - \cosh \gamma) a_3\\
	&= (e^{\beta + \varphi}\sinh \gamma - e^{2\beta} \sinh \gamma \cosh\gamma) a_1,
\end{alignedeqn}
where the second relation employs \cref{eqn:EO} and coincides with \cref{eqn:EQ}. Together with the coefficient of $\tau_0$ we are left with three independent equations for the four variables $a_0,a_1,a_3$ and $\varphi$.

The relation $SS_\ast = S_\ast$ is linear in $S_\ast$ and can therefore not determine the overall normalization of the coefficients $a_\mu$. For the normalization we employ the knowledge that $S_\ast$ has one eigenvalue one and the other eigenvalue zero. This requires the additional condition
\begin{equation}\label{eqn:ES}
	\Tr S_\ast
	= 1.
\end{equation}
With 
\begin{equation}\label{eqn:ET}
	a_0
	= \frac{1}{2}
\end{equation}
one finds
\begin{equation}\label{eqn:EU}
	a_1
	= \pm 
	\frac{e^{-\beta}}{2 \csqrt{e^{-2\beta} + \sinh^2\gamma \, e^{2\beta}}}
\end{equation}
and 
\begin{alignedeqn}\label{eqn:EV}
	e^\varphi
	&= e^\beta\cosh \gamma + \frac{e^{-\beta}}{2a_1}\\
	&= e^\beta\cosh \gamma \pm \csqrt{e^{-2\beta} + \sinh^2\gamma \,  e^{2\beta}},
\end{alignedeqn}
in accordance with \cref{eqn:EH}. The advantage of the use of the relations \labelcref{eqn:EI,eqn:ES} for more complicated situations is the possibility to exploit directly symmetries of $S$ and $S_\ast$.

\section{Normalized classical wave function and non-linear unitary evolution}
\label{app:non-linear unitary evolution}

If $S_{\tau\rho}$ is not an orthogonal matrix the norm of the wave function, $\sum_\tau \tilde q_\tau^2$, will not be conserved. In this case $\tilde q^2_\tau(t)$ can not be associated with a probability. We have, however, still the relations \labelcref{eqn:87,eqn:90} (for invertible $S$) which yield for the product $\bar q_\tau \tilde q_\tau$ the identity (no sum over $\tau$ here)
\begin{equation}\label{eqn:46F}
	\bar q_\tau(t + \epsilon) \, \tilde q_\tau (t + \epsilon)
	= S^{-1}_{\rho\tau} \, S_{\tau\sigma} \, \bar q_\rho(t) \, \tilde q_\sigma(t).
\end{equation}
If the product $\bar q_\tau \tilde q_\tau$ remains positive for all $\tau$ and all $t$, and if the initial and final wave functions are normalized such that $Z = 1$, we can define probabilities by \cref{eqn:22A}. This allows us to define the normalized classical wave function by
\begin{equation}\label{eqn:46H}
	q_\tau(t)
	= \csqrt{\bar q_\tau(t) \, \tilde q_\tau(t)} \, \sign\bigl(\tilde q_\tau(t)\bigr).
\end{equation}
For the particular case $\bar q = \tilde q$ one has $q = \tilde q$. The normalized classical wave function $q$ obeys
\begin{equation}\label{eqn:155A}
	\sum_\tau q^2_\tau
	= \sum_\tau p_\tau
	= 1.
\end{equation}

Since the length of the vector $\{q_\tau\}$ is fixed for all $t$,
\begin{equation}\label{eqn:46I}
	\sum_\tau \, q_\tau (t)q_\tau (t)
	= \sum_\tau \bar q_\tau(t) \, \tilde q_\tau(t)
	= 1,
\end{equation}
the time evolution is now unitary, with a normalized rotation matrix $R _{\tau\rho}$,
\begin{equation}\label{eqn:46J}
	q_\tau (t + \epsilon)
	= R^{(n)}_{\tau\rho} \, q _{\rho}(t).
\end{equation}
While $R^{(n)}$ is an orthogonal matrix, it is no longer independent of the wave function. What is specific for quantum systems is the linearity of the evolution law, i.e. the fact that $R$ is independent of the wave function. Unitarity of the evolution can be realized in a much wider context. In fact, it always holds if probabilities $p_\tau$ are represented as the squared elements of a real wave function.

Classical Ising-spin systems are examples where $S_{\tau\rho}$ is not orthogonal, but $\bar q_\tau q_\tau$ remains positive for all $\tau \text{ and } t$. \Cref{eqn:46H} applies and we can describe the time evolution by a non-linear rotation $R^{(n)}_{\tau\rho}$ which depends on the wave function. In this sense a generic classical statistical system can be viewed as a non-linear generalization of quantum mechanics where the Hamiltonian depends on the wave function. Again this raises the interesting question under which conditions the evolution of subsystems can become linear and therefore describe quantum systems.

\section{Step evolution operators for simple memory materials}
\label{app:step evolution operators}

In this appendix we discuss properties of step evolution operators for memory materials, concentrating on a small number of local states $N$. We investigate regular positive $N \times N$-matrices with more than one eigenvalue $|\lambda| = 1$, while all other eigenvalues have absolute value smaller than one. We proceed by specifying the eigenvalues and construct positive matrices that realize these eigenvalues. The key restriction arises from the requirement that all matrix elements are positive or vanish. The simplest example for $N = 2$ has eigenvalues $\lambda \in \{1,-1\}$. The only positive matrix obeying this condition is $\tau_1$. This constitutes a unique jump operator.

\subsection{Traceless \texorpdfstring{$4 \times 4$}{4 x 4}-matrices}

For $N = 4$ we first consider $\lambda \in \{a,-a,b,-b\}$, where $a$ or $b$ may be imaginary. From $\Tr S = \sum\lambda_i = 0$ we conclude that positivity requires all diagonal elements to vanish, $S_{ii} = 0$. For this case one finds
\begin{equation}\label{eqn:F1}
	\det (S-\lambda)
	= \det S - B\lambda - A\lambda^2 + \lambda^4,
\end{equation}
with 
\begin{alignedeqn}\label{eqn:F2}
	A
	&= S_{12} \, S_{21} + S_{13} \, S_{31} + S_{14} \, S_{41}\\
	&\hphantom{{}=}+ S_{23} \, S_{32} + S_{24} \, S_{42} + S_{34} \, S_{43},
\end{alignedeqn}
and 
\begin{alignedeqn}\label{eqn:F3}
	B
	&= S_{12} \, S_{23} \, S_{31} + S_{12} \, S_{24} \, S_{41} + S_{13} \, S_{32} \, S_{21}\\
	&\hphantom{{}=}+ S_{13} \, S_{34} \, S_{41} + S_{14} \, S_{42} \, S_{21} + S_{14} \, S_{43} \, S_{31}\\
	&\hphantom{{}=}+ S_{23} \, S_{34} \, S_{42} + S_{24} \, S_{43} \, S_{32}.\\
\end{alignedeqn}
The requirement
\begin{equation}\label{eqn:F4}
	\det (S - \lambda_i)
	= 0
\end{equation}
for all $\lambda_i$ amounts to $(a,b \neq 0)$
\begin{equation}\label{eqn:F5}
	A
	= a^2 + b^2,
	\quad
	B
	= 0.
\end{equation}
From $A^\ast = A$, $A\geq 0$ we conclude that one of the largest eigenvalues must be real and we take $a = 1$. The eigenvalue $b$ may be real or purely imaginary, $b \neq 0$. One therefore has
\begin{equation}\label{eqn:F5A}
	\Det S
	= b^2,
	\quad
	A
	= 1 + b^2.
\end{equation}
For positive $S$ the condition $B = 0$ requires that each one of the eight terms in \cref{eqn:F3} has to vanish.

For $a = 1$, $b = i$ one needs $A = 0$. Each one of the six terms in \cref{eqn:F2} has to vanish. The conditions $A = 0$ and $B = 0$ can be obeyed only if each column and each row has a unique non-zero element. Indeed, without loss of generality we can take $S_{12} \neq 0$ such that $S_{21} = 0$. Then invertibility of $S$ requires that either $S_{31}$ or $S_{41}$ differs from zero, and we take without loss of generality $S_{31} \neq 0$, $S_{13} = 0$. From $B = 0$ one infers $S_{23} = 0$, such that invertibility requires $S_{24} \neq 0$, $S_{42} = 0$. In turn, $B = 0$ now needs $S_{41} = 0$ and therefore $S_{43} \neq 0$, $S_{43} = 0$. Finally, $B = 0$ requires $S_{32} = 0$, $S_{14} = 0$.

With $\det S = -S_{12} \, S_{24} \, S_{43} \, S_{31} = -1$ the normalization of the product of the nonzero elements is fixed. The most general positive $4 \times 4$-matrices with eigenvalues $\lambda = (1,-1,i,-i)$ are therefore characterized by three positive numbers $S_{12},S_{24},S_{43}$, with $S_{31} = (S_{12} \, S_{24} \, S_{43})^{-1}$, or suitable permutations of columns and rows of these matrices. Rotation matrices are realized only for $S_{12} = S_{24} = S_{43} = S_{31} = 1$, and we recover the step evolution operators of the unique jump chains.

We next consider $a = 1$, $-1 < b^2\leq 1$, such that $A = 1 + b^2 > 0$, $\Det S = b^2 > -1$. As a first example, one may achieve $B = 0$ by 
\begin{equation}\label{eqn:F6}
	S_{31}
	= S_{41}
	= S_{32}
	= S_{42}
	= 0,
\end{equation}
such that 
\begin{alignedeqn}\label{eqn:F7}
	A
	&= x + y,
	\quad\Det S
	= xy,\\
	x
	&= S_{12} \, S_{21},
	\quad
	y
	= S_{34} \, S_{43}.
\end{alignedeqn}
With $x + y = 1 + b^2$, $x y = b^2$ one needs $b^2 > 0$ and 
\begin{equation}\label{eqn:F8}
	x + y
	= 1 + xy.
\end{equation}
The solutions require $x = 1$ or $y = 1$. We may take
\begin{equation}\label{eqn:F9}
	S_{12}S_{21}
	= 1,
	\quad
	S_{34}S_{43}
	= b^2.
\end{equation}
The four elements $S_{13}, S_{14}, S_{23}, S_{24}$ remain free.

Another, perhaps more interesting example for $B = 0$ is
\begin{equation}\label{eqn:F10}
	S_{14}
	= S_{23}
	= S_{32}
	= S_{41}
	= 0,
\end{equation}
where
\begin{equation}\label{eqn:F11}
	A
	= S_{12} \, S_{21} + S_{13} \, S_{31} + S_{24} \, S_{42} + S_{34} \, S_{43}
\end{equation}
and 
\begin{alignedeqn}\label{eqn:F12}
	\det S
	&= S_{12} \, S_{21} \, S_{34} \, S_{43} + S_{13} \, S_{31} \, S_{24} \, S_{42}\\
	&- S_{12} \, S_{24} \, S_{43} \, S_{31} - S_{13} \, S_{34} \, S_{42} \, S_{21}.
\end{alignedeqn}
For 
\begin{equation}\label{eqn:F13}
	b^2
	= A - 1
	= \Det S
\end{equation}
one can now achieve positive or negative values. (The previously discussed limit $b^2 = -1$ can be realized for $S_{13} = S_{34} = S_{42} = S_{21} = 0$ or $S_{12} = S_{24} = S_{43} = S_{31} = 0$.) There are plenty of choices of positive step evolution operators that lead to $|b| < 1$. For a particular case we consider 
\begin{alignedeqn}\label{eqn:F14}
	S_{12}
	&= S_{24}
	= S_{43}
	= S_{31}
	= w,\\
	S_{13}
	&= S_{34}
	= S_{42}
	= S_{21}
	= u,
\end{alignedeqn}
where 
\begin{equation}\label{eqn:F15}
	A
	= 4wu,
	\quad\det S
	= -(w^2 - u^2)^2.
\end{equation}
The elements $u$ and $w$ have to obey 
\begin{equation}\label{eqn:F16}
	1 + 2w^2u^2 - u^4 - w^4 - 4wu
	= 0,
\end{equation}
which is solved by
\begin{equation}\label{eqn:F17}
	b
	= i(1 - 2u).
\end{equation}
(The degenerate limit $b = 0$ is reached for $u = w = 1/2$.) For small $w\ll u$ the setting is close to the unique jump chain \labelcref{eqn:GN}. The eigenvectors to the eigenvalues $\lambda = 1,-1,i(1 - 2u)$, $-i(1 - 2u)$ are
\begin{equation}\label{eqn:F18}
	\begin{pmatrix} 1\\1\\1\\1 \end{pmatrix},
	\quad
	\begin{pmatrix} 1\\-1\\-1\\1 \end{pmatrix},
	\quad
	\begin{pmatrix} 1\\i\\-i\\-1 \end{pmatrix},
	\quad
	\begin{pmatrix} 1\\-i\\i\\-1 \end{pmatrix}.
\end{equation}

The examples \labelcref{eqn:F6} or \labelcref{eqn:F10}, with the restrictions on the non-zero elements as discussed above, realize all the eigenvalues $\lambda = (1,-1,b,-b)$. This demonstrates that many $4 \times 4$-matrices can have two different largest eigenvalues $\lambda = \pm 1$. These matrices do not need to be unique jump operators.

\subsection{Matrices close to unique jump operators}

We may investigate more generally the $4 \times 4$-matrices close to the unique jump operators with eigenvalues
\begin{equation}\label{eqn:EB1}
	\lambda
	\in \{1,-1,b_1,b_2\},
	\quad|b_1|
	\leq 1,
	\quad|b_2|
	\leq 1.
\end{equation}
For this purpose, we make the ansatz
\begin{equation}\label{eqn:EB2}
	S
	= \hat S + T,
\end{equation}
with 
\begin{equation}\label{eqn:EB3}
	\hat S_{13}
	= \hat S_{34}
	= \hat S_{42}
	= \hat S_{21}
	= 1,
\end{equation}
and $\hat S_{ij} = 0$ otherwise. The matrix $T$ is considered to be small -- for $T = 0$ the step evolution operator $S$ is a unique jump operator. An expansion linear in $T$ yields 
\begin{alignedeqn}\label{eqn:EB4}
	\det S
	&= -1 - \tr(\hat S^{-1} \, T)\\
	&= -1 - (T_{13} + T_{34} + T_{42} + T_{21})\\
	&= -b_1 \, b_2,
\end{alignedeqn}
while the quantities $A$ and $B$ in \cref{eqn:F2,eqn:F3} read
\begin{alignedeqn}\label{eqn:EB5}
	A
	&= T_{31} + T_{43} + T_{24} + T_{12},\\
	B
	&= T_{14} + T_{41} + T_{23} + T_{32}.
\end{alignedeqn}
We also have
\begin{equation}\label{eqn:EB6}
	D
	= \tr S
	= T_{11} + T_{22} + T_{33} + T_{44}
	= b_1 + b_2.
\end{equation}

Since $S$ is close to $\hat S$, the eigenvalues of $S$ have also to be close to the ones of $\hat S$, and we take
\begin{equation}\label{eqn:EB7}
	b_1
	= i + c_1,
	\quad
	b_2
	= -i + c_2,
\end{equation}
and therefore (in lowest order)
\begin{equation}\label{eqn:EB8}
	b_1 + b_2
	= c_1 + c_2,
	\quad
	b_1b_2
	= 1 - i (c_1 - c_2).
\end{equation}
This yields the relations
\begin{equation}\label{eqn:EB9}
	D
	= T_{11} + T_{22} + T_{33} + T_{44}
	= c_1 + c_2,
\end{equation}
and 
\begin{equation}\label{eqn:EB10}
	E
	= T_{13} + T_{34} + T_{42} + T_{21}
	= -i (c_1 - c_2).
\end{equation}
One concludes that $c_1 + c_2$ is real while $c_1 - c_2$ is purely imaginary. This yields 
\begin{equation}\label{eqn:EB11}
	b_1
	= i + \frac{1}{2} \, D + \frac{i}{2} \, E,
	\quad
	b_2
	= -i + \frac{1}{2} \, D - \frac{i}{2} \, E.
\end{equation}
From 
\begin{equation}\label{eqn:EB13}
	|b_1|^2
	= |b_2|^2
	= \Bigl(1 + \tfrac{E}{2}\Bigr)^2 + \frac14 D^2
	= 1 + E\leq 1
\end{equation}
we infer
\begin{equation}\label{eqn:EB14}
	E\leq 0.
\end{equation}
(Note that the sign of $T_{13}$, $T_{34}$, $T_{42}$ and $T_{21}$ is not fixed for positive $S$, since these elements add to the ones in \cref{eqn:EB3}.)

The condition $\det(S - \lambda) = 0$ for the four eigenvalues $\lambda \in \bigl\{+1,-1,i (1 + E/2),-i (1 + E/2)\bigr\}$ yields constraints on the matrix $T$. In linear order in $T$ one has
\begin{equation}\label{eqn:EB15}
	\det(S - \lambda)
	= \det S - B \, \lambda - A \, \lambda^2 - D \, \lambda^3 + \lambda^4.
\end{equation}
The conditions $\det(S - \lambda) = 0$ are obeyed for 
\begin{equation}\label{eqn:EB16}
	B + D
	= 0,
	\quad
	A
	= 1 + \det S
	= -E.
\end{equation}
In linear order in $T$ these are the only conditions for $S$ to have two eigenvalues $\lambda = 1$ and $\lambda = -1$. One finds a large number of matrices that differ from unique jump operators only in order $T$ and maintain two largest eigenvalues $\lambda = \pm 1$. The matrices \labelcref{eqn:F14} are part of this family, with $T_{12} = T_{24} = T_{43} = T_{31} = w$, $T_{13} = T_{34} = T_{42} = T_{21} = -w$.

\subsection{Unitary subsystems}

For $E < 0$ the asymptotic evolution in the bulk proceeds in the eigenspace of the eigenvalues $\lambda = \pm 1$. The eigenvectors to the other eigenvalues approach zero asymptotically. We are interested to construct the effective two-state system corresponding to the two largest eigenvalues.

Let us first compute the eigenvectors to the two largest eigenvalues. For $\lambda = \pm 1$ one finds to lowest order, respectively,
\begin{alignedeqn}\label{eqn:EB18}
	v_+
	&= \begin{pmatrix}
		1\\
		1 + x_2\\
		1 + x_2 + x_3 + x_4\\
		1 + x_2 + x_4
	\end{pmatrix},\\
	v_-
	&= \begin{pmatrix}
		1\\
		-(1 + y_2)\\
		-(1 + y_2 + y_3 + y_4)\\
		1 + y_2 + y_4
	\end{pmatrix}
\end{alignedeqn}
where
\begin{alignedeqn}\label{eqn:EB19}
	x_i
	&= \sum_j T_{ij},\\
	y_2
	&= T_{21} - T_{22} - T_{23} + T_{24},\\
	y_3
	&= T_{31} - T_{32} - T_{33} + T_{34},\\
	y_4
	&= -(T_{41} - T_{42} - T_{43} + T_{44}).
\end{alignedeqn}
The asymptotic wave function will be a linear combination of $v_+$ and $v_-$,
\begin{equation}\label{eqn:EB20}
	\tilde q_\ast(t)
	= w_+(t) \, v_+ + w_-(t) \, v_-.
\end{equation}
The four-component wave function will therefore be reduced to an effective two-component wave function $q_r(t)$.

We are interested in the evolution of this subsystem and want to construct an effective $2\times 2$ step evolution operator $S_r$, such that 
\begin{equation}\label{eqn:EB21}
	\tilde q_r(t + \epsilon)
	= S_r \, \tilde q_r(t).
\end{equation}
For this purpose we first have to specify a suitable basis for the reduced system. Let us assume that we are mainly interested in observables that can be computed from the first two components $\tilde q_1$ and $\tilde q_2$ of the wave function. It will then be useful to define the reduced system such that 
\begin{equation}\label{eqn:EB22}
	\tilde q_{r,1}(t)
	= \tilde q_1(t),
	\quad\tilde q_{r,2}(t)
	= \tilde q_2(t).
\end{equation}
The other two components $\tilde q_3(t)$ and $\tilde q_4(t)$ can be expressed by \cref{eqn:EB20} in terms of $\tilde q_1(t)$ and $\tilde q_2(t)$. This procedure will define the reduced step evolution operator $S_r$.

We first express $\tilde q_{\ast1}$ and $\tilde q_{\ast2}$ in terms of $w_+$ and $w_-$
\begin{equation}\label{eqn:EB23}
	\tilde q_{\ast1}
	= w_+ + w_-,
	\quad\tilde q_{\ast2}
	= (1 + x_2) \, w_+ - (1 + y_2) \, w_-.
\end{equation}
This relates $\tilde q_r$ and $w_\pm$,
\begin{equation}\label{eqn:EB24}
	\tilde q_r
	= Qw,
	\quad
	w= \begin{pmatrix}
		w_+\\
		w_-
	\end{pmatrix},
	\quad
	Q
	= \begin{pmatrix}
		1 & 1\\
		1 + x_2 & -1 - y_2
	\end{pmatrix}.
\end{equation}
We can invert \cref{eqn:EB24},
\begin{equation}\label{eqn:EB25}
	w
	= Q^{-1}\tilde q_r,
	\quad
	Q^{-1}
	= \frac{1}{2} \begin{pmatrix}
		1 - \frac{x_2 - y_2}{2} & 1 - \frac{x_2 + y_2}{2}\\
		1 + \frac{x_2 - y_2}{2} & -1 + \frac{x_2 + y_2}{2}
	\end{pmatrix},
\end{equation}
or
\begin{alignedeqn}\label{eqn:EB26}
	w_+
	&= \frac{1}{2}\Bigl\{\Bigl(1 - \tfrac{x_2}{2}\Bigr)(\tilde q_{r_1} + \tilde q_{r_2}) + \frac{y_2}{2}
	(\tilde q_{r1} - \tilde q_{r2})\Bigr\},\\
	w_-
	&= \frac{1}{2}\Bigl\{\Bigl(1 - \tfrac{y_2}{2}\Bigr)
	(\tilde q_{r1} - \tilde q_{r2}) + \frac{x_2}{2}(\tilde q_{r1} + \tilde q_{r2})\Bigr\}.\\
\end{alignedeqn}
From \cref{eqn:EB20} we can express $\tilde q_{\ast3}$ and $\tilde q_{\ast4}$ in terms of $w_\pm$ and therefore in terms of $\tilde q_{r1}$ and $\tilde q_{r2}$,
\begin{alignedeqn}\label{eqn:EB27}
	\tilde q_{3\ast}
	&= \frac{1}{2}\bigl\{(x_3 + x_4 - y_3 - y_4)\tilde q_{r1}\\
	&\hphantom{{}=} + (2 + x_3 + x_4 + y_3 + y_4)\tilde q_{r2}\bigr\},\\[1ex]
	\tilde q_{4\ast}
	&= \frac{1}{2}\bigl\{(2 + x_2 + x_4 + y_2 + y_4)\tilde q_{r1}\\
	&\hphantom{{}=} + (x_2 + x_4 - y_2 - y_4)\tilde q_{r2}\bigr\}
\end{alignedeqn}

These results can be inserted into the evolution equation for $\tilde q_r$
\begin{alignedeqn}\label{eqn:EB28}
	\tilde q_{r1}(t + \epsilon)
	&= T_{11}\tilde q_{r1}(t) + T_{12} \, \tilde q_{r2}(t)\\
	&\hphantom{{}=} + (1 + T_{13}) \, \tilde q_{\ast3}(t) + T_{14} \, \tilde q_{\ast4}(t),\\[1ex]
	\tilde q_{r2}(t + \epsilon)
	&= (1 + T_{21}) \, \tilde q_{r1}(t) + T_{22} \, \tilde q_{r2}(t)\\
	&\hphantom{{}=} + T_{23} \, \tilde q_{\ast3}(t) + T_{24}(t) \, \tilde q_{\ast4}(t).
\end{alignedeqn}
One extracts the step evolution operator $S_r$ of the subsystem in \cref{eqn:EB21}, where
\begin{equation}\label{eqn:E29}
	S_r
	= \begin{pmatrix}
		T_{11} + T_{14} + \gamma_- & 1 + T_{12} + T_{13} + \gamma_+\\
		1 + T_{21} + T_{24} & T_{22} + T_{23}
	\end{pmatrix}
\end{equation}
with 
\begin{alignedeqn}\label{eqn:E30}
	&\gamma_\pm
	= \frac{1}{2}\bigl[x_3 + x_4 \pm (y_3 + y_4)\bigr],\\
	&\gamma_+
	= T_{31} + T_{34} + T_{42} + T_{43},\\
	&\gamma_-
	= T_{32} + T_{33} + T_{41} + T_{44}.
\end{alignedeqn}

We can write $S_r$ in the form 
\begin{equation}\label{eqn:E30a}
	S_r
	= \tau_1 + V,
\end{equation}
where 
\begin{alignedeqn}\label{eqn:E31}
	V_{11}
	&= T_{11} + T_{33} + T_{44} + T_{14} + T_{41} + T_{32},\\
	V_{12}
	&= T_{12} + T_{13} + T_{31} + T_{34} + T_{42} + T_{43},\\
	V_{21}
	&= T_{21} + T_{24},\\
	V_{22}
	&= T_{22} + T_{23}.
\end{alignedeqn}
We observe
\begin{equation}\label{eqn:E32}
	V_{12} + V_{21}
	= A + E
	= 0
\end{equation}
and 
\begin{equation}\label{eqn:EB33}
	V_{11} + V_{22}
	= B + D
	= 0.
\end{equation}
These relations ensure that the eigenvalues of $S_r$ are $\lambda = \pm 1$ and $S^2_r = 1$.

For the case of a positive matrix $S$ the condition $B + D = 0$ requires
\begin{alignedeqn}\label{eqn:EB34}
	T_{11}
	&= T_{22}
	= T_{33}
	= T_{44}
	= T_{14}\\
	&= T_{41}
	= T_{23}
	= T_{32}
	= 0.
\end{alignedeqn}
In this case the diagonal elements of $V$ vanish. Up to rescalings the step evolution operator of the reduced system is a unique jump operator. This is perhaps not surprising since the only positive $2 \times 2$-matrix with eigenvalues $\pm 1$ is given by
\begin{equation}\label{eqn:EB35}
	S_r
	= \begin{pmatrix}
		0 & 1/a\\
		a & 0
	\end{pmatrix}.
\end{equation}

The precise form of the reduced step evolution operator $S_r$ depends on the basis that we choose for parametrizing the degrees of freedom of the subsystem. A change of basis for the subsystem results in a transformation $S_r \to S^\prime_r = DS_rD^{-1}$. For example, if we choose the eigenfunctions $w_+$ and $w_-$ as basis functions the reduced step evolution operator becomes $S^\prime_r = \tau_3$. This demonstrates that the reduced step evolution operator no longer needs to be positive.
\end{appendices}
\end{multicols}

\begin{multicols}{2}[{\printbibheading[title={References}]}]
	\printbibliography[heading=none]
\end{multicols}

\end{document}